\documentclass[11pt]{article}

\usepackage{jheppub}
\usepackage[usenames,dvipsnames,table]{xcolor}
\usepackage{graphicx,amsmath,amssymb,amsthm,multirow,array,bm,bbm,esint,slashed,hyperref}
\usepackage{xfrac,faktor}
\usepackage[mathscr]{eucal}
\usepackage[bbgreekl]{mathbbol}
\usepackage{subfig}
\usepackage{comment}
\usepackage{soul}

\newcommand\beq{\begin{equation}}
\newcommand\eeq{\end{equation}}

\title{
	AdS$_3$ gravity and random CFT
	}

\author[a]{Jordan Cotler}
\author[b]{and Kristan Jensen}
\affiliation[a]{Stanford Institute for Theoretical Physics, Stanford University, Stanford, CA 94305, USA}
\affiliation[b]{Department of Physics and Astronomy, San Francisco State University, San Francisco, CA 94132, USA}

\emailAdd{jcotler@stanford.edu}
\emailAdd{kristanj@sfsu.edu}

\abstract{
We compute the path integral of three-dimensional gravity with negative cosmological constant on spaces which are topologically a torus times an interval. These are Euclidean wormholes, which smoothly interpolate between two asymptotically Euclidean AdS$_3$ regions with torus boundary. From our results we obtain the spectral correlations between BTZ black hole microstates near threshold, as well as extract the spectral form factor at fixed momentum, which has linear growth in time with small fluctuations around it. The low-energy limit of these correlations is precisely that of a double-scaled random matrix ensemble with Virasoro symmetry. Our findings suggest that if pure three-dimensional gravity has a holographic dual, then the dual is an ensemble which generalizes random matrix theory.}

\begin{document}

\maketitle

\section{Introduction}

There has long been hope that simple models of quantum gravity with negative cosmological constant in less than four spacetime dimensions exist as consistent quantum mechanical models. It has recently been demonstrated that Jackiw-Teitelboim (JT) gravity~\cite{Jackiw:1984je, Teitelboim:1983ux, Jensen:2016pah, Maldacena:2016upp, Engelsoy:2016xyb, Stanford:2017thb}, a two-dimensional model of dilaton gravity, is just such a consistent theory~\cite{Saad:2019lba}. Moreover, as one expects for consistent theories of AdS quantum gravity, it has a holographic dual. Unlike standard examples of AdS/CFT, the holographic dual is not a quantum mechanical system in one lower dimension. Rather its dual is a statistical ensemble of large random Hermitian matrices $H$ \cite{Saad:2019lba}.

The holographic dictionary equates the $n$-point ensemble average of $\text{tr}(e^{-\beta_i H})$ in the dual matrix model to the JT path integral with $n$ boundaries of renormalized lengths $\beta_i$: 
\begin{equation*}
\includegraphics[scale=.16]{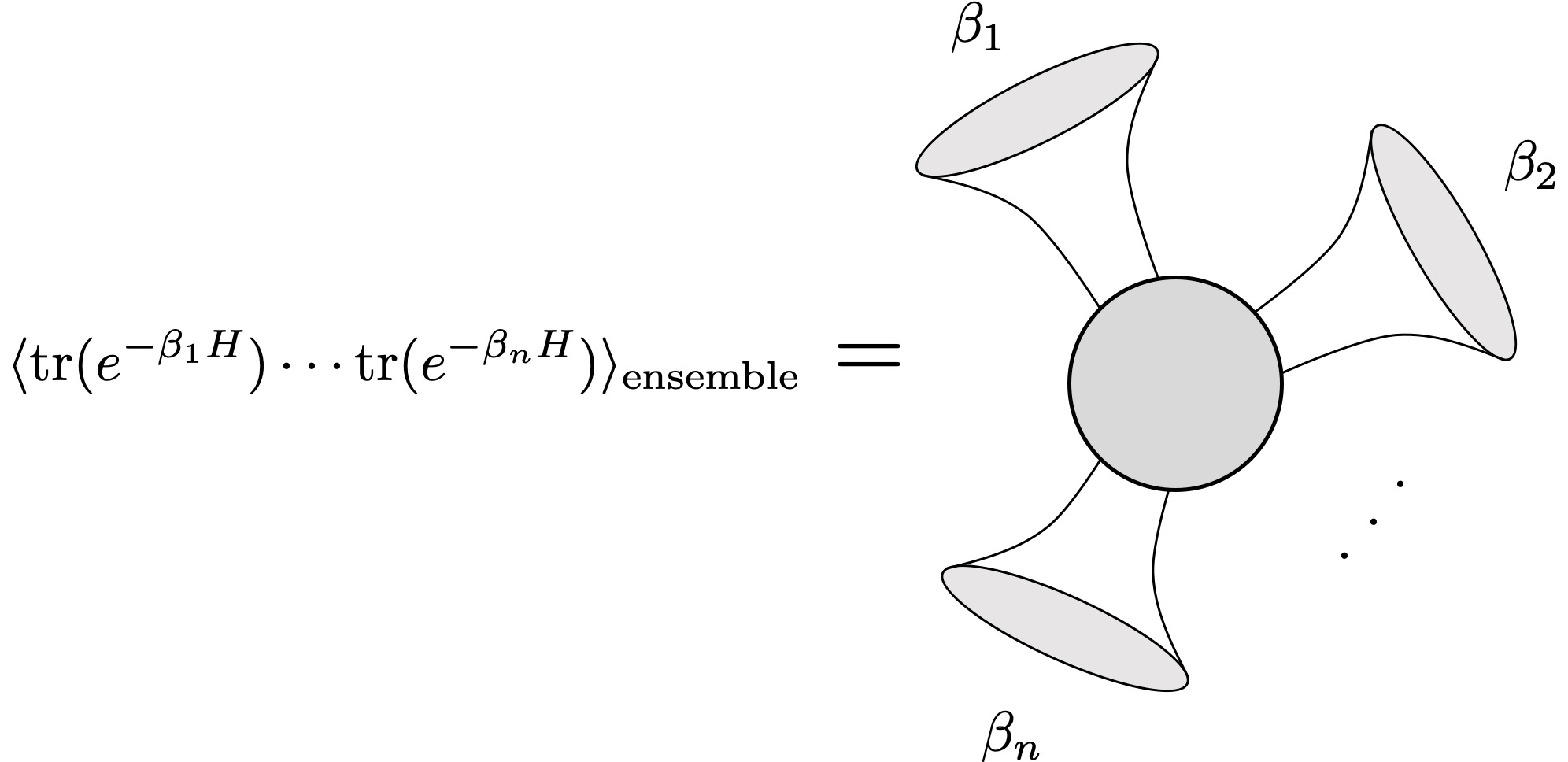}
\end{equation*}
The solid disk indicates a sum over all geometries which fill in the boundary circles. The right-hand side is known only as an asymptotic series in a genus expansion with effective string coupling $g_s$, while the left-hand side also has a genus expansion. Thanks to the topological recursion of matrix integrals~\cite{mirzakhani2007simple, mirzakhani2007weil, eynard2007invariants, eynard2007weil}, Saad, Shenker, and Stanford \cite{Saad:2019lba} have demonstrated that this equation holds to all orders in the genus expansion and for all $n$, and it is in this sense that JT gravity is dual to a certain matrix ensemble.

A striking feature of JT gravity is that it includes a sum over wormhole geometries which smoothly connect multiple asymptotic regions \cite{Saad:2018bqo, Saad:2019lba}. For example, in the genus expansion, the path integral with two boundaries is in pictures given by
\begin{equation*}
\includegraphics[width = \textwidth]{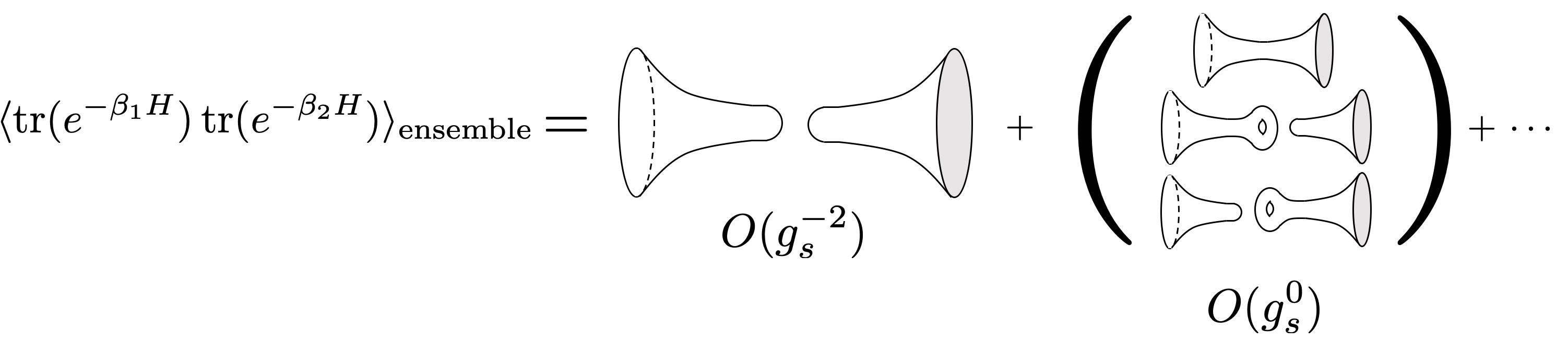}
\end{equation*}
The leading contribution of $O(g_s^{-2})$ comes from two disconnected disks, while at $O(g_s^0)$ both disconnected geometries \emph{and} connected ones contribute. In the matrix model this sum equals the complete two-point function $\langle \text{tr}\left(e^{-\beta_1 H}\right)\text{tr}\left(e^{-\beta_2 H}\right)\rangle_{\rm ensemble}$\,. Its connected part is, in the genus expansion, the sum over geometries which connect the two boundaries, and is dominated by the leading connected geometry,
\begin{equation*}
\includegraphics[scale = .17]{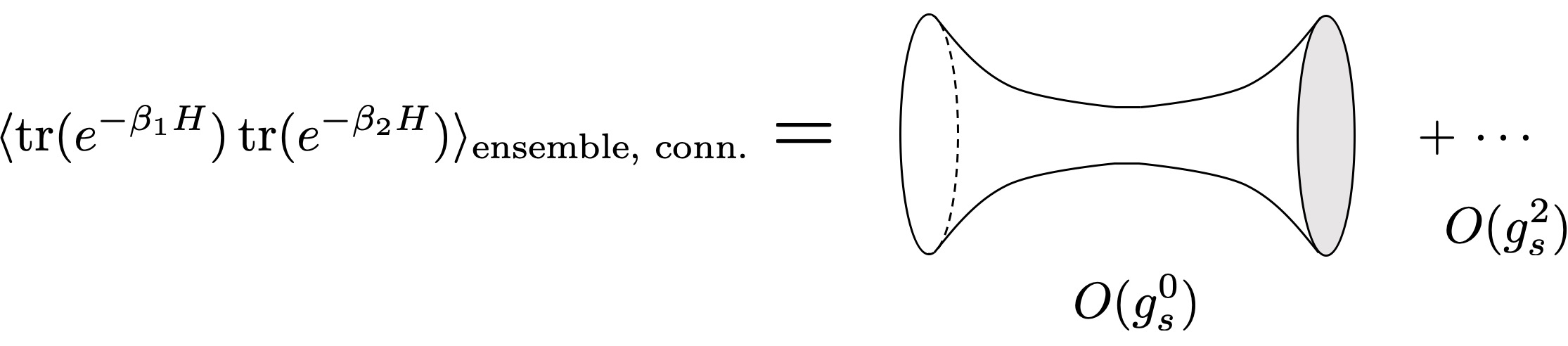}
\end{equation*}
 that is, by a Euclidean wormhole. 
 
There is a long history dating back to Coleman concerning Euclidean wormholes, whether or not they exist, and if they do what they imply for unitarity of Lorentzian quantum gravity \cite{coleman1988black, giddings1988loss, giddings1989baby}. In the context of AdS/CFT there is an additional complication, namely a tension between the existence of wormholes and the standard holographic dictionary.\footnote{It is worth noting that all but one example of Euclidean wormholes in string theory are known~\cite{maldacena2004wormholes,hertog2019euclidean} to be unstable (the only exception we know of is the AdS$_5$ wormhole of~\cite{maldacena2004wormholes}), so concerns about this tension may ultimately be much ado about nothing in standard AdS/CFT. }  Under that dictionary, the dual to gravity on a Euclidean wormhole would be a local Euclidean conformal field theory (CFT) on a disconnected space, while the connectedness of the bulk implies that dual CFT correlation functions do not factorize across the components. But these two statements contradict each other. However, in JT gravity we do not have the standard holographic dictionary: the dual description is not a single quantum mechanical theory, but an ensemble thereof (regarding $H$ as a quantum mechanical Hamiltonian), which is enough to alleviate this tension, as correlations across boundaries are induced by an ensemble average rather than nonlocal interactions. The insertion of a nearly AdS$_2$ boundary is associated with the insertion of $\text{tr}\left( e^{-\beta H}\right)$, i.e.~the thermal partition function $Z(\beta)$, into the dual matrix integral. As such it is perfectly consistent that the connected two-point function of $Z(\beta)$, the very thing that tells us that $Z(\beta)$ is not a $c$-number but a random variable, is equated in JT with a sum over wormholes that signal an ensemble average in the dual description.

The study of JT gravity has proven fruitful from a variety of perspectives, ranging from the resolution of old puzzles regarding the lack of decoupling of nearly extremal black holes in string theory \cite{Jensen:2016pah, nayak2018dynamics} to the Sachdev-Ye-Kitaev model \cite{kitaev, Maldacena:2016hyu, Kitaev:2017awl} to even toy models for de Sitter quantum gravity \cite{maldacena2019two, cotler2019low, cotler2019emergent} (see also \cite{anninos2019sitter}). In this article we turn instead to pure quantum gravity in three dimensions, which has long been a workhorse in the study of simple models of gravity. 

Classical ``AdS$_3$ gravity'' is an exactly soluble system: all solutions to Einstein's equations are locally AdS$_3$ and so one can find the phase space of classical solutions \cite{Witten:1988hc}. The classical model has no propagating degrees of freedom, but it does have edge modes (generated by large diffeomorphisms) with marginal interactions, in addition to moduli and a sum over topologies. However it is not yet known if the model is truly consistent or not.

Some years ago, Maloney and Witten~\cite{Maloney:2007ud} computed the leading contribution to the torus partition function of AdS$_3$ gravity, meaning the gravity path integral over metrics whose conformal boundary is a torus of complex structure $\tau$. Strictly speaking they did not compute the complete path integral; rather they summed over saddle points of this kind, and metrics continuously connected to them. These geometries all have the topology of a disk times a circle, equivalently solid tori.

The density of states unearthed by the Maloney and Witten is not that of a unitary, compact dual CFT. Their result has a large spectral gap between the vacuum and the BTZ threshold, above which the spectrum is continuous. More worrisome, the putative density of states is negative very near its spectral edge~\cite{Maloney:2007ud, keller2015poincare, benjamin2019light, alday2019rademacher}.  For this reason it has generally been concluded that AdS$_3$ gravity is not dual to a two-dimensional CFT, and perhaps is not a consistent theory of gravity.

This conclusion is perhaps premature. The sum over spacetimes which asymptote to a torus includes more than a sum over solid tori, and perhaps these non-saddle point contributions render the AdS$_3$ density of states non-negative~\cite{MTWIP}.  And while a compact CFT does not have a continuous density of states by definition, perhaps we should interpret this as a clue rather than a problem. That is, perhaps AdS$_3$ gravity is a consistent theory of gravity, dual not to a CFT but instead to a statistical ensemble, as in the holographic duality between JT gravity and a matrix model.

The point of this paper is to investigate this possibility. We do so by studying 3d gravity on spacetimes that are topologically a torus times an interval. These are Euclidean wormholes that smoothly connect two asymptotic regions with torus boundary. This is the simplest setting after solid tori for which we can reliably compute the gravitational path integral. What is this result dual to? The natural guess is that it is a contribution to the connected two-point function of torus partition functions,
\beq
\label{E:twoPoint}
	\langle Z(\tau_1)Z(\tau_2)\rangle_{\rm conn} = Z_{\mathbb{T}^2\times I}(\tau_1,\tau_2) + \cdots\,,
\eeq
which would in general be a sum over geometries which smoothly connect the two boundaries. On the left-hand side, we have allowed for brackets indicating a suitable ensemble average. If AdS$_3$ gravity is dual to a single theory rather than a ensemble, then the torus partition function is a $c$-number and the right-hand side vanishes. Conversely, if the right-hand side is nonzero, then we infer that the dual, if it exists, is some sort of statistical ensemble.

We exactly compute the contribution from the torus times an interval. It is
\beq
\label{E:mainResult}
	Z_{\mathbb{T}^2\times I}(\tau_1,\tau_2) = \frac{1}{2\pi^2}\,Z_0(\tau_1)Z_0(\tau_2) \sum_{\gamma\in PSL(2;\mathbb{Z})} \frac{\text{Im}(\tau_1)\text{Im}(\gamma \tau_2)}{|\tau_1+\gamma\tau_2|^2}\,, \quad Z_0(\tau) = \frac{1}{\sqrt{\text{Im}(\tau)}\,|\eta(\tau)|^2}\,.
\eeq
Here $Z_0(\tau)$ is the partition function of a non-compact boson with $\eta(\tau)=q^{1/24} \prod_{n=1}^{\infty}(1-q^n)$  the Dedekind eta function, $q=e^{2\pi i \tau}$, and $\gamma$ is a modular transformation which acts as $\gamma \tau = \frac{a\tau+b}{c\tau+d}$ for $ad-bc = 1$.
 
Unlike JT gravity, it is not known if AdS$_3$ gravity has a coupling constant which suppresses fluctuations of topology. As such we do not presently have a principled reason to expect our result to always be a good approximation to the complete sum over connected spacetimes in~\eqref{E:twoPoint}.  However, there is a kinematic limit in which JT gravity, plus a $U(1)$ gauge theory, gives a controlled approximation to AdS$_3$ gravity~\cite{ghosh2019universal}. This is the simultaneous limit of low temperature and fixed, large angular momentum $J$. The effective genus expansion parameter in this regime is $g_s \approx e^{-2\pi \sqrt{ c|J|/6}}$ for $c=3/(2G)$ the Brown-Henneaux central charge. Therefore we expect our result to be a good approximation in that limit.

So the wormhole amplitude~\eqref{E:mainResult} gives a strong piece of evidence that, if AdS$_3$ gravity has a holographic dual, then the dual is a statistical ensemble.

The computation that leads to~\eqref{E:mainResult} is non-trivial. There is no smooth saddle point of AdS$_3$ gravity with the topology $\mathbb{T}^2\times I$, and so the usual program of finding saddles and summing up fluctuations does not work. Indeed, the final result~\eqref{E:mainResult} does not depend on the gravitational coupling, and so has no saddle point approximation. 

Our route to~\eqref{E:mainResult} is therefore somewhat indirect. Our starting point is the first-order formulation of gravity in Lorentzian signature, on the annulus times time. The action of gravity in first-order form has a single time derivative, and so is in Hamiltonian form, meaning that in the quantum theory one is integrating over trajectories in a phase space rather than in a configuration space. Furthermore, the time components of the gravitational variables act as Lagrange multipliers, enforcing Gauss' Law constraints on the spatial components. On the annulus we can solve these constraints exactly, and so we arrive at a constrained system. Quantizing the constrained system results in a Hilbert space of gauge-invariant ``wormhole states.'' We define what we mean by Euclidean quantum gravity by the analytic continuation of this constrained system to imaginary time, and upon compactifying imaginary time, the Euclidean path integral is automatically a trace over this Hilbert space.

The constrained system does not admit wormhole saddles, but it does have the next best thing. There are wormhole configurations which are, in a precise sense, nearly saddle points. These configurations are ``constrained instantons'' in the language of~\cite{Affleck:1980mp}: they extremize the action subject to two further constraints, whereby we fix the energy and momentum encoded in the boundary stress tensor. These configurations are characterized by a spatial metric and spatial spin connection (since in order to arrive at the constrained system we had to integrate out the time components of the gravitational fields).  They are the dominant off-shell configurations of 3d gravity which contribute to the $\mathbb{T}^2 \times I$ amplitude.  See~\cite{cotler2020gravitational} for further discussion on constrained instantons in the context of Einstein gravity.

Of course we want to do more than find special field configurations: we want the complete path integral. To evaluate it we use the machinery of our earlier work~\cite{Cotler:2018zff}. After some simplification, the configurations that solve the Gauss Law constraints are characterized by four moduli and four chiral edge modes, two on each boundary. These edge modes are, at fixed Euclidean time, reparameterizations of a circle, more precisely elements of the quotient space $\faktor{\text{Diff}(\mathbb{S}^1)}{U(1)}$, and may be thought of as large gauge transformations. These fields are weighted by an action, previously studied by Alekseev and Shatashvili in a rather different context in the 80's \cite{Alekseev:1988ce}. This action, which we call Alekseev-Shatashvili theory, is a two-dimensional version of the Schwarzian action which weights the large diffeomorphisms of JT gravity. The integrals over these Alekseev-Shatashvili modes are one-loop exact, and the ensuing results are stitched together by a moduli space integral. The measure on this moduli space is determined by the symplectic structure of the constrained phase space. We arrive at the summand of~\eqref{E:mainResult} by performing the moduli space integral.

This procedure closely mirrors the construction of the ``double trumpet'' path integral in JT gravity in~\cite{Saad:2019lba}, in which two Schwarzian path integrals are stitched together by a moduli space integral.  In AdS$_3$, the analogue of the ``trumpet'' in JT gravity is two copies of the Alekseev-Shatashvili model.

The computation we just outlined is what we find for configurations where the spatial and temporal circles on one boundary interpolate through the bulk to the spatial and temporal circles on the other boundary. There are an infinite number of equivalent but topologically distinct configurations where, say, the spatial circle on one boundary interpolates to the temporal circle on the other. Summing over these contributions results in the modular sum in~\eqref{E:mainResult}.

Having outlined how we find the wormhole amplitude, let us describe a few of its basic properties. It is easy to verify that it is invariant under independent modular transformations $\tau_1 \to \frac{a_1\tau_1+b_1}{c_1\tau_1+d_1}$ and $\tau_2\to \frac{a_2\tau_2+b_2}{c_2\tau_2+d_2}$. As for the modular sum, it is logarithmically divergent. Fortunately the divergence is rather simple: it is an additive constant, independent of $\tau_1$ and $\tau_2$. Thus, as long as we study the dependence of $Z_{\mathbb{T}^2\times I}$ on its arguments rather than its absolute value, we are on firm footing. 

AdS$_3$ gravity is equipped with two copies of the Virasoro algebra that act as asymptotic symmetries on each boundary. The wormhole partition function is organized into contributions from Virasoro primaries and their descendants.  The contribution from the primaries $Z^P$ comes from taking the complete result~\eqref{E:mainResult} and stripping off the infinite products in the Dedekind eta functions,
\beq
	Z^P(\tau_1,\tau_2) = \frac{1}{2\pi^2\sqrt{\text{Im}(\tau_1)\text{Im}(\tau_2)}}\sum_{\gamma\in PSL(2;\mathbb{Z})}\frac{\text{Im}(\tau_1)\text{Im}(\gamma\tau_2)}{|\tau_1+\gamma\tau_2|^2}\,.
\eeq
It may be further decomposed into contributions from states of fixed spins on $s_1$ and $s_2$ on the two boundaries by Fourier transforming with respect to the real parts of $\tau_1$ and $\tau_2$.  At fixed spin and low temperature we find
\beq
\label{E:preRMT}
	Z^P_{s_1,s_2}(\beta_1,\beta_2) = \frac{1}{2\pi} \frac{\sqrt{\beta_1\beta_2}}{\beta_1+\beta_2}  \,e^{- E_{s_1} \beta_1-E_{s_2}\beta_2}\left(  \delta_{s_1,s_2} + O\left( \frac{1}{\beta}\right)\right)\,, \quad E_s =2\pi\left(  |s| - \frac{1}{12}\right)\,.
\eeq
Here $\text{Im}(\tau_1) = \beta_1$ and $\text{Im}(\tau_2) = \beta_2$, and we are studying the leading behavior as $\beta_1,\beta_2\to\infty$ with the ratio $\beta_1/\beta_2$ fixed. $E_s$ is the minimum energy of a black hole of spin $s$. 

In fact this limiting behavior is precisely related to random matrix theory. The spectrum of Virasoro primaries in two-dimensional CFT on the torus is characterized by a Hamiltonian $H$ and a commuting momentum $P$. One can simultaneously diagonalize the two, in which case $H$ is characterized by Hermitian blocks $H_s$ corresponding to states of fixed spin or momentum $s$. Instead of fixed matrices characterizing the spectrum of a particular CFT, if these blocks were large random matrices in a double-scaling limit, then a universal result in random matrix theory informs us that, after accounting for the threshold energy for black holes of fixed spin $s$, we would have
\beq
	\left\langle \text{tr}\left( e^{-\beta_1 H_{s_1}}\right)\text{tr}\left( e^{-\beta_2H_{s_2}}\right)\right\rangle_{\text{ensemble, conn.}} = \frac{1}{2\pi} \frac{\sqrt{\beta_1\beta_2}}{\beta_1+\beta_2}  \, e^{- E_{s_1} \beta_1-E_{s_2}\beta_2}\,\delta_{s_1,s_2} + \hdots\,,
\eeq
where the dots indicate genus corrections. This is of course exactly the leading low temperature limit of our wormhole amplitude in~\eqref{E:preRMT}.

Now consider the full amplitude, restoring the infinite products in the Dedekind eta functions. The full 2-point spectral statistics near threshold is determined by those of the primaries due to the Virasoro symmetry. These statistics match exactly with those expected from a random matrix theory ansatz with Virasoro symmetry, which we discuss later.

So at least in this limit, our AdS$_3$ results are related to random matrix theory. The low-temperature limit zooms in on the low-energy limit. In gravity the low-energy states are the microstates of BTZ black holes of fixed spin near threshold. Another way of stating our result is that near threshold (and to the extent that our result is a good approximation to the complete gravity answer), the 2-point fluctuation statistics of these black hole microstates is described by double-scaled random matrix theory.

In fact, from~\eqref{E:twoPoint} and~\eqref{E:preRMT} we may extract a spectral form factor 
\beq
	\langle Z^P_{s_1}(\beta + i T)Z^P_{s_2}(\beta-i T)\rangle_{\rm conn} = \frac{T}{4\pi \beta}\,e^{-E_{s_1}\beta_1-E_{s_2}\beta_2}\,\delta_{s_1,s_2} + O(T^{-1})\,,
\eeq
which grows linearly at late Lorentzian time $T$ and fixed spin. This ``ramp'' is, in random matrix theory, a consequence of eigenvalue repulsion, and it leads to linear growth with a universal slope that exactly matches our gravitational result \cite{Cotler:2016fpe, Saad:2018bqo, Saad:2019lba, eynard2015random}.

While our AdS$_3$ computation is related to random matrix theory, let us be clear: we are not claiming that AdS$_3$ gravity is dual to an ensemble of random matrices. It seems more likely to us that AdS$_3$ gravity is dual to an ensemble which generalizes random matrix theory (and in particular incorporates modular invariance), from which we may sample CFT partition functions. We refer to this ensemble, whatever it may ultimately be, as ``random CFT.'' (For very recent studies of ensembles of free CFT see~\cite{1800406,1800422}.) 

The remainder of this manuscript is organized as follows. In Section~\ref{S:preliminaries} we discuss some basic features of Hamiltonian, or phase space path integrals, as well as set up the basics of AdS$_3$ gravity in first-order form. The computation of the wormhole path integral in~\eqref{E:mainResult} may be found in Section~\ref{S:wormhole}. We Fourier transform to fixed spin and organize the modular sum at fixed spin as a Poincar\'e series in Section~\ref{S:ModularSum}, and show how this result is related to (and goes beyond) random matrix theory in Section~\ref{S:RMT}. We conclude with a Discussion in Section~\ref{S:discussion}. Some technical results are relegated to the Appendix.

\section{Preliminaries}
\label{S:preliminaries}

In this Section we set up the computation of the path integral of three-dimensional Euclidean gravity on $\mathbb{T}^2\times I$. This computation is rather delicate and requires some preparation before diving in. We begin with the first-order formulation for gravity with negative cosmological constant in Lorentzian signature, which in second-order formalism is described by the action
\beq
\label{E:AdS3action}
	S = \frac{1}{16\pi G}\int d^3x \sqrt{-g}(R+2)\,,
\eeq
up to a boundary term. Here and henceforth we set the AdS radius to unity.

In the first-order formalism, the gravity path integral is in fact a phase space path integral with constraints, analogous to Yang-Mills theory in Hamiltonian form. So we begin in Subsection~\ref{S:phaseSpace} with a review of phase space path integrals, using Yang-Mills and Chern-Simons theory as examples. We then move on to review the first-order formulation of AdS$_3$ gravity in Subsection~\ref{S:firstOrder} and carefully continue to imaginary time in Subsection~\ref{S:gravitycontinuation}. We conclude in Subsection~\ref{S:AS} with a discussion of the edge modes of AdS$_3$ gravity, large diffeomorphisms, whose contributions to the path integral are weighted by an action analogous to the Schwarzian action appearing in JT gravity.

\subsection{Phase space path integrals and ``constrain first'' quantization}
\label{S:phaseSpace}

Phase space or Hamiltonian path integrals with constraints naturally arise in field theory and gravity. By phase space path integral, we mean a model whose action has a single time derivative and one integrates over trajectories in phase space rather than in configuration space. For instance, the quantum mechanics of a particle with position $q^i(t)$ and Lagrangian $L = \frac{m\dot{q}^2}{2}-V(q)$ can equivalently be described in terms of a path integral over $q^i(t)$ and $p_j(t)$ with action $L' = p_i \dot{q}^i - H(p,q)$ with $H(p,q)$ the Hamiltonian.

The phase space path integral enjoys some relative advantages over the usual Feynman path integral, which will be of use to us in our study of Euclidean wormholes. The biggest one is that it allows us to perform an analysis at all! Our wormholes are not saddle points of the full gravity action, and so they are inaccessible by a standard perturbative treatment. Now in a Hamiltonian path integral for gauge theory or gravity, the time components of the gauge field or metric act as Lagrange multipliers enforcing Gauss' Law constraints.  In a sense that will be more clear later, our wormholes are instead near-saddles, extremizing the gravity action in all but two directions of field space. In the language of~\cite{Affleck:1980mp} they are constrained instantons. 

The phase space path integral description of gauge theories and gravity is an example of constrained quantization.  We refer interested readers to~\cite{jackiw1993constrained, faddeev1980gauge}. We focus on two examples: Yang-Mills theory, and the Chern-Simons path integral.

Our initial discussion will be pedagogical; our goal is to contextualize some technical aspects of our gravity computations. The reader who wishes to skip to a summary of this discussion can go straight to Subsection~\ref{S:stock}. 

\subsubsection{Yang-Mills in Hamiltonian form}

Ordinary Yang-Mills theory
\beq
	Z = \int \frac{[dA_{\mu}]}{\text{gauge}} \,e^{iS_{\rm YM}}\,, \qquad S_{\rm YM} = - \frac{1}{4g^2}\int d^dx \, \text{tr}(F^2)\,,
\eeq
may be recast as a phase space path integral. One integrates in electric fields $E^i$ so that
\beq
	Z = \int \frac{[dA_i][dE^i][dA_0]}{\text{gauge}} \, e^{iS'_{\rm YM}}\,, \quad 	S'_{\rm YM} =  \int d^dx \,\text{tr}\left( E^i \dot{A}_i - \frac{g^2}{2} E^2- \frac{1}{4g^2} F_{ij}F^{ij} + A_0 D_i E^i\right)\,,
\eeq
with $F_{ij}$ the field strength of $A_i$ and $D_i$ the gauge covariant derivative. 
This action is of the usual one-time-derivative term appropriate for a phase space action, of the schematic form
\beq
\label{E:phaseSpaceS}
	S' = \int dt \left( p_i \dot{q}^i - H(p,q) + \lambda_a \,\mathcal{C}^a(p,q)\right)\,,
\eeq
where $\lambda_a$ are Lagrange mutlipliers enforcing the vanishing of the constraints $\mathcal{C}^a = 0$. In the Hamiltonian formulation of Yang-Mills theory, $A_0$ acts as a Lagrange multiplier enforcing Gauss' Law, $A_i$ is conjugate to $E^i$, and there is the usual Hamiltonian $\sim \text{tr}(E^2+B^2)$. The action is gauge-invariant with the electric field transforming in the adjoint representation of the gauge symmetry, and by $\frac{1}{\text{gauge}}$ we refer to the division by the gauge symmetry.

One route to quantization is a ``constrain first'' approach in which one first integrates out $A_0$ with a flat measure. The residual path integral is performed over configurations of $A_i$ and $E^j$ that respect Gauss' Law. What happens next depends on the gauge-fixing, though of course the end result does not. 

Suppose we impose Coulomb gauge $\partial_i A_i = 0$. Then the Faddeev-Popov ghosts do not couple to $A_0$ and the residual integral is performed over configurations exactly satisfying $D_iE^i = 0$. If we instead impose Lorentz gauge $\partial_{\mu}A^{\mu}=0$, then the Faddeev-Popov ghosts have a linear coupling to $A_0$ and the Gauss' Law involves ghost bilinears. For definiteness consider the original path integral in Coulomb gauge, which reads
\begin{align}
\begin{split}
\label{E:YMreducedZ}
	Z &= \int [dA_i][dE^i] [dA_0] [d\bar{c}][dc] \, e^{iS_{\rm YM}'+i \int d^dx \,\text{tr}(\bar{c}\partial_i D^ic)} \prod_x \delta(\partial_i A_i)
	\\
	& = \int \frac{[dA_i][dE^i]}{\text{gauge}} \, e^{i\int d^dx \,\text{tr}(E^i\dot{A}_i - H(A,E))}\prod_x \delta(D_iE^i)\,.
\end{split}
\end{align}
After integrating out $A_0$ the residual integral is performed over the ghosts, as well as over configurations of $A_i$ and $E^j$ which obey Gauss' law. Since the ghost action does not include time derivatives, $c$ and $\bar{c}$ do not have conjugate momenta and the ghost integral is best understood as part of the measure for $A_i$ and $E^j$. At fixed time these configurations, modulo the gauge symmetry, parameterize a phase space as we presently demonstrate. From the $p\dot{q}$ term we extract a putative symplectic form on this space,
\beq
	\omega = \int d^{d-1}x\, \text{tr}(dE^i \wedge dA_i)\,,
\eeq
where by $d$ we mean a formal variation in the space of field configurations rather than the exterior derivative. The integral is taken over a constant-time slice. It is easy to verify that $\omega$ is gauge-invariant on account of Gauss' Law. In order for $\omega$ to be a symplectic form it must be closed and non-degenerate. Clearly $d\omega = 0$, and as for being non-degenerate, the zero eigenvalues of $\omega$ correspond to gauge-variations. That is, $\omega$ is non-degenerate on the space of constrained $A_i$ and $E^j$ modulo the gauge symmetry, and is therefore a symplectic form.

A nice feature of this approach is that it is easy to pass over to the operator formalism. One promotes $A_i$ and $E^j$ to operators with canonical commutation relations; the classical Hamiltonian becomes the quantum Hamiltonian (up to the usual ordering prescription); and one gets the Hilbert space of gauge-invariant states, parameterized by wavefunctionals of $(A_i,E^j)$ that satisfy Gauss' Law. 

\subsubsection{Chern-Simons theory}
\label{S:chernSimons}

Another example of a phase space path integral is pure $G_k$ Chern-Simons theory for a compact gauge group $G$ on a closed space $\Sigma_g$ times time \cite{Elitzur:1989nr}. This example is particularly useful to keep in mind, given the close relationship between three-dimensional gravity and Chern-Simons theory. The Chern-Simons path integral reads
\beq
	Z = \int \frac{[dA_{\mu}]}{\text{gauge}} \, e^{i S_{\rm CS}}\,, \qquad S_{\rm CS} = -\frac{k}{4\pi} \int \text{tr}\left( A \wedge dA + \frac{2}{3}A\wedge A \wedge A\right)\,,
\eeq
with $A$ anti-Hermitian and the trace taken in the fundamental representation of $G$. Unlike Yang-Mills theory, we do not have to integrate in degrees of freedom to put the path integral into Hamiltonian form. We simply separate time from space, $A = A_0 dt + A_i dx^i$, and then the Chern-Simons action reads
\beq
\label{E:CSaction}
	S_{\rm CS} = -\frac{k}{4\pi} \int d^3x \,\varepsilon^{ij}\text{tr}\left( -A_i \dot{A}_j + A_0 F_{ij}\right) \,,
\eeq
which is of the one-time-derivative form appropriate for a phase space action. The $p\dot{q}$ term implies that $A_i$ is conjugate to $\varepsilon^{ij} A_j$, there is no Hamiltonian, and $A_0$ enforces the Gauss' Law constraint. 

The story from here parallels the one for Yang-Mills theory, although there is a simpler route to the constrained system. One may obtain the classical phase space directly, and then quantize it. The classical constraint is simply $F_{ij}=0$ which may be solved locally as $A_i = \tilde{g}^{-1} \partial_i \tilde{g}$ where $\tilde{g}$ is a group-valued field which need not be periodic around cycles of $\Sigma_g$.  The effective action is simply the first term of~\eqref{E:CSaction}, which upon substituting in this expression for $A_i$ only depends on the holonomies of $A$ around the cycles of $\Sigma_g$. From the $p\dot{q}$ term we extract the symplectic form on the phase space,
\beq
	\omega = -\frac{k}{4\pi}\int d^2x \,\varepsilon^{ij}\text{tr}(dA_i \wedge dA_j)\,,
\eeq
which if we wished could be written in terms of the holonomies. Written this way it is clear that it satisfies all of the requirements of a symplectic form. It is manifestly closed; it is gauge-invariant on account of the Gauss' Law constraint; the trace form is negative semi-definite (since $A$ is anti-Hermitian and $G$ is compact); and the zero eigenvalues of $\omega$ correspond to pure gauge fluctuations. So $\omega$ is closed and non-degenerate on the space of constrained $A_i$ modulo the gauge symmetry, and the residual integral is performed over trajectories in this phase space. In this instance the Chern-Simons path integral reduces to quantum mechanics (with $H=0$) for the holonomies with a $2g$-dimensional Hilbert space. 

The quantization is richer on a space with boundary, for which the Hilbert space becomes infinite-dimensional. Consider the disk times time, perhaps with a puncture in the interior. In this case the basic result is that the Chern-Simons theory is equivalent to a chiral WZW model on the boundary, i.e.~to current algebra. The Hilbert space is isomorphic to a single highest weight module of a $G_{k'}$ Kac-Moody symmetry, where the weight is determined by the puncture. (Here we have allowed that $k$ and $k'$ differ by some renormalization, as in $k'=k+2$ for $G=SU(2)$.) The infinite-dimensional phase space comes from the fact that locally one may decompose $A_i = \tilde{g}^{-1} \partial_i \tilde{g}$, where $\tilde{g}$'s non-periodicity around the boundary circle is fixed by the conjugacy class of the puncture. The phase space and Chern-Simons action only depend on the boundary value of $\tilde{g}$ (modulo the redundancy $\tilde{g}(x,t) \to h(t) \tilde{g}(x,t)$ introduced by parameterizing $A_i$ in terms of $\tilde{g}$).

Another basic example is quantization on the annulus times time. In this instance Chern-Simons theory is equivalent to a full non-chiral WZW model on the boundary. The right-movers live on one boundary, the left-movers on the other, and the Hilbert space decomposes into a sum of tensor products of the form $\mathcal{H}_{\lambda}\otimes \mathcal{H}_{\lambda^*}$, with $\mathcal{H}_{\lambda}$ a (right-moving) highest weight module of the Kac-Moody symmetry in a representation $\lambda$ and $\mathcal{H}_{\lambda^*}$ the (left-moving) module in the conjugate representation. To arrive at this conclusion, one decomposes $A_i = \tilde{g}^{-1} \partial_i \tilde{g}$, where $\tilde{g}= e^{\lambda(y)x} g$ with $g$ single-valued. The physical degrees of freedom are a single-valued $G$-field, $g$ on each boundary, and a quantum mechanical holonomy characterized by $\lambda$. 

\subsubsection{Continuation to imaginary time and Hilbert space trace}
\label{S:continuation}

Now let us continue to imaginary time, $t=-i\tau$. In Euclidean signature we still want for $A_0$ to act as a Lagrange multiplier and, for Yang-Mills, we want the real part of the Euclidean action to be bounded below. These considerations fix $A_i$ and $E^i$ to continue with no factors of $i$, while
\beq
	A_0 dt \to A_{\tau} d\tau\,,
\eeq
i.e.~$A_0 \to i A_{\tau}$. So we integrate over the Euclidean gauge field $A = A_{\tau} d\tau + A_i dx^i$ with real contours for the $A_{\mu}$. For Yang-Mills theory the Euclidean phase space action is
\beq
	S_{\rm E,YM}' = \int d^dx \,\text{tr}\left( i E^i \partial_{\tau}A_i -\frac{g^2}{2}E^2 - \frac{1}{4g^2}F_{ij}F^{ij} + i A_{\tau} D_i E^i\right)\,.
\eeq
Compactifying Euclidean time with periodicity $\beta$, the path integral performed with this action and periodic boundary conditions gives us a trace over the physical Hilbert space $\mathcal{H}$ obtained after imposing Gauss' Law, $Z_E = \text{tr}_{\mathcal{H}}(e^{-\beta H})$. 

Similarly, in Chern-Simons theory, we arrive at a Euclidean action
\beq
	S_{\rm E,CS} = \frac{ik}{4\pi}\int d^3x \,\varepsilon^{ij}\text{tr}\left( -A_i \partial_{\tau} A_j + A_{\tau} F_{ij}\right)\,.
\eeq
Again this continuation guarantees that $A_{\tau}$ acts as a Lagrange multiplier. The Euclidean path integral again has the interpretation of a Hilbert space trace, now over the Hilbert space of states on $\Sigma_g$.

When there is a boundary we have seen that the Hilbert space is enlarged from finite-dimensional to infinite-dimensional. For example, on the disk times time, the Hilbert space is a single irreducible module of a Kac-Moody algebra i.e.~that of a chiral WZW model; and on the annulus times time it is the Hilbert space of the full $G_{k
'}$ non-chiral WZW model. 

Consider the disk times time. Upon continuation, one may arrange for the boundary to be a torus of complex structure $\tau$. The Euclidean path integral is the torus partition function of a $G_k$ chiral WZW model, a character of a highest weight module of the Kac-Moody symmetry. For the annulus times time, upon continuation, one equivalently has the Chern-Simons path integral on a torus times an interval. One may arrange for the boundary where the right-movers reside to be a torus of complex structure $\tau_1$, and the other $\tau_2$. For an appropriate choice of boundary conditions, the Euclidean path integral then becomes a generalized WZW torus partition function, where the left- and right-movers are at different temperatures:
\beq
	Z_{A\times \mathbb{S}^1}(\tau_1,\bar{\tau}_2) = \text{tr}_{\mathcal{H}_{\rm WZW}}\left(q_1^{L_0} \bar{q}_2^{\bar{L}_0}\right)\,,
\eeq
with $q_1 = e^{2\pi i \tau_1}$ and $q_2 = e^{2\pi i \tau_2}$. 

\begin{figure}[t]
\begin{center}
\includegraphics[scale=.25]{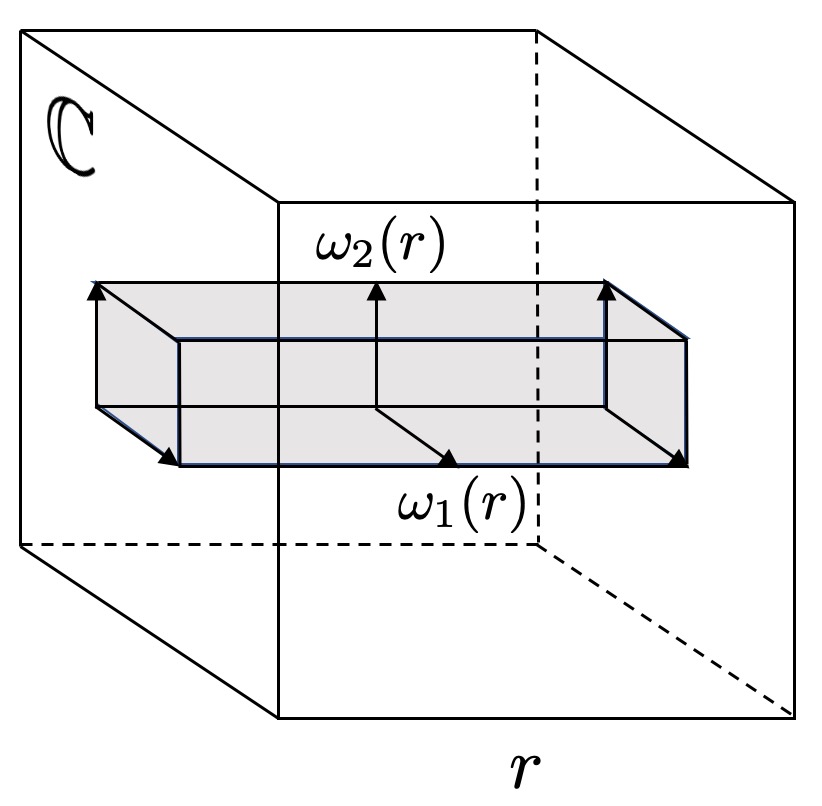}
\end{center}
\caption{\label{F:WZWmodular} Chern-Simons theory on $\mathbb{T}^2\times I$ is modular invariant, even when the boundary tori have different complex structures.  Here $r$ is the coordinate along the interval, and $(\omega_1(r), \omega_2(r))$ are the lattice vectors corresponding to the torus as a function of $r$.}
\end{figure}

The usual torus partition function of a WZW model, i.e.~with $\tau_1=\tau_2=\tau$, is modular invariant. This generalized partition function is also modular invariant even when $\tau_1\neq \tau_2$,
\beq
\label{E:genModularInvariance}
	Z_{A\times\mathbb{S}^1}\left(\frac{a\tau_1+b}{c\tau_1+d},\frac{a\bar{\tau}_2+b}{c\bar{\tau}_2+d}\right) = Z_{A\times\mathbb{S}^1}(\tau_1,\bar{\tau}_2)\,, \qquad ad-bc= 1\,.
\eeq
This invariance is geometrized in the Chern-Simons computation. See Fig.~\ref{F:WZWmodular}. The three-dimensional space is topologically $\mathbb{T}^2\times I$. The usual arguments for modular invariance of 2d CFT on the torus can then be adapted here. Recall that the torus is a quotient of the complex plane $\mathbb{T}^2 =  \faktor{\mathbb{C}}{\mathbb{Z}\times\mathbb{Z}}$ generated by a basis of lattice vectors $(\omega_1,\omega_2)$ with $\tau = \frac{\omega_2}{\omega_1}$. For $\mathbb{T}^2\times I$ we should regard the basis vectors $\omega_1$ and $\omega_2$ as smoothly varying along the interval. But we may describe the same torus with any basis we wish, and any two bases are related by a modular transformation, which then acts simultaneously on $\tau_1$ and $\tau_2$. 

This last example of $G_k$ Chern-Simons theory on $\mathbb{T}^2\times I$ is particularly relevant for us. It is a prototype for the Euclidean wormholes of AdS$_3$ gravity that we will study, which are also topologically $\mathbb{T}^2\times I$, smoothly connecting two asymptotic regions with torus boundary.

There is another useful fact to glean from Fig.~\ref{F:WZWmodular}. The bulk orientation induces opposite orientations on the two boundaries. In our AdS$_3$ analysis we wish to study wormholes where both boundaries have the same orientation. In practice this can be accomplished by parity-flipping boundary $2$. With respect to the new orientation both boundaries are endowed with modes of the same chirality. At the level of the partition function we do this by reinterpreting its complex structure as $\tau_2 \to -\bar{\tau}_2$, i.e. by flipping the real part of $\tau_2$ while preserving the imaginary part. The ensuing partition function is $Z'(\tau_1,\tau_2) = Z_{A\times \mathbb{S}^1}(\tau_1,-\tau_2)$, and by~\eqref{E:genModularInvariance} it obeys $Z'(\gamma\tau_1,\gamma^{-1}\tau_2) = Z'(\tau_1,\tau_2)$ for $\gamma\in SL(2;\mathbb{Z})$.

\subsubsection{Taking stock}
\label{S:stock}

Let us summarize a few lessons from the phase space path integral form of gauge theory that will appear in our AdS$_3$ analysis.
\begin{enumerate}
	\item The time component of the gauge field acts as a Lagrange multiplier, enforcing the Gauss' Law constraint. For Chern-Simons theory the constraint is simply that the spatial field strength vanishes,  and the residual integral is performed over flat spatial connections, locally $A_i = G^{-1} \partial_i G$. The residual integral only depends on topological data (holonomies around contractible cycles) and boundary values of $G$.
	\item After imposing the constraints and modulo the gauge symmetry, the residual space of field configurations is, at constant time, a phase space with a symplectic form determined by a single-time-derivative term $\sim p\dot{q}$ in the effective action. This gives us a canonical measure for the residual path integral, on each time slice given by the Pfaffian of the symplectic form of that phase space.
	\item The continuation to imaginary time is simply $t\to -i \tau$ and $A_t \to i A_{\tau}$, i.e.
	\beq
	\label{E:A0wick}
		A_0 dt \to A_{\tau} d\tau\,,
	\eeq
	with a real integration contour for $A_{\tau}$. This guarantees that $A_{\tau}$ continues to act as a Lagrange multiplier in the Euclidean theory. Upon compactifying Euclidean time, the Euclidean path integral then has the interpretation of a trace over the Hilbert space of states obtained after enforcing the Gauss' Law constraint.
	\item Finally, in Chern-Simons theory on $\mathbb{T}^2\times I$ we have encountered a prototype of the Euclidean wormholes we will study in 3d gravity. We can clearly tune the complex structures $\tau_1$ and $\tau_2$ of the boundary tori to be different from one another, but they are connected through the bulk. The space is equivalent to an annulus times a circle, which continues in real time to an annulus times time. The Hilbert space is that of a $G$ WZW model, where the right-movers reside on one boundary and the left-movers on the other. The Euclidean path integral is a trace over that Hilbert space, $\text{tr}_{\mathcal{H}_{\rm WZW}}\left( q_1^{L_0} \bar{q}_2^{\bar{L}_0}\right)$ with $q_i = e^{2\pi i \tau_i}$, and it is modular invariant. 
\end{enumerate}

All of these lessons correspond to aspects of our gravity computation. In the first-order formalism the gravitational action is already in Hamiltonian form. The time components of the dreibein $e^A_0$ and spin connection $\omega^A{}_{B0}$ act as Lagrange multipliers, enforcing curvature and torsion constraints. The continuation to imaginary time parallels~\eqref{E:A0wick},
\beq
	e^A_0 dt \to e^A_{\tau} d\tau\,, \qquad \omega^A{}_{B0}dt \to \omega^A{}_{B\tau}d\tau\,,
\eeq
and the local Lorentz symmetry remains $SO(2,1)$. This continuation is essentially fixed by discrete symmetries and the requirement that $e^A_{\tau}$ and $\omega^A{}_{B\tau}$ continue to act as Lagrange multipliers.  However, this means that we continue to integrate over Lorentzian-signature metrics even though we are in imaginary time. This statement is initially puzzling, but we can made our peace with it on two counts.  First, we require this continuation in order to have a Hilbert space interpretation in real time. Second, it has a parallel in JT gravity, where much more is known about the gravity path integral. The analogue there is the integration over imaginary values of the dilaton, which enforces the constant curvature constraint. 

Finally, our wormholes are topologically $\mathbb{T}^2\times I$, or equivalently $A\times \mathbb{S}^1$. These configurations are not saddle points of the complete gravity action, however we can fully enumerate them, after integrating out $e^A_{\tau}$ and $\omega^A{}_{B\tau}$. As in the Chern-Simons case, we may independently dial the complex structures of the boundary tori, and we expect on general grounds for the gravity path integral to be modular invariant under simultaneous modular transformations acting on $\tau_1$ and $\tau_2$. This is a strong consistency condition which is obeyed by our result. 

\subsection{First order formulation of AdS$_3$ gravity}
\label{S:firstOrder}

Three-dimensional gravity with negative cosmological constant is often said to be classically equivalent to a Chern-Simons theory~\cite{Achucarro:1987vz}. (For statements going beyond classical physics see~\cite{Witten:2007kt, Kim:2015qoa}.) This statement needs additional qualifiers in order to be true. More precisely, AdS$_3$ gravity on spacetimes of the topology disk times time (i.e.~global AdS$_3$ and spacetimes continuously connected to it) is equivalent to a topological sector of $SO(2,2)$ Chern-Simons theory. This equivalence holds in the quantum theories to all orders in perturbation theory but, as we will see, it does not hold on other spacetimes. In other words there is only a perturbative equivalence between three-dimensional gravity and a Chern-Simons theory, but they are non-perturbatively different theories.

Let us review the usual classical equivalence, which is seen as follows. Let $M,N$ be spacetime indices. Consider the first-order formulation, where we decompose the metric into a dreibein $e^A_M$ through $g_{MN} = \eta_{AB}e^A_M e^B_N$, where $A,B$ are flat indices which are raised and lowered with the Minkowski metric $\eta_{AB}$. We introduce a spin connection $\omega^A{}_{BM}$ satisfying $\omega_{(AB)M} = 0$. The gravity action in first-order variables is
\beq
\label{E:FOaction}
	S = -\frac{1}{16\pi G}\int \varepsilon_{ABC}e^A \wedge \left( d\omega^{BC} +\omega^B{}_D \wedge \omega^{DC} +\frac{1}{3 }e^B \wedge e^C\right) + (\text{boundary term})\,,
\eeq
where $e^A = e^A_M dx^M$ and $\omega^A{}_B =\omega^A{}_{BM}dx^M$ are one-forms. Notably the action only has a single time derivative, and each component of the dreibein and spin connection only appears once. So we are dealing with a phase space path integral with constraints. To see the relation to a Chern-Simons theory, one then groups the first order variables into the vector-valued one-forms
\beq
\label{E:fromeToA}
	A^A_M = \frac{1}{2}\varepsilon^{ABC} \omega_{BCM} + e^A_M \,, \qquad \bar{A}^A_M = \frac{1}{2}\varepsilon^{ABC}\omega_{BCM}-e^A_M\,,
\eeq
and introduces generators $J_A$ and $\bar{J}_A$ in the fundamental representation of $\mathfrak{sl}(2;\mathbb{R})$ satisfying
\beq
	[J_A,J_B] = \varepsilon_{ABC}J^C\,, \qquad \text{tr}(J_A J_B) = \frac{1}{2}\eta_{AB}\,,
\eeq
and similarly for $\bar{J}_A$, with $\varepsilon_{012} = -1$. Define the algebra-valued one-forms $A = A^A J_A$ and $\bar{A} = \bar{A}^A\bar{J}_A$. Note that $A$ and $\bar{A}$ are independent real fields. When we require an explicit form of the generators, we use
\beq
	J_0 = -\frac{i}{2}\sigma_2 \,, \qquad J_1 = \frac{1}{2}\sigma_1\,, \qquad J_2 = \frac{1}{2}\sigma_3\,.
\eeq
Then up to a boundary term a short computation shows that the action~\eqref{E:AdS3action} may be written as a difference of Chern-Simons terms,
\beq
	S = -\frac{k}{4\pi}\int (I[A]-I[\bar{A}])\,, \qquad k=\frac{1}{4G}\,, \qquad I[A] = \text{tr}\left(A\wedge dA+\frac{2}{3}A\wedge A\wedge A\right)\,.
\eeq

Since $A$ and $\bar{A}$ are linear combinations of the dreibein and spin connection, it follows that the classical equations of motion for the first-order variables are simply those for $A$ and $\bar{A}$, the flatness conditions
\beq
	F = dA + A \wedge A = 0\,, \qquad \bar{F} = d\bar{A} + \bar{A} \wedge \bar{A} = 0\,.
\eeq
Further, on a classical solution, infinitesimal diffeomorphisms and local Lorentz rotations act in the same way as infinitesimal $\mathfrak{sl}(2;\mathbb{R})\times \mathfrak{sl}(2;\mathbb{R})$ gauge transformations.\footnote{Let $\xi^M$ be an infinitesimal diffeomorphism, and $v^A{}_B$ an infinitesimal Lorentz rotation. Then, on-shell, these variations act on $A$ and $\bar{A}$ in the same way as infinitesimal $\mathfrak{sl}(2;\mathbb{R})\times\mathfrak{sl}(2;\mathbb{R})$ gauge transformations with gauge parameters $\Lambda = A_M \xi^M + v$ and $\bar{\Lambda} = \bar{A}_M\xi^M+v$ respectively,}

So on-shell, $A$ and $\bar{A}$ appear to be $\mathfrak{sl}(2;\mathbb{R})\times\mathfrak{sl}(2;\mathbb{R})$ gauge fields, and linearized diffeomorphisms and local Lorentz rotations act as linearized gauge transformations. Of course we would like to see that the converse is true, i.e.~that linearized gauge transformations are in one-to-one correspondence with linearized diffeomorphisms/rotations. But this is only the case if $(A_M-\bar{A}_M)^A$ is non-degenerate as a $3\times 3$ matrix. Eqn.~\eqref{E:fromeToA} implies $\frac{(A_M-\bar{A}_M)^A}{2} = e^A_M$, so this is simply the condition that the spacetime metric is non-degenerate.

To sum up, we see that AdS$_3$ gravity is on-shell equivalent to $\mathfrak{sl}(2;\mathbb{R})\times\mathfrak{sl}(2;\mathbb{R})$ Chern-Simons theory at the level of the equations of motion and linearized gauge symmetries, so long as the spacetime metric is non-degenerate.  Of course the quantum theories are rather different.

\subsubsection{Global AdS$_3$ and boundary conditions}

The simplest solution to Einstein's equations on a spacetime with the topology of a disk times time is global AdS$_3$, parameterized by the dreibein
\beq
	e^0 = \cosh(\rho) dt\,, \qquad e^1 = \sinh(\rho)dx \,, \qquad e^2 = d\rho\,,
\eeq
with $x\sim x+2\pi$. This spacetime has a conformal boundary as $\rho\to\infty$. Upon solving for the spin connection so that the torsion vanishes, the combinations $A$ and $\bar{A}$ are
\beq
\label{E:globalAdS}
	A = \frac{1}{2}\begin{pmatrix} d\rho & -e^{-\rho}(dx+dt) \\ e^{\rho}(dx+dt)& -d\rho\end{pmatrix}\,, \quad \bar{A} = \frac{1}{2}\begin{pmatrix} -d\rho & e^{\rho}(dx-dt) \\ -e^{-\rho}(dx-dt) & d\rho\end{pmatrix} \,.
\eeq
Since $F = \bar{F}=0$ we may write $A = G^{-1}dG$ and $\bar{A} = \bar{G}^{-1}d\bar{G}$ for some (possibly multi-valued) $SL(2;\mathbb{R})$ elements $G$ and $\bar{G}$. One representative is
\beq
	G = e^{(x+t) J_0}e^{\rho J_2}\,, \qquad \bar{G} = e^{-(x-t) J_0}e^{-\rho J_2}\,.
\eeq
This $G$ and $\bar{G}$ are double-valued, with $G(x+2\pi,t,\rho) = - G(x,t,\rho)$ and similarly for $\bar{G}$. Recall that $PSL(2;\mathbb{R})$ is the quotient of $SL(2;\mathbb{R})$ by its $\mathbb{Z}_2$ center $\{I,-I\}$, and that  $SO(2,2)$ is its double cover, the quotient of $SL(2;\mathbb{R})\times SL(2;\mathbb{R})$ by the $\mathbb{Z}_2$ subgroup $\{(I,I),(-I,-I)\}$. Then $G$ and $\bar{G}$ are single-valued only as elements of either $PSL(2;\mathbb{R})\times PSL(2;\mathbb{R})$ or $SO(2,2)$, and not of any cover thereof. So, to the extent that the Chern-Simons description has a good candidate for the global form of the gauge group, it must be either $PSL(2;\mathbb{R})\times PSL(2;\mathbb{R})$ \cite{Castro:2011iw} or $SO(2,2)$~\cite{Cotler:2018zff}.

In fact nonlinear classical equivalence selects $SO(2,2)$. The isometry group of global AdS$_3$ is $SO(2,2)$, since the combined transformation $x\to x+2\pi$ and $t\to t$, acts as the identity. It is easy to verify that these nonlinear isometries act on $A$ and $\bar{A}$ as $SO(2,2)$ gauge transformations.

$SO(2,2)$ has a fundamental group isomorphic to $\mathbb{Z}\times \mathbb{Z}$, and it is worth noting that the $SO(2,2)$ field $(G,\bar{G})$ parameterizing global AdS$_3$ has a non-trivial winding number around the spatial circle. Configurations with other winding numbers have curvature singularities in the interior. So first-order AdS$_3$ gravity (on the disk times time) is only classically equivalent to a particular winding sector of $SO(2,2)$ Chern-Simons theory.

In order to fully specify the classical theory we must also supply a variational principle and add requisite boundary terms to the Chern-Simons action. We impose the standard Brown-Henneaux boundary conditions on the metric, which when translated to the first-order variables become 
\begin{align}
\label{E:AdS3BC}
	A = \frac{1}{2}\begin{pmatrix} d\rho  & 0 \\ e^{\rho}(dx+dt)   & -d\rho \end{pmatrix}+ O(e^{-\rho})\,, \qquad 
	\bar{A} = \frac{1}{2}\begin{pmatrix} -d\rho & e^{\rho}(dx-dt)\\ O(e^{-\rho}) & d\rho\end{pmatrix}+O(e^{-\rho})\,,
\end{align}
as one approaches a conformal boundary at $\rho\to\infty$. These boundary conditions clearly allow global AdS$_3$, and crucially, an infinite-dimensional phase space of solutions connected to it parameters by the action of large diffeomorphisms on global AdS$_3$.

These boundary conditions are consistent with a good variational principle for $e$ and $\omega$, equivalently $A$ and $\bar{A}$, only if we add a boundary term to the Chern-Simons action which we describe momentarily.

\subsubsection{Going off-shell}

In this manuscript we are interested not in classical physics but the complete gravity path integral. Our approach is to take the first-order path integral as fundamental, and in Section~\ref{S:wormhole} we will see just what a Chern-Simons description does and does not get correct.

In the original first-order action $e^A_0$ and $\omega^A{}_{B0}$ appear linearly and we proceed by treating them as Lagrange multipliers. Equivalently, we take the combinations $A_0$ and $\bar{A}_0$ to be Lagrange multipliers. So we separate time from space $x^{\mu}=(t,x^i)$ with $i=1,2$ the spatial directions, and decompose $A$ and $\bar{A}$ into their temporal and spatial parts as $A = A_0 dt + A_i dx^i$, $\bar{A}= \bar{A}_0dt+\bar{A}_idx^i$. Then
\beq
\label{E:AdS3Action}
	S = \frac{k}{4\pi} \int d^3x \,\varepsilon^{ij}\text{tr}\left( -A_i \dot{A}_j + A_0F_{ij}\right) - (A\to \bar{A}) + S_{\rm bdy}\,.
\eeq
Here $S_{\rm bdy}$ is a boundary term required in order for there to be a variational principle consistent with the boundary conditions~\eqref{E:AdS3BC}. It is given by
\beq
\label{E:bdyTerm}
	S_{\rm bdy} =- \frac{k}{4\pi} \int d^2x\Big( \text{tr}(A_x^2) + \text{tr}(\bar{A}_x^2) \Big)\,.
\eeq
This action is indeed of the form~\eqref{E:phaseSpaceS} with a Hamiltonian only coming from the boundary term, and a symplectic form coming from the $p\dot{q}$ term as
\beq
\label{E:CSsymplectic}
	\omega = -\frac{k}{4\pi}\int d^2x \,\varepsilon^{ij} \text{tr}\left( dA_i\wedge dA_j\right) - (A\to\bar{A})= -\frac{k}{4\pi}\int d^2x \,\varepsilon^{ij}\varepsilon_{ABC} de^A_i \wedge d\omega^{BC}{}_j\,,
\eeq
taken over a constant time slice. $A_0$ and $\bar{A}_0$ appear as Lagrange multipliers enforcing the Gauss' Law constraints. 

From here the path integral quantization that ensues closely parallels that of Chern-Simons theory as described in Subsection~\ref{S:chernSimons}. One solves the flatness constraints to obtain the classical phase space, which one then quantizes using the action~\eqref{E:AdS3Action} and symplectic structure~\eqref{E:CSsymplectic}. The step that is different from Chern-Simons theory is that we integrate over inequivalent metrics, not inequivalent gauge fields. For 3d gravity on the torus times interval, equivalently annulus times time, this difference boils down to a different field range for the wormhole moduli as well as a different integration measure over the moduli space.

\subsubsection{Continuing to imaginary time}
\label{S:gravitycontinuation}

Now we continue to imaginary time. The story here parallels the continuation of gauge theory to imaginary time $t=-iy$\footnote{Here and henceforth we refer to Euclidean time as $y$, reserving $\tau$ for the torus complex structure.} in Subsection~\ref{S:continuation}. We require that the time components of the dreibein and spin connection, equivalently $A$ and $\bar{A}$, continue to act as Lagrange multipliers, and that the spatial metric is real. These considerations fix $e^A_0 \to i e^A_{y}$ and $\omega^A{}_{B0} \to i \omega^A{}_{By}$, with the spatial components not picking up any factors of $i$. Equivalently,
\beq
	A^A_0dt \to A^A_{y} dy\,, \qquad \bar{A}^A_0 dt \to \bar{A}^A_{y}dy\,,
\eeq
with real integration contours for $e^A_{y}$ and $\omega^A{}_{By}$, equivalently $A^A_{y}$ and $\bar{A}^A_{y}$. 

Relatedly, in our continuation we take the local rotations to remain valued in $SO(2,1)$ rather than $SO(3)$. We do this because we are interested in a Hilbert space of the Lorentzian signature theory, whose local Lorentz invariance is $SO(2,1)$ rather than local rotational invariance $SO(3)$. Consequently we sum over Lorentzian signature metrics $g_{MN} = \eta_{AB}  e^A_M e^B_N$ even though we are in imaginary time.

Let us be more explicit. Consider Global AdS$_3$,~\eqref{E:globalAdS}. Continuing $t = - i y$, we have
\beq
\label{E:globalAdS2}
	A = \frac{1}{2}\begin{pmatrix} d\rho & -e^{-\rho} d\bar{z} \\ e^{\rho} d\bar{z} & -d\rho\end{pmatrix}\,, \qquad \bar{A}= \frac{1}{2}\begin{pmatrix} -d\rho & e^{\rho} dz \\ -e^{-\rho} dz & d\rho\end{pmatrix}\,, \qquad z = x + i y\,,
\eeq
and the resulting line element
\beq
	ds^2 = \cosh^2(\rho) dy^2 + \sinh^2(\rho)dx^2 +d\rho^2
\eeq
is Euclidean.  We may further identify $z \sim z+2\pi n + 2\pi m \tau$ so that the boundary is a torus of complex structure $\tau$. Note that $A_y$ and $\bar{A}_y$ are pure imaginary. In going off-shell, we integrate over $A$ and $\bar{A}$ subject to the boundary conditions
\beq
	A = \frac{1}{2} \begin{pmatrix} d\rho & 0 \\ e^{\rho} d\bar{z} & -d\rho\end{pmatrix} + O(e^{-\rho}) \,, \qquad \bar{A} = \frac{1}{2}\begin{pmatrix} - d\rho& e^{\rho} dz \\ 0 & d\rho\end{pmatrix} + O(e^{-\rho})\,,
\eeq
and we integrate over \emph{real} fluctuations. That is, we fix $A_y$ and $\bar{A}_y$ (equivalently $e^A_y$ and $\omega^A{}_{By}$) to be pure imaginary near the boundary, and integrate over a real contour for them in the interior. 

This is analogous to what one does in JT gravity, where one fixes the dilaton to be a real constant on the boundary and integrates over imaginary fluctuations in the bulk.

\subsubsection{What about $SL(2;\mathbb{C})$?}

Clearly we must be careful about what we mean by Euclidean quantum gravity. Let us briefly comment on another possible definition, namely the sum over real Euclidean metrics. In the first-order formalism we would have a dreibein $e^A_M$ and spin connection $\omega^A{}_{BM}$ with $\omega_{(AB)M} = 0$, where now we raise and lower flat indices with the Euclidean metric $\delta_{AB}$ and mod out by local $SO(3)$ rotations. The Euclidean action is the same as~\eqref{E:FOaction}, except now $\varepsilon_{ABC}$ is the antisymmetric invariant tensor of $SO(3)$ instead of $SO(2,1)$. The spin connection appears quadratically and can be integrated out, resulting in Euclidean gravity in the second-order formalism.

On-shell, there is again a story about classical equivalence with a Chern-Simons theory. One can group the dreibein and spin connection into an $\mathfrak{sl}(2;\mathbb{C})$ gauge field, on which linearized diffeomorphisms and local rotations act in the same way as infinitesimal $\mathfrak{sl}(2;\mathbb{C})$ gauge transformations. The gravity action equals a Chern-Simons action for this gauge field. Nonlinearly, for Euclidean global AdS$_3$, the $PSL(2;\mathbb{C})$ isometries of EAdS$_3$ act on this gauge field as $PSL(2;\mathbb{C})$ gauge transformations.

However, unlike in Lorentzian signature where classical AdS$_3$ gravity is equivalent to a winding sector of $SO(2,2)$ Chern-Simons theory, $PSL(2;\mathbb{C})$ is simply connected and so has no winding sectors. 

In any case, the reader may wonder why we do not start with this definition. The reason is simple: in this definition, while $e^A_M$ and $\omega^B{}_{CM}$ appear linearly in the action (for fixed $M$), the action is real and so they do not act as Lagrange multipliers with a real contour of integration. 

\subsection{Alekseev-Shatashvili theory}
\label{S:AS}

AdS$_3$ gravity has no propagating degrees of freedom, but it does have edge modes. These are sometimes called boundary gravitons. They are generated by acting on a spacetime with large diffeomorphisms and Lorentz transformations (meaning they do not die off at the boundary, and so are not part of the gauge symmetry) that preserve the boundary conditions.

In our previous work~\cite{Cotler:2018zff}, we obtained the effective action for these large gauge transformations for global AdS$_3$. Our approach was to use the classical equivalence between AdS$_3$ gravity on the solid cylinder $D\times\mathbb{R}$ and a particular winding sector of $SO(2,2)$ Chern-Simons theory, and to then quantize using the Chern-Simons description. In this case the Chern-Simons phase space coincides with the phase space of 3d gravity on the disk, and so both lead to the same quantum theory. Our goal for this Subsection is to summarize that effective action, its path integral, and a deformation of it which will appear in Section~\ref{S:wormhole}.

Let us briefly summarize the Chern-Simons quantization, and refer the reader to~\cite{Cotler:2018zff} for more details. We integrated out the time components $A_0$ and $\bar{A}_0$, which enforced that the spatial gauge field was flat. Parameterizing it as
\beq
\label{E:flatness}
	A_i = G^{-1} \partial_i G\,, \qquad \bar{A}_j = \bar{G}^{-1} \partial_j \bar{G}\,,
\eeq
the Chern-Simons action becomes a chiral $SO(2,2)$ WZW action for $(G,\bar{G})$. Our decomposition~\eqref{E:flatness} introduces a gauge symmetry, having nothing to do with the original $SO(2,2)$ gauge symmetry, instead a redundancy under
\beq
\label{E:SOquotient}
	G(t,x,\rho) \to h(t)G(t,x,\rho)\,, \qquad \bar{G}(t,x,\rho) \to \bar{h}(t) \bar{G}(t,x,\rho)\,,\qquad (h,\bar{h})\in SO(2,2)\,,
\eeq
which clearly leads to the same gauge field. So we identify these configurations in the path integral over $G$ and $\bar{G}$. Decomposing
\beq
	G = e^{\phi J_0} e^{\lambda J_2} e^{\psi(J_1-J_0)}\,, \qquad \bar{G} = e^{-\bar{\phi}J_0} e^{-\bar{\lambda}J_2} e^{\bar{\psi}(J_1+J_0)}\,,
\eeq
the AdS$_3$ boundary conditions~\eqref{E:AdS3BC} fix $(\lambda,\psi,\bar{\lambda},\bar{\psi})$ near the boundary in terms of $(\phi,\bar{\phi})$ which remain finite on the boundary. The winding condition implies that $(\lambda,\psi,\bar{\lambda},\bar{\psi}$) are single-valued, but $\phi$ and $\bar{\phi}$ are not periodic, instead obeying
\begin{align}
\begin{split}
\label{E:DiffS1}
	\phi(x+2\pi,t) &= \phi(x,t) + 2\pi\,, \qquad \phi' >0\,, 
	\\
	\bar{\phi}(x+2\pi,t) &= \bar{\phi}(x,t) + 2\pi\,, \qquad \bar{\phi}' >0\,,
\end{split}
\end{align}
with $' = \partial_x$. That is, $\phi$ and $\bar{\phi}$ are, at fixed time, diffeomorphisms of the circle, $\text{Diff}(\mathbb{S}^1)$. Translated into an identification on $\phi$ and $\bar{\phi}$,~\eqref{E:SOquotient} means that we identify
\beq
\label{E:SL2}
	\tan\left( \frac{\phi(x,t)}{2}\right) \sim\frac{a(t) \tan\left(\frac{\phi(x,t)}{2}\right)+b(t)}{c(t)\tan\left(\frac{\phi(x,t)}{2}\right)+d(t)} \,, \quad \tan\left(\frac{\bar{\phi}(x,t)}{2}\right)\sim\frac{\bar{a}(t) \tan\left(\frac{\bar{\phi}(x,t)}{2}\right)+\bar{b}(t)}{\bar{c}(t)\tan\left(\frac{\bar{\phi}(x,t)}{2}\right)+\bar{d}(t)}\,, 
\eeq
where\footnote{Note that in $PSL(2;\mathbb{R})$ we identify $(a,b,c,d)$ with $(-a,-b,-c,-d)$, while in $SO(2,2)$, we identify $(a,b,c,d;\bar{a},\bar{b},\bar{c},\bar{d})$ with $(-a,-b,-c,-d,-\bar{a},-\bar{b},-\bar{c},-\bar{d})$. But the fractional linear transformation in~\eqref{E:SL2} is the same for $(a,b,c,d)$ and $(-a,-b,-c,-d)$. So the quotient is effectively by two independent copies of $PSL(2;\mathbb{R})$, rather than by $SO(2,2)$.}
\beq
	 ad-bc = \bar{a}\bar{d}-\bar{b}\bar{c} = 1\,.
\eeq
Thus, at fixed time, $\phi$ and $\bar{\phi}$ are elements of the quotient space $\faktor{\text{Diff}(\mathbb{S}^1)}{PSL(2;\mathbb{R})}.$ 

With these substitutions the chiral WZW action then simplifies to the desired boundary effective action for the edge modes $\phi$ and $\bar{\phi}$,\footnote{This result was anticipated in~\cite{Barnich:2017jgw}. Its quadratic approximation was independently arrived at in~\cite{Haehl:2018izb} as an effective field theory for 2d CFTs dominated by exchange of the identity operator and its Virasoro descendants.}
\beq
	S[\phi,\bar{\phi}] = S_-[\phi] + S_+[\bar{\phi}]\,,
\eeq
with
\beq
	S_{\pm}[\phi] = - \frac{C}{24\pi}\int d^2x \left( \frac{\phi''\partial_{\pm}\phi'}{\phi'^2} - \phi' \partial_{\pm}\phi\right)\,, \quad C = \frac{3}{2G}\,, \quad \partial_{\pm} = \frac{1}{2}\left( \partial_x \pm \partial_t\right)\,.
\eeq
The quantity $C$ is the Brown-Henneaux central charge, and $C\gg 1$ is the weak coupling limit both for AdS$_3$ gravity and for this model. This action is Lorentz-invariant, in fact conformally invariant, despite not being manifestly so. Being linear in time derivatives, it is already in Hamiltonian form. The time derivative term tells us that the symplectic form is
\beq
\label{E:KKomega}
	\omega = \frac{C}{48\pi} \int_0^{2\pi} dx \left( \frac{d\phi'\wedge d\phi''}{\phi'^2 }- d\phi \wedge d\phi'- (\phi\to\bar{\phi})\right)\,,
\eeq
which is also what we get from the bulk symplectic form~\eqref{E:CSsymplectic} upon rewriting $A$ in terms of $\phi$ and $\bar{\phi}$. At the quantum mechanical level, the AdS$_3$ path integral is then
\beq
	Z_{D\times\mathbb{R}} = \int \prod_t \left( \frac{[d\phi(t)][d\bar{\phi}(t)]}{PSL(2;\mathbb{R})\times PSL(2;\mathbb{R})} \,\text{Pf}(\omega(t))\right) e^{i S}\,.
\eeq
The notation indicates that the measure of integration is, at fixed time, over the space $\faktor{\text{Diff}(\mathbb{S}^1)}{PSL(2;\mathbb{R})} \times \faktor{\text{Diff}(\mathbb{S}^1)}{PSL(2;\mathbb{R})}$ with the symplectic measure inherited from~\eqref{E:KKomega}.

So large gauge transformations are weighted by the action $S$, and in the quantum theory we integrate over them. 

We see that the action and path integral chirally factorize into models for $\phi$ and $\bar{\phi}$. The classical chiral action for $\phi$ (or $\bar{\phi})$ was not new. It first appeared in a paper of Alekseev and Shatashvili \cite{Alekseev:1988ce}, and for this reason we refer to it as Alekseev-Shatashvili theory. The point is that the integration space of $\phi$ at fixed time is a coadjoint orbit of the Virasoro group. See~\cite{Oblak:2016eij, Cotler:2018zff} for a practical primer to coadjoint orbits. Coadjoint orbits of Lie groups $G$ are phase spaces with a $G$-invariant symplectic form, and under certain conditions one can consider the quantum mechanics of trajectories in the phase space, the quantization of the phase space. This quantization generally produces a quantum mechanics whose Hilbert space is a single irreducible representation of $G$. There is some freedom in the choice of Hamiltonian, but the natural ones correspond to elements of the Lie algebra of $G$. Building upon previous work, Alekseev and Shatashvili obtained the classical action corresponding to the quantization of the coadjoint orbit $\faktor{\text{Diff}(\mathbb{S}^1)}{PSL(2;\mathbb{R})}$ of the Virasoro group, which for a Hamiltonian corresponding to the generator $L_0$ coincides with the action $S_{+}$ above. (The action $S_-$ is its chiral conjugate, with a Hamiltonian corresponding to $\bar{L}_0$.)

There is a close relationship between this model and the Schwarzian path integral describing Euclidean JT gravity on the hyperbolic disk. The field of the Schwarzian path integral is an element of the same phase space $\phi \in \faktor{\text{Diff}(\mathbb{S}^1)}{PSL(2;\mathbb{R})}$. It may be thought of as a phase space integral, $Z_{\rm disk} \sim \int dxdp \,e^{-\beta H}$. The Alekseev-Shatashvili model is simply the quantization of this phase space, and AdS$_3$ gravity on the disk times time is two decoupled copies of this quantum mechanics, one right-moving and one left-moving.

To investigate the Hilbert space of the Alekseev-Shatashvili model we continue to imaginary time $t=-iy$ and put the model on a torus of complex structure $\tau$ by identifying $z=x+iy\sim z+2\pi \tau$ (recall that $x$ is already periodic with $x\sim x+2\pi$). The ensuing Euclidean action is
\beq
\label{E:ASeuclidean}
	S_E = \frac{C}{24\pi} \int d^2x \left( \frac{\phi''\partial \phi'}{\phi'^2} - \phi' \partial \phi + \frac{\bar{\phi}''\bar{\partial}\bar{\phi}'}{\bar{\phi}'^2} - \bar{\phi}'\bar{\partial}\bar{\phi}\right)\,.
\eeq
The boundary conditions for $\phi$ and $\bar{\phi}$ are that they wind once around the spatial cycle, and are periodic around the other,\footnote{$\phi$ and $\bar{\phi}$ depend on both $z$ and $\bar{z}$. We are only writing out the dependence on $z$ to simplify the notation.}
\beq
\label{E:diffBC}
	\phi(z+2\pi n +2\pi m \tau) = \phi(z) + 2\pi n\,, \qquad \bar{\phi}(z+2\pi n + 2\pi m \tau) = \bar{\phi}+2\pi n\,.
\eeq
This corresponds in three dimensions to our continuation of gravity on global AdS$_3$, where the asymptotic geometry is Euclidean and the boundary is a torus of complex structure $\tau$. The bulk is a disk times a circle, and we call the path integral $Z(\tau)$. 

There is a unique solution to the equations of motion with these boundary conditions modulo the quotients, and at large $C$ we compute $Z(\tau)$ to one loop by expanding $\phi$ and $\bar{\phi}$ in fluctuations around the classical trajectory. The one-loop determinant must be evaluated with respect to the symplectic measure~\eqref{E:KKomega}, with the result
\beq
	Z_{\rm 1-loop}(\tau) = |\chi_{0,c}(\tau)|^2 = \left| q^{-\frac{c}{24}} \prod_{n=2}^{\infty}\frac{1}{1-q^n}\right|^2\,, \qquad c = C+13\,,
\eeq
which we recognize as the Virasoro character of the vacuum representation with a one-loop renormalization of the central charge by 13. (This one-loop renormalization can also be seen from a one-loop computation in the bulk~\cite{Giombi:2008vd}.) In other words, the path integral tells us that the Hilbert space of the mode is the vacuum representation of two copies of the Virasoro group. The holomorphic contribution comes from $\bar{\phi}$, and the antiholomorphic from $\phi$.

In our previous work~\cite{Cotler:2018zff} we showed that the torus partition function of the Alekseev-Shatashvili model is one-loop exact by a localization argument. In a sense the argument is the quantization of the argument of Stanford and Witten~\cite{Stanford:2017thb} that the Schwarzian path integral is one-loop exact. The gist is to exponentiate the symplectic measure with ghosts. The total action, including the ghost term, is invariant under a Grassmann-odd BRST supercharge $Q$. Because the integration space at fixed time is not only symplectic but K\"ahler with a metric invariant under the action of $L_0$ and $\bar{L}_0$, there is a $Q$-exact term constructed from the K\"ahler metric whose bosonic part is positive definite. Adding this term to the action with a large positive coefficient leaves the partition function invariant, but localizes the path integral. We then conclude that the exact Alekseev-Shatashvili path integral is 
\beq
	Z(\tau) = |\chi_{0,c}(\tau)|^2\,.
\eeq

In this computation the spatial circle of the boundary is contractible in the bulk. There are infinitely other configurations where other combinations of boundary cycles are contractible in the bulk, and the path integral on each is given by $Z(\gamma \tau)$ for some modular transformation $\gamma\in PSL(2;\mathbb{Z})$. Gravity sums over this choice, so that the complete disk times circle partition function is
\beq
\label{E:MW}
	Z_{D\times \mathbb{S}^1}(\tau) = \sum_{\gamma\in PSL(2;\mathbb{Z})/\Gamma_{\infty}}|\chi_{0,c}(\gamma \tau)|^2\,.
\eeq 
Here $\Gamma_{\infty}$ is the subgroup of $PSL(2;\mathbb{Z})$ generated by the $T$ transformation, which leaves the vacuum character invariant.

The path integral analysis complements that of Maloney and Witten. Using~\cite{Witten:1987ty} they performed a K\"ahler quantization of the phase space of 3d gravity on the cylinder, the same one we mentioned above, $\faktor{\text{Diff}(\mathbb{S}^1)}{PSL(2;\mathbb{R})}\times \faktor{\text{Diff}(\mathbb{S}^1)}{PSL(2;\mathbb{R})}$. The ensuing Hilbert space is the vacuum representation of Virasoro at some central charge $c$. Passing over to imaginary time and taking the boundary to be a torus of complex structure $\tau$, the Euclidean path integral is the vacuum character, and the sum over other configurations gives the same partition function~\eqref{E:MW}.

There is a one-parameter family of deformations of the model~\eqref{E:ASeuclidean} that will appear in our wormhole analysis. In Euclidean signature, the deformation of the right-moving part is labeled by a constant $\bar{b}^2>-1$ and is given by
\beq
	S_{\rm AS}= \frac{C}{24\pi} \int d^2x \left( \frac{\bar{\phi}''\bar{\partial}\bar{\phi}'}{\bar{\phi}'^2} + \bar{b}^2 \bar{\phi}' \bar{\partial}\bar{\phi}\right)\,.
\eeq
At constant time the field $\bar{\phi}$ is an element of the quotient $\faktor{\text{Diff}(\mathbb{S}^1)}{U(1)}$, meaning it obeys the same boundary conditions as in~\eqref{E:diffBC}, but is subject to a gauge symmetry
\beq
	\bar{\phi}(x,y) \sim \bar{\phi}(x,y) + \bar{a}(y)\,.
\eeq
This more general Alekseev-Shatashvili path integral,
\beq
	Z_{\rm AS}(\tau|\bar{b}) = \int \prod_y \left( \frac{[d\bar{\phi}(y)]}{U(1)} \,\text{Pf}(\omega(y))\right) e^{-S_{\rm AS}} \,,
\eeq
is also one-loop exact (in fact it is secretly a quadratic theory after a non-local field redefinition), giving an ordinary holomorphic Virasoro character
\beq
\label{E:holoCharacter}
	Z_{\rm AS}(\tau|\bar{b}) = \chi_{h,c}(\tau) = q^{h-\frac{c}{24}} \prod_{n=1}^{\infty} \frac{1}{1-q^n}\,,
\eeq
where the renormalized central charge $c$ and scaling weight $h$ are
\beq
	c = C+1 \,, \qquad h = \frac{c-1}{24} + \frac{C\bar{b}^2}{24}\,.
\eeq
A left-moving mode with some $b$ would give rise to the antiholomorphic character with
\beq
	Z_{\rm AS}^*(\bar{\tau}|b) = \chi_{\bar{h},c}^*(\bar{\tau}) = \bar{q}^{\bar{h} - \frac{c}{24}}\prod_{n=1}^{\infty} \frac{1}{1-\bar{q}^n}\,, \qquad \bar{h} = \frac{c-1}{24} + \frac{Cb^2}{24}\,.
\eeq

In our wormhole analysis, we can imagine cutting the wormhole into two ``trumpets.'' We will see that each trumpet is endowed with two Alekseev-Shatashvili modes of this sort, one holomorphic and with one antiholomorphic, and characterized by some $b$ and $\bar{b}$. The total contribution from the trumpet is then a non-holomorphic character with scaling weights $(h,\bar{h})$ as above. In a picture, we have
\begin{equation*}
\includegraphics[width=5.5in]{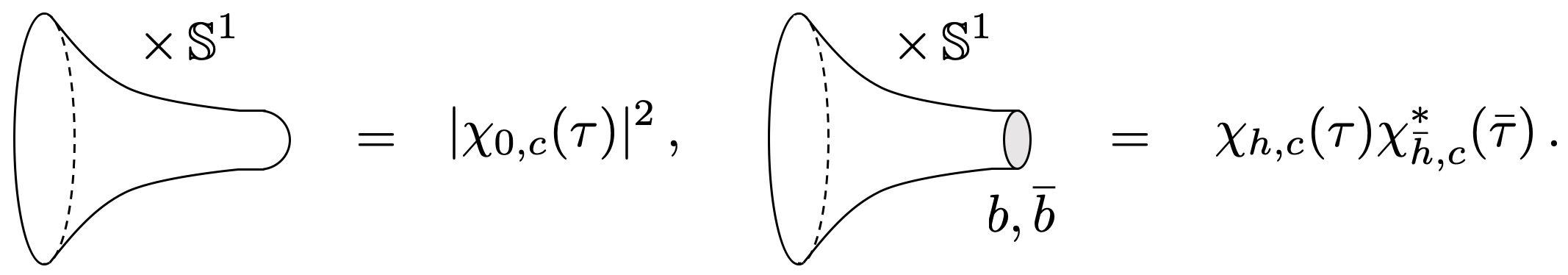}
\end{equation*}

\subsection{Spectral form factor in random matrix theory}
\label{S:SFF}

The spectral form factor lies at the focal point of the connection between our AdS$_3$ analysis and random matrix theory.  Here we provide a brief review of the spectral form factor in random matrix theory and gravity.  (See \cite{eynard2015random} for a modern review of random matrix theory.)

We begin with random matrix theory. Suppose we have an ensemble of $d \times d$ Hamiltonians, that is, a probability distribution $P(H)$ over the space of $d \times d$ Hermitian matrices.  Each $H$ has $d$ eigenvalues $E_1,...,E_d$, and we can write the averaged density of states as
\begin{align}
	\rho(E) &= \int dH \, P(H) \, \frac{1}{d} \sum_{i=1}^d \delta(E - E_i) = \left\langle \frac{1}{d} \sum_{i=1}^N \delta(E - E_i)  \right\rangle_{\text{ensemble}}\,.
\end{align}
Due to the ensemble average, $\rho(E)$ will often be a smooth function of $E$ instead of a sum of delta functions.

The density of states encodes the probability of finding an eigenvalue at energy $E$.  We can also consider two-point correlations such as
\beq
	\rho(E, E') = \left\langle \frac{1}{d^2} \sum_{i,j=1}^d \delta(E - E_i) \delta(E' - E_j) \right\rangle_{\text{ensemble}}
\eeq
which encodes the joint probability of finding an eigenvalue at energy $E$ and another eigenvalue at energy $E'$.  For ensembles of sufficiently ``generic'' Hamiltonians, the energy eigenvalues will repel one another.  Furthermore, in small, contiguous energy windows away from the edge of the spectrum, the pair correlations of eigenvalues are equivalent to the pair correlations of a classical one-dimensional Coulomb gas with logarithmic repulsion at finite temperature.  Thus the eigenvalues form a crystal in one dimension, which can be more rigid or more floppy depending on the strength of the level repulsion.  The long-range level repulsion is captured by a term in $\rho(E, E')$ of the form $\frac{1}{d^2 (E - E')^2}$.

The spectral form factor is the Fourier transform
\beq
	d^2 \int dE \, dE' \, \rho(E,E') \, e^{- i (E - E')T} = \left\langle \sum_{j,k=1}^{d} e^{-i(E_j - E_k)T}\right\rangle_{\text{ensemble}} =  \big\langle \text{tr}(e^{-i H T})\,\text{tr}(e^{i H T}) \big\rangle_{\text{ensemble}}\,.
\eeq
Due to the inverse square repulsion $\frac{1}{d^2 (E - E')^2}$, the spectral form factor will contain term linear in time $\sim T$.  Often it is convenient to consider a finite-temperature analog of the spectral form factor, namely
\beq
\label{eq:finiteTSFF}
	\big\langle \text{tr}\big(e^{-(\beta + i T) H}\big)\, \text{tr}\big(e^{- (\beta - i T) H }\big) \big\rangle_{\text{ensemble}}\,.
\eeq
This will also contain linear growth in time $\sim T$, called the ramp.

The time-dependence of the (finite-temperature) spectral form factor is as follows.  At time $T = 0$, its value is $\sim d^2$, which then decays to zero either exponentially or polynomially depending on the details of the ensemble.  Once the spectral form factor has decayed to an $O(1)$ value, the linear growth $\sim T$ (with small fluctuations around it) dominates until times of $O(d)$.  As stated above, this linear growth is due to long-range level-level repulsion.  At $T \sim O(d)$, the spectral form factor is probing short distances on the scale of the average nearest-neighbor level spacing, which causes the spectral form factor to become constant (with small fluctuations).

The spectral form factor has been computed for many random matrix theories and disordered theories. Using matrix fat graphs, the diagrams which correspond to the ramp correspond to discretized wormhole-like geometries of the connected part $\big\langle \text{tr}(e^{-i H t})\,\text{tr}(e^{i H t}) \big\rangle_{\text{ensemble},\,\text{conn.}}$ \cite{brezin1993universality}.  More recently, the initial decay and subsequent ramp of the spectral form factor was computed analytically and numerically in the SYK model \cite{Cotler:2016fpe, Saad:2018bqo}.  Here the ramp manifested as bulk gravitational configurations corresponding to a Euclidean wormhole.  Perhaps most interestingly, the spectral form factor can be computed in JT gravity, dual to a random matrix theory \cite{Saad:2019lba}.  One subtlety is that this random matrix theory is double scaled, i.e. $d \to \infty$, but in such a way that $e^{S_0}$ (where $S_0$ is a large genus expansion parameter) takes on the role of $d$ in the spectral form factor and other spectral correlators.  Again the ramp corresponds to Euclidean wormhole configurations in the gravity picture.

It is worth noting several aspects of the spectral form factor that are relevant to our analysis in the present paper.  First, suppose we can consider an ensemble of Hamiltonians with a common symmetry that allows us to block-diagonalize each Hamiltonian as
\begin{equation*}
	H = \begin{pmatrix}
	H_1 & & & \\
	& H_2 & & \\
	& & \ddots &
	\end{pmatrix}
\end{equation*}
Then generically, each block will behave like an independent random matrix.  That is, the eigenvalues within a particular $H_j$ will experience level repulsion, but there will not be level repulsion between eigenvalues of $H_j$ and eigenvalues of an $H_k$ for $j \not = k$.  As such, it is natural to compute the spectral form factor for each block independently, i.e.
\begin{equation*}
	\big\langle \text{tr}\big(e^{-(\beta + i T) H_j}\big)\, \text{tr}\big(e^{- (\beta - i T) H_j }\big) \big\rangle_{\text{ensemble}}\,.
\end{equation*}
Such spectral form factors will contain a late-time ramp $\sim T$.

One of the more notable features of the spectral form factor is that it is not a self-averaging quantity \cite{prange1997spectral}.  In particular, if we sample a random $\widetilde{H}$ from our ensemble $H$, the quantity 
\begin{equation}
	\label{eq:singleHamSFF1}
	\text{tr}\big(e^{-(\beta + i T) \widetilde{H}}\big)\, \text{tr}\big(e^{- (\beta - i T) \widetilde{H} }\big) 
\end{equation}
for large $d$ will not approximate the spectral form factor at late times in terms of having a large percent error.  Eqn.~\eqref{eq:singleHamSFF1} will have a ramp, but with large fluctuations that are the size of the height of the ramp itself.  Similarly the plateau at times $t \gtrsim O(d)$ is swamped by large fluctuations.  Only the early-time behavior corresponding to the decay of the spectral form factor is self-averaging.

If we take a single Hamiltonian (without regard to any ensemble) and compute Eq.~\eqref{eq:singleHamSFF1}, we will find similar behavior, i.e. the ramp and plateau will be present but nearly swamped out by large fluctuations.  It is only by ensemble averaging that these fluctuations become suppressed.

In this paper we effectively compute the connected part of the spectral form factor in pure AdS$_3$ gravity at late times (but before the plateau). On the boundary we have Virasoro symmetry, including a conserved, commuting Hamiltonian and momentum. The Virasoro symmetry implies that many states at a given energy and momentum are descendants of primary states, with quantum numbers fixed by those of the primaries. So we will work with primary states in sectors of fixed momentum.  Now the key question is whether the spectral form factor in AdS$_3$ gravity is more like Eq.~\eqref{eq:singleHamSFF1} with a single Hamiltonian, having a ramp with large fluctuations, or instead is like Eq.~\eqref{eq:finiteTSFF} which is an ensemble average and thus has a ramp with suppressed fluctuations.  We will provide evidence for the latter.  Furthermore, while we provide strong evidence that AdS$_3$ gravity is an ensemble-averaged theory, it is only resembles the random matrix theory of a single block-diagonal Hamiltonian in the large-time limit, with the blocks corresponding to sectors of fixed momentum. Outside of this regime, there is an intriguing departure from the discussion above. Pure AdS$_3$ gravity is, at its simplest, a matrix model with infinitely many non-independent matrices. More broadly, we conjecture that AdS$_3$ gravity is an ensemble from which one draws CFT partition functions, rather than quantum mechanical Hamiltonians, and that his ensemble becomes double-scaled random matrix theory in a certain limit.

\section{Wormholes}
\label{S:wormhole}

In this Section we compute the path integral of AdS$_3$ gravity for the torus times an interval. In our Hamiltonian framework it is more useful to think of the space as an annulus times a circle. See~\cite{scarinci2013universal, henneaux2020asymptotic} for previous work on the classical phase space of AdS$_3$ gravity on the annulus.

Let us begin by setting up the conventions, boundary conditions, and solving the constraints. We let $x\sim x+2\pi$, $y\sim y+2\pi$ parameterize a torus and $\rho\in \mathbb{R}$ be a radial coordinate which parameterizes the interval. The two conformal boundaries are approached as $\rho \to \pm \infty$. In this setting the asymptotically AdS$_3$ boundary conditions are that, as $\rho\to\infty$, the combinations $A$ and $\bar{A}$ defined in~\eqref{E:fromeToA} approach
\beq
\label{E:boundary1}
	A = \frac{1}{2}\begin{pmatrix}  d\rho & 0 \\ e^{\rho}(dx+\bar{\tau}_1 dy) & -d\rho\end{pmatrix} + O(e^{-\rho})\,, \qquad \bar{A} = \frac{1}{2}\begin{pmatrix} -d\rho & -e^{\rho}(dx+\tau_1dy) \\ 0 & d\rho\end{pmatrix}  + O(e^{-\rho})\,.
\eeq
With this choice, the spacetime metric approaches
\beq
	ds^2 \approx \frac{e^{2\rho}}{4}|dx+\tau_1dy|^2 + d\rho^2\,,
\eeq
so that the conformal boundary is indeed a torus of complex structure $\tau_1$. Note that we are imposing boundary conditions so that the spatial part of the dreibein and spin connections are real, but the temporal components have an imaginary part. This is required so that the metric is Euclidean near the boundary. In the path integral we integrate over real fluctuations of all of the fields. Similarly we impose that near the other boundary $\rho\to-\infty$ we have
\beq
\label{E:boundary2}
	A = \frac{1}{2}\begin{pmatrix} d\rho & e^{-\rho} (dx+\bar{\tau}_2dy) \\ 0 & -d\rho\end{pmatrix} +O(e^{\rho})\,, \qquad \bar{A} = \frac{1}{2}\begin{pmatrix}  -d\rho & 0 \\ -e^{\rho} (dx+\tau_2 dy) & d\rho \end{pmatrix} + O(e^{\rho})\,,
\eeq
so that the conformal boundary is a torus of complex structure $\tau_2$. With these boundary conditions, the spatial and temporal circles on boundary 1 respectively interpolate to the spatial and temporal circles on boundary 2.

Let us call the path integral with these boundary conditions and bulk topology $Z(\tau_1,\tau_2)$. This object is not the complete gravity path integral on $\mathbb{T}^2\times I$, which we denote as $Z_{\mathbb{T}^2\times I}(\tau_1,\tau_2)$. The latter includes a sum of $PSL(2;\mathbb{Z})$ Dehn twists of the torus on boundary 1 relative to the torus on boundary 2. For instance, we ought to sum over bulk configurations in which the spatial circle on boundary 1 smoothly interpolates to the temporal circle on boundary 2. As we will see at the end of Subsection~\ref{S:theResult}, the partition function $Z(\tau_1,\tau_2)$ includes an infinite sum of these Dehn twists generated by the axial $T$ transformation $(\tau_1,\tau_2)\to (\tau_1,\tau_2+1)$. As a result the total wormhole amplitude is given by
\beq
\label{E:dehn}
	Z_{\mathbb{T}^2\times I}(\tau_1,\tau_2) = \sum_{\gamma \in PSL(2;\mathbb{Z})/\Gamma_{\infty}} Z(\tau_1,\gamma\tau_2)\,,
\eeq
where $\Gamma_{\infty}$ is the subgroup generated by $T$, i.e. the subgroup of modular transformations which preserves $\tau = i\infty$. 

The Euclideanized gravity action, including boundary terms is
\begin{align}
\begin{split}
\label{E:euclideanS}
	S_E & = -\frac{ik}{4\pi} \int d^3x \,\varepsilon^{ij}\text{tr}\left( -A_i \partial_y A_j + A_y F_{ij}\right) - (A\to\bar{A}) + S_{\rm bdy}\,,
	\\
	S_{\rm bdy} &= \frac{ik}{4\pi}\left(  \int_{\rho\to\infty} d^2x \,\text{tr}\left( \bar{\tau}_1 A_x^2 - \tau_1 \bar{A}_x^2\right) + \int_{\rho\to-\infty} d^2x \,\text{tr}\left( \bar{\tau}_2 A_x^2 - \tau_2 \bar{A}_x^2\right)\right)\,.
\end{split}
\end{align}
One may verify that these boundary terms and the boundary conditions are consistent with a good variational principle. 

Integrating out $A_y$ and $\bar{A}_y$ imposes the Gauss' Law constraints $F_{ij} = \bar{F}_{ij} = 0$, which are solved by
\beq
\label{E:solvedConstraints}
	A_i = \tilde{G}^{-1} \partial_i \tilde{G}\,, \qquad \bar{A}_i = \tilde{\bar{G}}^{-1} \partial_i \tilde{\bar{G}}\,,
\eeq
where $\tilde{G}$ and $\tilde{\bar{G}}$ are $SL(2;\mathbb{R})$-valued fields which may be multi-valued around the spatial circle but which are periodic in $y$. This non-periodicity may be parameterized as
\beq
	\tilde{G} = e^{\lambda(y) x} G\,, \qquad \tilde{\bar{G}} = e^{\bar{\lambda}(y)x } \bar{G}\,,
\eeq
with $G$ and $\bar{G}$ periodic fields in the trivial winding sector of $SL(2;\mathbb{R})$. It is simple to show that smoothness of the spatial metric requires that $\lambda$ and $\bar{\lambda}$ are ``spacelike'' vectors in $\mathfrak{sl}(2;\mathbb{R})$. 

By decomposing $A$ into $\tilde{G}$ and $\bar{A}$ into $\tilde{\bar{G}}$ we introduce a redundancy under
\beq
	\tilde{G}(x,y,\rho) \to h(y) \tilde{G}(x,y,\rho)\,, \qquad \tilde{\bar{G}}(x,y,\rho)\to  \bar{h}(y)\tilde{\bar{G}}(x,y,\rho)\,, 
\eeq
for $h$ and $\bar{h}$ elements of $SL(2;\mathbb{R})$, since both configurations parameterize the same $A_i$ and $\bar{A}_i$. This redundancy may be partially alleviated by fixing the non-periodicity of $\tilde{G}$ and $\tilde{\bar{G}}$ in group space. We pick
\beq
\label{E:tildeG}
	\tilde{G} = e^{b(y) x J_1}G\,, \qquad \tilde{\bar{G}} =e^{\bar{b}(y) x J_1} \bar{G}\,.
\eeq
The redundancy is then only under those $h$ and $\bar{h}$ which commute with the non-periodic parts, meaning under transformations $h = e^{a(y) J_1}$ and $\bar{h} = e^{\bar{a}(y) J_1}$. 

If we were quantizing Chern-Simons theory rather than gravity, then $b(y)$ and $\bar{b}(y)$ would parameterize the holonomies of $A$ and $\bar{A}$ respectively around the spatial circle at time $y$. These holonomies would be in the hyperbolic conjugacy class of $SL(2;\mathbb{R})$. Their interpretation in gravity will become clear in the next Subsection.

To solve the boundary conditions we find it convenient to decompose
\beq
\label{E:G}
	G = e^{\phi J_1} e^{\Lambda J_2} e^{\psi(J_1-J_0)}\,, \qquad \bar{G} = e^{\bar{\phi}J_1}e^{-\bar{\Lambda}J_2}e^{\bar{\psi}(J_1+J_0)}\,.
\eeq
The parameters $b(y)$ and $\bar{b}(y)$ appear together with the fields $\phi$ and $\bar{\phi}$ in the combinations
\beq
	\Phi (x,y,\rho)= b(y)x + \phi(x,y,\rho)\,, \qquad \bar{\Phi}(x,y,\rho) = \bar{b}(y) x + \bar{\phi}(x,y,\rho)\,.
\eeq
The residual redundancy we described above implies that we identify
\beq
\label{E:U1gauge}
	\phi(x,y,\rho) \sim \phi(x,y,\rho) + a(y)\,, \qquad \bar{\phi}(x,y,\rho) \sim \bar{\phi}(x,y,\rho) + \bar{a}(y)\,.
\eeq
Following our previous work~\cite{Cotler:2018zff}, the boundary conditions imply that, at large $\rho$, the fields $\Lambda$ and $\psi$ are fixed in terms of $\Phi$, and $\bar{\Lambda}$ and $\bar{\psi}$ in terms of $\bar{\Phi}$ as
\beq
\label{E:solvedBC}
	\Lambda \approx  \ln \left( \frac{e^{\rho}}{\Phi'}\right)\,, \qquad \psi \approx -\frac{e^{-\rho}\Phi''}{\Phi'}\,, \qquad \bar{\Lambda} \approx \ln \left( \frac{e^{\rho}}{\bar{\Phi}'}\right)\,, \qquad \bar{\psi} \approx -\frac{e^{-\rho}\bar{\Phi}''}{\bar{\Phi}'}\,,
\eeq
with $\Phi$ and $\bar{\Phi}$ finite as $\rho\to\infty$. We denote $\Phi_1 = \lim_{\rho\to\infty} \Phi$ and $\bar{\Phi}_1 = \lim_{\rho\to\infty} \bar{\Phi}$. Similar statements hold near the other boundary.

Plugging~\eqref{E:solvedConstraints},~\eqref{E:tildeG},~\eqref{E:G}, and the asymptotic profiles~\eqref{E:solvedBC} into the action~\eqref{E:euclideanS}, we arrive at 
\begin{align}
\begin{split}
\label{E:bdyAction1}
	S_E = \frac{C}{24\pi} \int d^2x &\left( \frac{\Phi_1''\partial_1 \Phi_1'}{\Phi_1'^2} + \frac{\bar{\Phi}_1''\bar{\partial}_1 \bar{\Phi}_1'}{\bar{\Phi}_1'^2} - \frac{i}{2}\Big( \bar{\tau}_1\Phi_1'^2+\phi_1'\partial_y \phi_1-\tau_1\bar{\Phi}_1'^2-\bar{\phi}_1'\partial_y \bar{\phi}_1\Big) \right.
	\\
	&\quad  + \left.\frac{\Phi_2''\partial_2 \Phi_2'}{\Phi_2'^2} + \frac{\bar{\Phi}_2''\bar{\partial}_2 \bar{\Phi}_2'}{\bar{\Phi}_2'^2}-\frac{i}{2}\Big( \bar{\tau}_2 \Phi_2'^2 - \phi_2'\partial_y \phi_2 - \tau_2 \bar{\Phi}_2'^2 + \bar{\phi}_2'\partial_y \bar{\phi}_2\Big)\right)
	\\
	& \qquad \qquad - \frac{iC}{24}\int_0^{2\pi} dy\left( b^2 \partial_y Y - \bar{b}^2\partial_y \bar{Y}\right)\,.
\end{split}
\end{align}
In order to simplify this expression we have defined
\beq
	\partial_1 = -\frac{i}{2} (\bar{\tau}_1\partial_x + \partial_y)\,, \qquad \partial_2 = -\frac{i}{2}(\bar{\tau}_2\partial_x - \partial_y)\,,
\eeq
along with
\begin{align}
\begin{split}
\label{E:twists}
	Y(y) &= \frac{1}{2\pi b(y)} \int_0^{2\pi} dx(\phi_1(x,y)-\bar{\phi}_1(x,y))\,, 
	\\
	\bar{Y}(y) &= \frac{1}{2\pi \bar{b}(y)}\int_0^{2\pi} dx (\phi_2(x,y)-\bar{\phi}_2(x,y))\,.
\end{split}
\end{align}
One consistency check on the action~\eqref{E:bdyAction1} is that it is invariant under the gauge redundancy~\eqref{E:U1gauge}, which acts simultaneously on the $1$ and $2$ fields as
\beq
 	\phi_1(x,y)\sim \phi_1(x,y)+a(y)\,, \qquad \phi_2(x,y)\sim\phi_2(x,y)+a(y)\,,
\eeq
and similarly for the barred fields. The ``twist'' fields $Y$ and $\bar{Y}$ are in fact gauge-invariant. 

The effective action~\eqref{E:bdyAction1} is a bit complicated. We will simplify it shortly. For now, we note that it also has a single time derivative and therefore is in Hamiltonian form. Accordingly, the single time derivative term in the action determines a symplectic measure on the space of field configurations, which we will use to perform the path integral below.

\subsection{Representative wormholes}

\begin{figure}[t]
\begin{center}
\includegraphics[scale=.25]{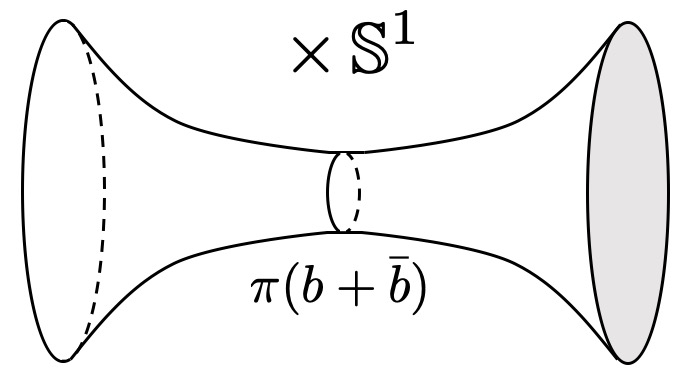}
\end{center}
\caption{\label{F:doubletrumpet} The wormhole in~\eqref{E:wormholeRep}. This is a bottleneck geometry, where the length of the bottleneck is $\pi|b+\bar{b}|$ and spinning fields are twisted by an amount determined by $b-\bar{b}$ after going around it.}
\end{figure}

Let us pause to consider some representative wormhole geometries before going on to compute the full path integral. Consider the configuration
\beq
	b(y) = b\,, \qquad \bar{b}(y) = \bar{b}\,, \qquad \phi=\bar{\phi} = 0\,, 
\eeq
with $\phi$ and $\bar{\phi}$ defined as in~\eqref{E:G}. This corresponds to $\phi_1=\phi_2=\bar{\phi}_1=\bar{\phi}_2=Y=\bar{Y}=0$. In fact this configuration is almost a saddle point of the constrained action~\eqref{E:bdyAction1} for all $b,\bar{b}$. After a shift of $\rho$, this configuration corresponds to
\beq
\label{E:wormholeRep}
	A_i dx^i = \frac{1}{2}\begin{pmatrix} d\rho & e^{-\rho}bdx \\ e^{\rho}b dx & -d\rho\end{pmatrix}\,, \qquad \bar{A}_i dx^i = \frac{1}{2}\begin{pmatrix} -d\rho & -e^{\rho}\bar{b}dx \\ -e^{-\rho}\bar{b} dx & d\rho\end{pmatrix}\,.
\eeq
Extracting the spatial dreibein, we arrive at a spatial metric (recall that in imposing the constraints we have integrated out the temporal component of the dreibein)
\beq
\label{E:spatialWormhole}
	ds^2_{\rm spatial} = \left( b\bar{b}\sinh^2(\rho) + \frac{(b+\bar{b})^2}{4}\right)dx^2 + d\rho^2\,.
\eeq
This is a bottleneck geometry, as in Fig.~\ref{F:doubletrumpet}. Clearly we must have $b\bar{b}>0$ in order for the wormhole to be smooth and non-singular. The bottleneck is characterized by a minimum length geodesic around the $x$-circle at $\rho=0$. It has length
\beq
	L = \pi |b+\bar{b}|\,,
\eeq
which gives an interpretation to the sum of $b$ and $\bar{b}$. 

To interpret the difference of $b$ and $\bar{b}$ we consider the spin connection. The wormhole described by~\eqref{E:wormholeRep} has a spin connection with some curvature. So the holonomy around some curve depends on the curve. However a natural way to describe the wormhole is the holonomy of the spin connection around a particular curve, the minimum length geodesic around the bottleneck. Recall that the spin connection is valued in $\mathfrak{sl}(2;\mathbb{R})=\mathfrak{so}(2,1)$. The holonomy in the two-dimensional representation is in the hyperbolic conjugacy class with
\beq
\label{E:wormholeHolo}
	\text{tr}\,\mathcal{P}\exp\left( \oint_{\rho=0} \omega\right) = 2\cosh\left( \frac{\pi(b-\bar{b})}{2}\right)\,.
\eeq
So fields with spin are twisted by an amount determined by $b-\bar{b}$ when going around the bottleneck.

Studying the spatial metric~\eqref{E:spatialWormhole} and holonomy~\eqref{E:wormholeHolo} we learn that these wormholes are characterized by arbitrary $b,\bar{b}\geq 0$. (The configurations with $b,\bar{b}\leq 0$ are equivalent to those with $b,\bar{b}\geq 0$.) Our continuation to imaginary time rears its head here. In our continuation we do not continue the gauge group for local Lorentz rotations from $SO(2,1)$ to $SO(3)$. This has the following consequence: the holonomy~\eqref{E:wormholeHolo} allows for arbitrarily large $b-\bar{b}$, whereas if the spin connection was for local $SO(3)$ rotations, its holonomies would be valued within a finite range.

The gravitational action~\eqref{E:bdyAction1} evaluated on~\eqref{E:spatialWormhole} is
\beq
	S_E = -\frac{i \pi C}{12}\left(  \bar{b}^2(\tau_1+\tau_2) - b^2(\bar{\tau}_1+\bar{\tau}_2)\right)\,.
\eeq
The fact that the action depends on $b$ and $\bar{b}$ tells us that the wormhole is a constrained saddle. It is a saddle only when we fix $b$ and $\bar{b}$, and the complete path integral will involve an integral over $b,\bar{b}\geq 0$.

This action also gives us another way to interpret the parameters $b$ and $\bar{b}$. The classical approximation to the path integral (upon fixing $b$ and $\bar{b}$) is
\beq
	e^{-S_E} = (q_1q_2)^{\frac{C\bar{b}^2}{24}} (\bar{q}_1\bar{q}_2)^{\frac{Cb^2}{24}}\,, \qquad q_i = e^{2\pi i \tau_i}\,.
\eeq
So $b$ and $\bar{b}$ correspond to the left- and right-moving energies respectively, $L_0 -\frac{c}{24} \sim C\bar{b}^2/24$ and $\bar{L}_0-\frac{c}{24} \sim Cb^2/24$, and both boundaries perceive the same left- and right-moving energies.

\subsection{Moduli space field range and measure}
\label{S:moduliMeasure}

The full set of constrained saddles is parameterized by constant $b,\bar{b}$ as above, in addition to twists between the two boundaries:
\beq
\label{E:wormholeRep2}
	b(y) = b\,, \quad \bar{b}(y) = \bar{b}\,, \quad \phi =b\gamma(\rho) =  b(\alpha(\rho)+\beta(\rho))\,, \quad \bar{\phi} = \bar{b}\bar{\gamma}(\rho) = \bar{b}(\alpha(\rho)-\beta(\rho))\,,
\eeq
where $\alpha$ and $\beta$ are finite at infinity with boundary values $\alpha_i$ and $\beta_i$ for $i=1,2$. The fields $\alpha$ and $\beta$ are related to the twists $Y$ and $\bar{Y}$ defined in~\eqref{E:twists} by
\beq
\label{E:generalTwists}
	Y = \alpha_1 -\alpha_2 + \beta_1- \beta_2\,, \qquad \bar{Y} = \alpha_1-\alpha_2 - \beta_1+\beta_2 \,.
\eeq
More precisely, the most general saddles are characterized by constant $b,\bar{b}$, and the boundary values of $\alpha$ and $\beta$. These are the wormhole moduli. As we will see, $\alpha_1$ and $\beta_1$ parameterize a spacetime translation on the boundary 1, while $\alpha_2$ and $\beta_2$ parameterize a translation on boundary 2. Common translations are trivial, and lead to the same configuration. Only relative translations are physical, and these are precisely the twists.

We already found above that the zero modes $b,\bar{b}$ are non-negative. Our goal in this Subsection is to obtain the field ranges of the twists and the measure on the moduli space.

As in~\eqref{E:wormholeRep} the wormhole configurations correspond to
\begin{align}
\begin{split}
\label{E:twistedWormhole}
	A_i dx^i& = \frac{1}{2}\begin{pmatrix} d\rho & e^{-\rho}b(dx+d\alpha+d\beta) \\ e^{\rho}b(dx+d\alpha+d\beta) & -d\rho\end{pmatrix}\,, 
	\\
	\bar{A}_i dx^i &= \frac{1}{2}\begin{pmatrix} -d\rho & -e^{\rho}\bar{b}(dx+d\alpha-d\beta) \\ -e^{-\rho}\bar{b}(dx+d\alpha-d\beta) & d\rho\end{pmatrix}\,.
\end{split}
\end{align}
Near both boundaries, the spatial geometry is approximately
\beq
	ds^2_{\rm spatial} \approx \frac{e^{2|\rho|}b\bar{b}}{4}\left( dx + d\alpha+d\beta\right)(dx+d\alpha-d\beta) + d\rho^2\,.
\eeq
We then see that a shift of $\alpha_1$ corresponds to a spatial translation on boundary 1, and a shift of $\alpha_2$ to a spatial translation on boundary 2. So $\alpha$ is compact with periodicity $2\pi$. However, $\beta$ appears to be non-compact. As result the ``vector twist'' $\frac{Y+\bar{Y}}{2}=\alpha = \alpha_1-\alpha_2$ is compact and it seems the ``axial twist'' $ \frac{Y-\bar{Y}}{2}=\beta = \beta_1-\beta_2$ is non-compact. 

To interpret this result we temporarily go back to Lorentzian signature, where the boundaries are cylinders $\mathbb{S}^1\times \mathbb{R}$. The asymptotic metric is
\beq
	ds^2 \approx \frac{e^{2|\rho|}b \bar{b}}{4}(dx+dt+d\alpha+d\beta)(dx-dt+d\alpha-d\beta) + d\rho^2\,,
\eeq
and so shifts in $\beta$ correspond to time translations. So axial twists $\beta=\beta_1-\beta_2$  are relative time translations of the boundaries and are obviously non-compact. From the last line of~\eqref{E:bdyAction1} we extract the symplectic structure on the moduli space
\beq
	\omega_{\rm moduli} = \frac{C}{24}(db^2 \wedge dY - d\bar{b}^2 \wedge d\bar{Y}) = \frac{C}{24}\Big((db^2-d\bar{b}^2)\wedge d\alpha + (db^2+d\bar{b}^2)\wedge d\beta\Big)\,,
\eeq
and so a measure $db^2 d\bar{b}^2 d\alpha d\beta \,\text{Pf}(\omega_{\rm moduli})$. 

In Euclidean signature however, the periodicity of Euclidean time suggests that we may rotate the contour of the zero mode $\beta$ so that axial twists correspond to relative Euclidean time translation. If $\tau_1$ and $\tau_2$ are pure imaginary, then the statement is simply that we rotate $\beta_1$ and $\beta_2$ (and so the twist $\beta$) to be pure imaginary.

This idea is simple enough, but its execution is a little more tricky than one might think, on account of the independent boundary complex structures.

Consider the full spacetime dreibein near the boundary, including the temporal part anchored down as a boundary condition. From~\eqref{E:boundary1} and~\eqref{E:twistedWormhole} we have at large positive $\rho$
\begin{align}
\begin{split}
	e^+ &= e^0 + e^1 \approx \frac{e^{\rho}b}{2}(dx+\bar{\tau}_1 dy+d\gamma)\,,
	\\
	e^- &= e^0-e^1 \approx - \frac{e^{\rho}\bar{b}}{2}(dx+\tau_1dy+d\bar{\gamma} )\,,
\end{split}
\end{align}
and at large negative $\rho$,
\begin{align}
\begin{split}
	e^+ & \approx \frac{e^{-\rho}\bar{b}}{2}(dx+\tau_2 dy+d\bar{\gamma})\,,
	\\
	e^- & \approx -\frac{e^{-\rho}b}{2}(dx+\bar{\tau}_2dy+d\gamma)\,,
\end{split}
\end{align}
with $e^2 = d\rho$ in both regions. We proceed by parameterizing the translations $\gamma$ and $\bar{\gamma}$ near the boundaries as
\begin{align}
\begin{split}
\label{E:gammaSplit}
	\gamma&\approx  \alpha(\rho) + \beta(\rho) \times \begin{cases} \bar{\tau}_1\,, & \rho\to \infty\, \\ \bar{\tau}_2\,, & \rho \to -\infty\,\end{cases} \,,
	\\
	\gamma&\approx \alpha(\rho) +\beta(\rho)\times \begin{cases} \tau_1\,,  & \rho \to\infty \, \\ \tau_2\,, & \rho \to-\infty \,\end{cases}
\end{split}
\end{align}
with $\alpha(\rho)$ and $\beta(\rho)$ real. This definition for $\alpha$ is consistent with the one we used above, but the definition for $\beta$ is different, in particular it is not purely axial. Crucially we are allowing for the zero modes to acquire an imaginary part. With this decomposition, at large positive $\rho$ we have
\beq
	e^+ \approx \frac{e^{\rho}b}{2}(dx+d\alpha+\bar{\tau}_1 (dy+d\beta))\,, \qquad e^- \approx -\frac{e^{\rho}\bar{b}}{2}(dx+d\alpha + \tau_1(dy+d\beta))\,,
\eeq
and similarly at boundary 2. With this continuation shifts of $\alpha_{i}$ are spatial translations and shifts of $\beta_{i}$ are translations in imaginary time. So, as expected, the $\alpha_{i}$ and $\beta_{i}$ are compact with periodicity $2\pi$. 

Points in the moduli space are labeled not by the individual translations $\alpha_i$ and $\beta_j$ but by the relative translations $\alpha = \alpha_1-\alpha_2$ and $\beta=\beta_1-\beta_2$. The original twists $Y$ and $\bar{Y}$, however, are not functions on the moduli space: we have $Y = \alpha +(\bar{\tau}_1\beta_1-\bar{\tau}_2\beta_2)$ and $\bar{Y} = \alpha + (\tau_1\beta_1-\tau_2\beta_2)$ which only depends on $\beta$ when the two complex structures are aligned as $\tau_1=\tau_2=\tau$. 

Let us take this limiting case of $\tau_1 = \tau_2 = \tau$. Then from the last line of~\eqref{E:bdyAction1} we extract a symplectic form on moduli space
\beq
\label{E:limitingMeasure}
	\omega_{\rm moduli} = \frac{C}{24}\left( (db^2-d\bar{b}^2)\wedge d\alpha + (\bar{\tau} db^2-\tau d\bar{b}^2) \wedge d\beta\right) \,.
\eeq
This leads to a consistent symplectic measure on the moduli space $db^2 d\bar{b}^2d\alpha d\beta \,\text{Pf}(\omega_{\rm moduli}) \propto db^2 d\bar{b}^2 d\alpha d\beta \,\text{Im}(\tau)$. 

However, for more general $\tau_1$ and $\tau_2$, the $p\dot{q}$ term in~\eqref{E:bdyAction1} does not lead to a symplectic form since the twists are functions of the individual $\beta_i$ and not $\beta$. To illustrate the point, suppose that we define the temporal twist to only act on boundary $1$, i.e.~$\beta_2 = 0$ and $\beta = \beta_1$. Then we would have a putative symplectic form $\propto (db^2-d\bar{b}^2)\wedge d\alpha + (\bar{\tau}_1db^2-\tau_1 d\bar{b}^2)\wedge d\beta$ and arrive at a measure $\propto db^2 d\bar{b}^2 d\alpha d\beta \,\text{Im}(\tau_1)$. Clearly we would get a different answer if we defined our axial twist to only act on boundary $2$. 

So we are forced to analytically continue both $\omega_{\rm moduli}$ and the volume form when the complex structures are no longer aligned. We view this as part of the continuation to imaginary time. We define
\beq
	  \Omega = \frac{C}{24}\left( (db^2 - d\bar{b}^2)\wedge d\alpha + (\bar{\tau}_1 db^2-\tau_1 d\bar{b}^2) \otimes d\beta  - d\beta \otimes (\bar{\tau}_2 db^2 - \tau_2 d\bar{b}^2)\right)\,,
\eeq
which is a tensor on the moduli space and reduces to $\omega_{\rm moduli}$ in~\eqref{E:limitingMeasure} when $\tau_1=\tau_2$. $\Omega$ is non-degenerate and so defines a covariant measure,
\beq
\label{E:finalMeasure}
	db^2 d\bar{b}^2 d\alpha d\beta \sqrt{|\Omega|}\,, \qquad  \sqrt{|\Omega|} = \left(\frac{C}{24}\right)^2 2\sqrt{\text{Im}(\tau_1)\text{Im}(\tau_2)}\,,
\eeq
which is indeed a volume form on the moduli space. Up to the factor of $\sqrt{\text{Im}(\tau_1)\text{Im}(\tau_2)}$ this is two copies of the Weil-Petersson measure on the moduli of the hyperbolic cylinder.

To summarize, the moduli may be labeled by $(b,\bar{b},\alpha,\beta)$ with $b,\bar{b}\geq 0$. The spatial and temporal twists $\alpha$ and $\beta$ are compact zero modes of periodicity $2\pi$. The moduli space measure is given by~\eqref{E:finalMeasure}, and it will play a pivotal role in our analysis. In particular, its dependence on $\tau_1$ and $\tau_2$ ensures that the wormhole amplitude is modular invariant.

It is worth noting that if we were studying $SO(2,2)$ Chern-Simons theory instead of gravity, then $b,\bar{b}\geq 0$ correspond to hyperbolic holonomies and the twists $Y$ and $\bar{Y}$ would be non-compact. So the measure would be proportional to $db^2 d\bar{b}^2 dYd\bar{Y}$. This is a perturbative (1-loop) difference between Chern-Simons theory and gravity on the space $\mathbb{T}^2\times I$. However, we are of the opinion that it should be regarded as a non-perturbative difference between the two theories, since in gravity the wormhole is already a non-perturbative effect.

\subsection{The path integral}
\label{S:theResult}

We now compute the wormhole amplitude. Recall the boundary action~\eqref{E:bdyAction1}. We proceed by redefining the fields $\phi_i$ and $\bar{\phi}_i$ so that the twist fields $Y$ and $\bar{Y}$ are independent degrees of freedom. Given the subtleties about twist zero modes that we discussed in the last Subsection, we explicitly separate out the axial twist zero mode:
\beq
	\phi_1(x,y) \to \phi_1(x,y) + \alpha(y)+\beta(y)+\bar{\tau}_1 \beta\,,  \qquad
	\bar{\phi}_1(x,y)  \to \bar{\phi}_1(x,y) + \alpha(y) - \beta(y) +\tau_1\beta\,, 
\eeq
while leaving $\phi_2$ and $\bar{\phi}_2$ alone. Here $\alpha(y)$ is an arbitrary function of Euclidean time with $\alpha \sim \alpha + 2\pi$, while $\beta(y)$ is non-compact and has no zero mode, i.e. $\int_0^{2\pi} dy \,\beta(y) = 0$. The cost of performing this redefinition is that we must introduce gauge symmetries. The redundancies~\eqref{E:U1gauge} are enhanced to 
\begin{align}
\begin{split}
	\phi_1(x,y) & \sim \phi_1(x,y)+a_1(y)\,, \qquad \phi_2(x,y)\sim \phi_2(x,y) + a_2(y)\,,
	\\
	\bar{\phi}_1(x,y) & \sim \bar{\phi}_1(x,y)+\bar{a}_1(y)\,, \qquad \bar{\phi}_2(x,y) \sim \bar{\phi}_2(x,y) + \bar{a}_2(y)\,.
\end{split}
\end{align}
The action~\eqref{E:bdyAction1} becomes
\begin{align}
\begin{split}
\label{E:bdyAction2}
	S_E  = \frac{C}{24\pi} \int d^2x& \left( \frac{\Phi_1''\partial_1 \Phi_1'}{\Phi_1'^2} + \frac{\bar{\Phi}_1''\bar{\partial}_1 \bar{\Phi}_1'}{\bar{\Phi}'^2} -\frac{i}{2}\left( \bar{\tau}_1 \Phi_1'^2+\phi_1'\partial_y \phi_1 - \tau_1 \bar{\Phi}_1'^2 - \bar{\phi}_1'\partial_y \bar{\phi}_1\right)\right.
	\\
	 & \quad + \left.\frac{\Phi_2''\partial_2 \Phi_2'}{\Phi_2'^2} + \frac{\bar{\Phi}_2''\bar{\partial}_2 \bar{\Phi}_2'}{\bar{\Phi}_2'^2} - \frac{i}{2}\left( \bar{\tau}_2 \Phi_2'^2 - \phi_2 \partial_y \phi_2 - \tau_2 \bar{\Phi}_2'^2 + \bar{\phi}_2\partial_y \bar{\phi}_2\right)\right)
	 \\
	 & \quad \qquad - \frac{iC}{24}\int_0^{2\pi} dy \left( (b^2-\bar{b}^2) \partial_y \alpha + (b^2+\bar{b}^2)\partial_y \beta\right)\,.
\end{split}
\end{align}

Our next step is to integrate out the twists $\alpha(y)$ and $\beta(y)$, which only appear in the last line of the action through a $p\dot{q}$ term. Clearly $\alpha$ and $\beta$ act as Lagrangian multipliers enforcing that $b(y)$ and $\bar{b}(y)$ are constants. But let us proceed slowly and compute the exact effect, including the normalization. 

Consider a simpler version of the problem at hand. Let $p(y)$ be a non-compact variable with some range and $q(y)$ compact with periodicity $2\pi$, and an action $S_E = i c \int_0^{2\pi} dy \,p\partial_y q$. We may expand $p$ and $q$ into Fourier and winding modes
\beq
	p(y) =p+ \sum_{m=\neq 0}\tilde{p}_m e^{imy} \,, \qquad q(y) = q + w y + \sum_{m\neq 0}\tilde{q}_m e^{imy}\,,
\eeq
where $(p,q)$ are the zero modes, $w$ is a winding number of $q$ around the thermal circle, and the Fourier expansion coefficients satisfy $\tilde{X}_m^* = \tilde{X}_{-m}$. So we can eliminate the negative modes and regard the $\tilde{X}_{m>0}$ as independent complex variables. Because the action is of the $p\dot{q}$ form there is a symplectic structure $\omega =c \,dq \wedge dp$ on the space of zero modes, and on the space of nonzero Fourier modes. At fixed $m$ we have 
\beq
	\omega_m = c\, dq_m \wedge dp_{-m} \,,
\eeq
so that the integration measure for fixed $m>0$ is $d^2q_m d^2p_m \,\text{Pf}(\omega_m) = d^2q_m d^2p_m c^2$. In this parameterization the action becomes
\beq
	S_E = 2\pi c \sum_{m>0} m( p_m q_m^*-p_m^*q_m) + 2\pi i w  c p\,.
\eeq
Then we have as a general identity for any functional $\mathcal{F}(p(y))$,
\begin{align}
\begin{split}
	\int [dq][dp] e^{-S_E} \mathcal{F}(p(y)) &= \sum_{w=-\infty}^{\infty} \int (dq dp \,c)\prod_{m>0}\left( d^2q_md^2p_m \,c^2\right) e^{-S_E} \mathcal{F}(p(y))
	\\
	& = \prod_{m>0}\frac{1}{m^2}\int (dq dp\, c)  \,\mathcal{F}(p) \sum_{w=-\infty}^{\infty} e^{-2\pi i w c p}
	\\
	& = \frac{1}{2\pi} \int (dq dp\,c) \,\mathcal{F}(p) \sum_{n=-\infty}^{\infty} \delta(cp-n)\,,
	\\
	& = \sum_n \mathcal{F}(p_n)\,, \qquad p_n = \frac{n}{c}\,.
\end{split}
\end{align}
In going from the first line to the second we integrated out the Fourier expansion coefficients, which sets $p$ to constant at the cost of a Jacobian which we Zeta-regularize as
\beq
	\frac{1}{\text{det}'(|\partial_y|)} = \prod_{m>0}\frac{1}{m^2} =e^{-2\sum_{m>0}\ln m} \to e^{2\zeta'(0)} = \frac{1}{2\pi}\,.
\eeq
In going from the second line to the third we use the Poisson summation formula, $\sum_{w=-\infty}^{\infty} e^{-2\pi i w c p} = \sum_{n=-\infty}^{\infty} \delta(cp-n)$. Finally, in the last line we used that the field range of $q$ is $2\pi$ and the sum is taken over those values of $p_n$ which lie within the original integration range of the zero mode $p$. 

To summarize, the Fourier modes of $q$ set $p$ to be constant up to a Jacobian $\frac{1}{2\pi}$; the winding modes quantize $cp$\,; and there is a remaining integral over the zero modes of $q$ and $p$. 

Now we turn to the problem we really want to solve. Let us decompose $b(y)$, $\bar{b}(y)$, $\alpha(y)$, and $\beta(y)$ into Fourier and winding modes. We have
\begin{align}
\begin{split}
	b(y)^2 & = b^2 + \sum_{m\neq 0} \tilde{b}_m e^{i m y}\,, \qquad \qquad\, \bar{b}(y)^2 = \bar{b}^2 + \sum_{m\neq 0}\tilde{\bar{b}}_m e^{i m y}\,,
	\\
	\alpha(y) & = \alpha + w y+\sum_{m\neq 0}\tilde{\alpha}_m e^{imy} \,, \qquad \beta(y)  = \sum_{m\neq 0}\tilde{\beta}_m e^{i my}\,.
\end{split}
\end{align}
The zero modes $(b^2,\bar{b}^2,\alpha)$ along with $\beta$ defined above parameterize the moduli space, and $w$ is a winding number. Integrating out the Fourier modes of $\alpha_m$ and $\beta_m$ sets $b$ and $\bar{b}$ to be constant with a Jacobian of $\frac{1}{(2\pi)^2}$ since we have integrated out two sets of modes rather than one. This Jacobian is canceled by the $(2\pi)^2$ field range of the twist zero modes $\alpha$ and $\beta$. The sum over windings quantizes
\beq
\label{E:quantizedJ}
	\frac{C(\bar{b}^2-b^2)}{24} \in \mathbb{Z}\,.
\eeq
As we will see shortly this is the integer quantization of spin.

It remains to integrate over the fields $\phi_i$ and $\bar{\phi}_i$. Once $b$ and $\bar{b}$ are constant we may perform a field redefinition which simplifies the problem, namely
\begin{align}
\begin{split}
	\Phi_1 &= b x + \phi_1 \to b \phi_1 \,, \qquad \Phi_2 = b x + \phi_2 \to b \phi_2\,,
	\\
	\bar{\Phi}_1 & = \bar{b}x + \bar{\phi}_1 \to \bar{b}\bar{\phi}_1\,, \qquad \bar{\Phi}_2 = \bar{b}x+\bar{\phi}_2\to \bar{b}\bar{\phi_2}\,.
\end{split}
\end{align}
So defined the fields $\phi_1,\phi_2,$ etc., are at fixed time elements of $\faktor{\text{Diff}(\mathbb{S}^1)}{U(1)}$, meaning
\begin{align}
\begin{split}
	\phi_1(x+2\pi,y) &= \phi_1(x,y)+2\pi\,, 
	\\
	\phi_1(x,y+2\pi) &= \phi_1(x,y)\,, 
	\\
	 \phi_1(x,y)&\sim \phi_1(x,y) + \phi_1(x,y)+a_1(y)\,,
\end{split}
\end{align}
and similarly for the other fields. The remaining part of the action, the first two lines of~\eqref{E:bdyAction2}, then becomes four decoupled copies of the Alekseev-Shatashvili theory we discussed in Section~\ref{S:AS},
\begin{align}
\begin{split}
	S = \frac{C}{24}\int d^2x& \left( \frac{\phi_1''\partial_1 \phi_1'}{\phi_1'^2} + b^2 \phi_1'\partial_1 \phi_1 + \frac{\bar{\phi}_1''\bar{\partial}_1 \bar{\phi}_1'}{\bar{\phi}_1'^2} + \bar{b}^2 \bar{\phi}_1'\bar{\partial}_1\bar{\phi}_1\right.
	\\
	& \qquad +\left. \frac{\phi_2''\partial_2 \phi_2}{\phi_2'^2} + b^2 \phi_2'\partial_2 \phi_2 + \frac{\bar{\phi}_2''\bar{\partial}_2 \bar{\phi}_2'}{\bar{\phi}_2'^2} + \bar{b}^2 \bar{\phi}_2'\bar{\partial}_2 \bar{\phi}\right)\,.
\end{split}
\end{align}
The Alekseev-Shatashvili path integral over each mode is one-loop exact \cite{Cotler:2018zff}. The integrals over $\bar{\phi}_1$ and $\bar{\phi}_2$ produce holomorphic Virasoro characters~\eqref{E:holoCharacter} which we reprise as
\beq
\label{E:ZAS}
	Z_{\rm AS}(\tau|\bar{b}) = \chi_{h,c}(\tau)= q^{h-\frac{c}{24}}\prod_{n=1}^{\infty} \frac{1}{1-q^n}\,, \quad c=C+1\,, \quad	h = \frac{c-1}{24} + \frac{C\bar{b}^2}{24}\,,
\eeq
while the integrals over $\phi_1$ and $\phi_2$ produce antiholomorphic characters with $\bar{h} = \frac{c-1}{24} + \frac{Cb^2}{24}$. By~\eqref{E:quantizedJ} the spin is then quantized:
\beq
	h - \bar{h} \in \mathbb{Z}\,.
\eeq

Converting the integration over $b,\bar{b}\geq 0$ to an integral over $h,\bar{h}\geq \frac{c-1}{24}$, we then have
\begin{align}
\begin{split}
\label{E:preThere}
	Z(\tau_1,\tau_2) = 2\sqrt{\text{Im}(\tau_1)\text{Im}(\tau_2)} \int_{\frac{c-1}{24}}^{\infty} d h d\bar{h}& 	 \,Z_{\rm AS}^*(\bar{\tau}_1|b) Z_{\rm AS}(\tau_1|\bar{b}) Z_{\rm AS}^*(\bar{\tau}_2|b)Z_{\rm AS}(\tau_2|\bar{b}) 
	\\
	& \qquad \times  \sum_{n=-\infty}^{\infty} \delta(h-\bar{h}-n)\,.
\end{split}
\end{align}
Using the Poisson summation formula and~\eqref{E:ZAS} we can rewrite this as
\begin{align}
\begin{split}
\label{E:almostThere}
	Z(\tau_1,\tau_2) &=\frac{ 2\sqrt{\text{Im}(\tau_1)\text{Im}(\tau_2)}}{|\eta(\tau_1)|^2|\eta(\tau_2)|^2}\sum_{n=-\infty}^{\infty} \int_{\frac{c-1}{24}}^{\infty} dh d\bar{h} \,e^{2\pi i h (\tau_1+\tau_2+n)}e^{- 2\pi i \bar{h}(\bar{\tau}_1+\bar{\tau}_2+n)}
	\\
	& = \frac{1}{2\pi^2} Z_0(\tau_1)Z_0(\tau_2) \sum_{n=-\infty}^{\infty} \frac{\text{Im}(\tau_1)\text{Im}(\tau_2)}{|\tau_1+\tau_2+n|^2}=\sum_{n=-\infty}^{\infty} \tilde{Z}(\tau_1,\tau_2+n)\,,
\end{split}
\end{align}
where $Z_0(\tau) $ is the modular invariant partition function of a non-compact boson $ \frac{1}{\sqrt{\text{Im}(\tau)}|\eta(\tau)|^2}$. The sum over windings $n$ corresponds to a partial sum over Dehn twists, relative $T^n$ transformations of the boundaries. The basic object $\tilde{Z}$ is the gravity path integral on $\mathbb{T}^2\times I$ without any Dehn twist at all between the boundaries.

We are nearly finished. Before going on, we note that the more basic object $\tilde{Z}$ is invariant under simultaneous modular transformations,
\beq
	\tilde{Z}(\gamma \tau_1,\gamma^{-1}\tau_2) = \tilde{Z}(\tau_1,\tau_2)\,, \qquad \gamma \tau = \frac{a\tau+b}{c\tau+d}\,,
\eeq
for $\gamma\in PSL(2;\mathbb{Z})$. In fact our amplitude had to be invariant under this simultaneous transformation for the same reason why Chern-Simons theory on $\mathbb{T}^2\times I$ is as we discussed in Subsection~\ref{S:continuation}. This is a strong consistency check on our result. 

We note that had we not continued the axial twist as in Subsection~\ref{S:moduliMeasure} we would have instead gotten infinity times a non-modular-invariant result. So modular invariance in fact tells us that we had to continue the axial twist. Relatedly, modular invariance combined with the fact that the moduli space is symplectic for $\tau_1=\tau_2$ together fix the moduli space measure we landed on in~\eqref{E:finalMeasure}.\footnote{With the result~\eqref{E:almostThere} in hand, modular invariance alone implies that the measure must be proportional to $\sqrt{\text{Im}(\tau_1)\text{Im}(\tau_2)} \,G\left( \frac{\text{Im}(\tau_1)\text{Im}(\tau_2)}{|\tau_1+\tau_2|^2}\right)$ for some function $G$. However recall that for $\tau_1=\tau_2=\tau$ the measure was proportional to $\text{Im}(\tau)$. This fixes $G$ to be a constant, and so the measure in~\eqref{E:finalMeasure} is the unique one consistent with modular invariance and the aligned limit $\tau_1=\tau_2$.}

As we discussed before~\eqref{E:dehn}, the complete wormhole partition function $Z_{\mathbb{T}^2\times I}$ is given by the sum $\sum_{\gamma\in PSL(2;\mathbb{Z})/\Gamma_{\infty}}Z(\tau_1,\gamma\tau_2)$, corresponding to a sum over relative Dehn twists. This Maloney-Witten-like modular sum over $Z(\tau_1,\gamma\tau_2)$ then gives our main result,
\beq
\label{E:mainResult2}
	\boxed{ Z_{\mathbb{T}^2\times I}(\tau_1,\tau_2)= \frac{1}{2\pi^2}\, Z_0(\tau_1)Z_0(\tau_2) \sum_{\gamma\in PSL(2;\mathbb{Z})}\frac{\text{Im}(\tau_1)\text{Im}(\gamma\tau_2)}{|\tau_1+\gamma\tau_2|^2}\,}
\eeq

\section{The modular sum}
\label{S:ModularSum}

In the last Section we computed the path integral of AdS$_3$ gravity on $\mathbb{T}^2\times I$. The result is just above, in~\eqref{E:mainResult}, and is expressed as a modular sum over $PSL(2;\mathbb{Z})$. The goal of this Section is to compute the modular sum, and to process it into something more useful. Because $Z_0(\tau)$ is modular invariant, we need only compute the Poincar\'{e} series
\beq
\label{E:preSum}
	\mathcal{F}(\tau_1,\tau_2) = \sum_{\gamma\in PSL(2;\mathbb{Z})} \frac{\text{Im}(\tau_1)\text{Im}(\gamma\tau_2)}{|\tau_1 + \gamma \tau_2|^2}\,.
\eeq
This sum does not converge. However the divergence is particularly simple and it can be understood by a variant of Zeta regularization.

To simplify the notation, we henceforth work with complex variables $z$ and $w$ rather than $\tau_1$ and $\tau_2$, with $z=z_1+iz_2$ and $w=w_1+w_2$. Define a Zeta function
\beq
\label{E:sZeta1}
	\mathcal{F}_s(z,w)=	\sum_{\gamma\in PSL(2;\mathbb{Z})} \left( \frac{\text{Im}(z)\text{Im}(\gamma w)}{|z + \gamma w|^2} \right)^s\,,
\eeq
which converges for $\text{Re}(s) > 1$, and which becomes our sum at $s=1$.  Further as $s\to_+1$ we find that
\beq
	\mathcal{F}_s(z,w) = \frac{3}{s-1} + (\text{finite})\,.
\eeq
This demonstrates that the modular sum $\mathcal{F}(z,w)$ in~\eqref{E:preSum} diverges. Zeta-regularizing the sum with $s=1+\varepsilon$ and taking $\varepsilon\to 0$, there is a $1/\varepsilon$ pole with constant coefficient, independent of $z$ and $w$, and a finite remainder. This finite remainder is almost independent of the regulatory scheme. In Appendix~\ref{App:ZetaReg} we demonstrate that other Zeta-based regulatory schemes, which have the virtue of maintaining modular invariance, also lead to finite pieces which agree up to an additive constant. So it seems reasonable to define a renormalized version of the prefactor $\mathcal{F}(z,w)$ by this finite remainder, up to addition by a constant.

When it converges the Zeta-regularized sum $\mathcal{F}_s(z,w)$ is invariant under independent modular transformations, $\mathcal{F}_s(z,w) = \mathcal{F}_s(\gamma_1 z,w) = \mathcal{F}_s(z,\gamma_2w)$ and in particular under independent $T$ transformations, $z \to z+1$, and $w\to w+1$. As a result it has a well-defined Fourier series in the real parts of $z$ and $w$,
\beq
	\mathcal{F}_s(z,w) = \sum_{s_1,s_2=-\infty}^{\infty} e^{-2\pi i z_1 s_1 -2\pi i w_1 s_2} \widetilde{F}_{s,s_1,s_2}(z_2,w_2)\,.
\eeq
Because the divergence as $s\to 1$ is independent of $z$ and $w$, it follows that the double Fourier transform of the modular sum for $\mathcal{F}(z,w)$ converges, as long as at least one of the momenta $s_i$ is nonzero. The divergence only contributes to a constant divergence in the zero momentum subsector. So as long as we work with the Fourier transformed version of our sum, we can set $s=1$ from the start.

In the remainder of this Section we compute this double Fourier transform and analyze it in some detail.

\subsection{Fourier series expansion}

The Fourier coefficients in question are
\beq
	\widetilde{F}_{s_1,s_2} = \int_0^1 dz_1 \int_0^1 dw_1\,e^{2\pi i z_1 s_1 +2\pi i w_1 s_2} \mathcal{F}(z,w)\,.
\eeq
To evaluate $\widetilde{F}_{s_1, s_2}$, we make several simplifications enabled by the modular sum and the symmetries of $\mathcal{F}$.  Defining
\beq
	f(z,w) = \frac{\text{Im}(z) \text{Im}(w)}{|z + w|^2}\,,
\eeq
we can write $\mathcal{F}(z,w) = \sum_{\gamma \in PSL(2;\mathbb{Z})} f(z, \gamma w)$ and similarly
\beq
	\widetilde{F}_{s_1,s_2} = \sum_{\gamma \in PSL(2;\mathbb{Z})} \int_0^1 dz_1 \int_0^1 dw_1\,e^{2\pi i z_1 s_1 +2\pi i w_1 s_2} f(z,\gamma w)\,.
\eeq
Note that $f(z, \gamma w) = f(\gamma^{-1} z, w)$.  We can parameterize a general element $\gamma$ of $SL(2;\mathbb{Z})$ by an integer matrix
\begin{equation}
	\begin{pmatrix}
	a & & b \\ c & & d
	\end{pmatrix}\,, \qquad a d - b c = 1\,,
\end{equation}
which acts as $\gamma\tau =\frac{a \tau + b}{c \tau + d}$.  For $c$ and $d$ coprime, let us denote by $\gamma_{c,d}$ some element of $SL(2;\mathbb{Z})$ of the form
\begin{equation*}
\begin{pmatrix}
* && * \\ c && d
\end{pmatrix}\,.
\end{equation*}
There may be multiple such elements of this form (and there is guaranteed to be at least one), but for our purposes the specific choice will not matter.  Letting $(\mathbb{Z}/c\mathbb{Z})^*$ denote the residue classes mod $c$ which are multiplicatively invertible (more concretely, consider the set of integers from $1$ to $c-1$ which are coprime to $c$), we show in Appendix~\ref{App:FourierSum} that
\begin{align}
\begin{split}
\label{E:FourierFsimp2}
	\widetilde{F}_{s_1,s_2} &= \sum_{n \in \mathbb{Z}} \int_0^1 dz_1 \int_0^1 dw_1 \, e^{2 \pi i z_1 s_1 + 2\pi i w_1 s_2}f(z, w + n)  
	\\
	& \qquad + \sum_{c \geq 1,\,\,d \in (\mathbb{Z}/c\mathbb{Z})^*} \int_{-\infty}^\infty dz_1 \int_{-\infty}^{\infty} dw_1\,e^{2\pi i z_1 s_1 +2\pi i w_1 s_2} f(z,\gamma_{c,d} w)\,.
\end{split}
\end{align}
The first term equals
\begin{align}
\label{E:Fs1s2p1}
	\sum_{n \in \mathbb{Z}} \int_0^1 dz_1 \int_0^1 dw_1 \, e^{2 \pi i z_1 s_1 + 2\pi i w_1 s_2}f(z, w + n) = \frac{\pi z_2 w_2}{z_2 + w_2} \, e^{- 2\pi (z_2 + w_2) |s_1|} \,\delta_{s_1, s_2}\,.
\end{align}
Let us denote the second term in~\eqref{E:FourierFsimp2} by $\mathcal{G}_{s_1, s_2}(z_2, w_2)$.  Its general form is slightly complicated, although for $\text{sgn}(s_1) = -\text{sgn}(s_2)$ it simplifies to
\begin{align}
\label{E:Fs1s2p2}
	&\mathcal{G}_{s_1, s_2}^{\text{sgn}(s_1) \,=\, -\text{sgn}(s_2)}(z_2, w_2) =  
	\\	
	\nonumber
	& \qquad \quad \sum_{c \geq 1,\,\,d \in (\mathbb{Z}/c\mathbb{Z})^*}  \frac{\pi^2 \sqrt{z_2 w_2}}{c\sqrt{1 + c^2 z_2 w_2}} \, e^{2\pi i \left(\frac{d}{c} \, s_1 + \frac{d^{-1}}{c} \, s_2 \right)} \, e^{-2 \pi \sqrt{\frac{z_2}{w_2}(1 + c^2 z_2 w_2)} \frac{|s_1|}{c} - 2 \pi \sqrt{\frac{w_2}{z_2}(1 + c^2 z_2 w_2)} \frac{|s_2|}{c}}\,.
\end{align}
Here, $d^{-1}$ is any multiplicative inverse of $d$ mod $c$. In the remaining parts of this Section we will examine the salient features of the $c = 0$ contribution in~\eqref{E:Fs1s2p1}, and the $c \geq 1$ contributions packaged as $\mathcal{G}_{s_1, s_2}(z_2, w_2)$. 

\subsection{Zero momentum mode}
\label{S:zeroMomentum}
We begin by examining the zero momentum mode $\widetilde{F}_{0,0}$.  Eqn.'s~\eqref{E:Fs1s2p1} and~\eqref{E:Fs1s2p2} simplify to
\beq
	\widetilde{F}_{0,0} = \frac{\pi z_2 w_2}{z_2 + w_2} + \sum_{c \geq 1,\, d \in (\mathbb{Z}/c \mathbb{Z})^*}\frac{\pi^2 \sqrt{z_2 w_2}}{c\sqrt{(1 + c^2 z_2 w_2)}}\,.
\eeq
The second term is divergent, which we can see by rewriting it as
\beq
	\sum_{c = 1}^\infty \frac{\pi^2 \sqrt{z_2 w_2}}{c\sqrt{(1 + c^2 z_2 w_2)}} \, \varphi(c)\,,
\eeq
where $\varphi(c)$ is the Euler totient function which counts the number of positive integers less than or equal to $c$ which are coprime to $c$, i.e. the size of $(\mathbb{Z}/c\mathbb{Z})^*$. This sum is then
\beq
	\sum_{c = 1}^\infty \frac{\pi^2}{c^2} \, \varphi(c) \,+\, (\text{convergent})\,.
\eeq
Using
\beq
	\sum_{c = 1}^\infty \frac{\varphi(c)}{c^s} = \frac{\zeta(s-1)}{\zeta(s)}\,,
\eeq
we find that the source of the divergence of our sum is
\beq
	\sum_{c = 1}^\infty \frac{\pi^2}{c^2} \, \varphi(c) = 6 \, \zeta(1)\,,
\eeq
i.e., we run into the $s = 1$ pole of $\zeta(s)$.

As suggested earlier, we can regulate the divergence as follows.  Letting $\mathcal{F}_s(z,w) = \sum_{\gamma\in PSL(2;\mathbb{Z})} \left( \frac{\text{Im}(\tau_1)\text{Im}(\gamma\tau_2)}{|\tau_1 + \gamma \tau_2|^2} \right)^s$ as per Eqn.~\eqref{E:sZeta1}, we define
\beq
	\widetilde{F}_{s, s_1, s_2} = \int_0^1 dz_1 \int_0^1 dw_1 \, e^{2 \pi i z_1 s_1 + 2\pi i w_1 s_2} \mathcal{F}_s(z,w)\,.
\eeq
Computing $\widetilde{F}_{s,0,0}$, we find that divergent term we examined above is regularized to\footnote{We thank D.~Stanford for pointing out an error in a previous version of this equation.}
\beq
	\label{E:fourierDivergence}
	\pi \left(\frac{\Gamma(s-1/2)}{\Gamma(s)}\right)^2 (z_1 z_2)^{1-s} \sum_{c=1}^\infty \frac{\varphi(c)}{c^{2s}} \left( 1 + O(1/c^2) \right) = \frac{3}{s-1} \,+\,(\text{finite as }s\to 1)\,.
\eeq
It is natural to subtract off the $s = 1$ pole, and then take the limit as $s \to 1$.  In this way, we can regularize $\widetilde{F}_{0,0}$ so that it is finite.  Our regularization procedure is robust up to an additive constant which is independent of $z_2$ and $w_2$, i.e.~different ways of implementing the zeta-function regularization will differ by an additive constant.

Next, we will show that the the remaining Fourier coefficients $\widetilde{F}_{s_1, s_2}$ for at least one of $s_1, s_2$ non-zero are finite.  Furthermore, we establish that once we have regularized the divergence in $\widetilde{F}_{0, 0}$\,, the function $\mathcal{F}(z,w)$ itself is finite.

\subsection{Convergence of remainder of the Fourier series}

Here we provide bounds on $\widetilde{F}_{s_1, s_2}$ for at least one of $s_1, s_2$ nonzero, establishing that the modular sum in $\widetilde{F}_{s_1, s_2}$ converges. The $c = 0$ contribution~\eqref{E:Fs1s2p1} is manifestly finite.  For the $c \geq 1$ contributions, the sum over $c$ and $d$ is quite subtle.  Defining the Kloosterman sum
\begin{equation}
S(j,J,s) = \sum_{d \in (\mathbb{Z}/c\mathbb{Z})^*} \, e^{2 \pi i \left(j\frac{d}{c} + J \frac{d^{-1}}{c} \right)}\,,
\end{equation}
we can write after some simplifications
\begin{align}
\begin{split}
\label{E:generalGs}
	\mathcal{G}_{s_1,s_2}(z_2,w_2) =&\sum_{c = 1}^\infty \frac{\pi \sqrt{z_2 w_2}}{c\sqrt{(1 + c^2 z_2 w_2)}} \, S(s_1, s_2, c) \,e^{- 2 \pi z_2 |s_1|} 
	\\
	& \qquad \qquad \qquad \times \int_{-\infty}^\infty dx \, 
e^{-\frac{2\pi i}{B x + i \,\text{sgn}(s_1) c w_2}\frac{s_1}{c} + 2 \pi i B x \frac{s_2}{c}} \frac{1}{x^2 + 1}\,,
\end{split}
\end{align}
for $B = \sqrt{\frac{w_2}{z_2}(1 + c^2 z_2 w_2)}$. The above expression is upper bounded by
\beq
\label{E:upperbound1}
	\mathcal{G}_{s_1,s_2}(z_2,w_2)\leq\,\pi^2 \, e^{- 2\pi z_2 |s_1|}\,\sum_{c=1}^\infty \frac{S(s_1, s_2, c)}{c^2}\,.
\eeq
The Weil bound tells us that
\begin{equation}
|S(j,J,c)| \leq \sqrt{\text{gcd}(j,J,c)} \sqrt{c} \, \tau(c) 
\end{equation}
where $\tau(c)$ is the number of positive divisors of $c$.  Since $\tau(c) \ll c^{\delta}$ for any constant $\delta > 0,$ and since for at least one of $j,J$ nonzero we have $\text{gcd}(j,J,c) \leq \text{max}(|j|,|J|)$, we can further upper bound~\eqref{E:upperbound1} by
\begin{equation}
\label{E:upperbound2}
	\mathcal{G}_{s_1,s_2}(z_2,w_2)\leq\,\sqrt{\text{max}(|s_1|, |s_2|)}\, \pi^2 \, e^{- 2\pi z_2 |s_1|}\,\sum_{c=1}^\infty \frac{1}{c^{3/2 - \delta}}
\end{equation}
which is finite.  Therefore Eqn.~\eqref{E:upperbound1} converges, and so $\widetilde{F}_{s_1,s_2}$ is well-defined (with the exception of $\widetilde{F}_{0,0}$ which we already dealt with).

Furthermore, Eqn.~\eqref{E:upperbound2} implies that $\mathcal{G}_{s_1, s_2}(z_2, w_2) \leq C_{s_1, s_2}\,e^{- 2 \pi z_2 |s_1|}$ where $C_{s_1, s_2}$ is a constant depending on $s_1, s_2$.  Since $\mathcal{G}_{s_1, s_2}(z_2, w_2) = \mathcal{G}_{s_2, s_1}(w_2, z_2)$ by symmetry, we additionally have $\mathcal{G}_{s_1, s_2}(z_2, w_2) \leq C_{s_1, s_2}\,e^{- 2 \pi w_2 |s_2|}$.  Taken together, we have
\begin{align}
\begin{split}
\mathcal{G}_{s_1, s_2}(z_2, w_2) &\leq C_{s_1, s_2} \, e^{- 2 \pi \,\text{max}\{z_2 |s_1|, w_2 |s_2|\}} 
\\
&\leq C_{s_1, s_2} \, e^{- \pi z_2 |s_1| - \pi w_2 |s_2| }
\end{split}
\end{align}
where we have used $\text{max}(a,b) \geq \frac{1}{2}(a+b)$. Examining the bound on $\mathcal{G}_{s_1, s_2}(z_2, w_2)$ alongside~\eqref{E:Fs1s2p1}, we find that $\widetilde{F}_{s_1,s_2}$ is exponentially suppressed in $|s_1|$ and $|s_2|$.  So once we have regularized the zero mode, the Fourier series expansion of $\mathcal{F}(z,w)$ converges.

\subsection{Summary}

We have decomposed $\mathcal{F}(z,w)$ into a Fourier expansion in $z_1$ and $w_1$ with Fourier components $\widetilde{F}_{s_1,s_2}$. The zero mode $\widetilde{F}_{0,0}$ is divergent, which we showed how to Zeta-regularize.  All of the remaining Fourier coefficients $\widetilde{F}_{s_1,s_2}$ are finite, and upon regularizing $\widetilde{F}_{0,0}$ the entire Fourier series for $\mathcal{F}(z,w)$ converges. 

The Fourier coefficients are given by
\beq
\label{E:fourierCoefficients}
	\widetilde{F}_{s_1,s_2}(z_2,w_2) = \frac{\pi z_2 w_2}{z_2+w_2} \, e^{-2\pi (z_2+w_2)|s_1|}\delta_{s_1,s_2} + \mathcal{G}_{s_1,s_2}(z_2,w_2)\,.
\eeq
The first term accounts for all of the terms in the modular sum generated by $\tau\to \tau+1$ while the second term $\mathcal{G}_{s_1,s_2}$ includes all terms in the modular sum with at least one $S$ transformation. In general $\mathcal{G}_{s_1,s_2}$ is given by~\eqref{E:generalGs}, and when the signs of $s_1$ and $s_2$ are opposite it simplifies to~\eqref{E:Fs1s2p2}.

In the next Section, we will show how our the Fourier coefficients $\widetilde{F}_{s_1,s_2}$ allow us to relate our Euclidean wormhole computation in AdS$_3$ to the calculation of a ramp in the spectral form factor of a putative dual ``random CFT.''  Our results will generalize the analysis of the spectral form factor in random matrix theory.

\section{Beyond random matrices}
\label{S:RMT}

Equipped with the Fourier series decomposition of $\mathcal{F}(\tau_1, \tau_2)$ defined in~\eqref{E:preSum} in~\eqref{E:fourierCoefficients}, we will now analyze the wormhole amplitude $Z_{\mathbb{T}^2\times I}(\tau_1,\tau_2)$ at fixed spins and low temperature. We will show how to extract a spectral form factor from the amplitude, and provide evidence that, if AdS$_3$ gravity has a dual, then the dual is an ensemble which provides a generalization of random matrix theory.

\subsection{Low temperature, long times}

From the results of Section~\ref{S:ModularSum} we can write the amplitude~\eqref{E:mainResult2} as
\beq
\label{E:Z12v1}
Z_{\mathbb{T}^2\times I}(\tau_1, \tau_2) = \frac{1}{2\pi^2} \, Z_0(\tau_1) Z_0(\tau_2) \sum_{s_1, s_2 = -\infty}^\infty e^{- 2 \pi i \text{Re}(\tau_1) s_1 - 2\pi i \text{Re}(\tau_2) s_2} \widetilde{F}_{s_1,s_2}(\text{Im}(\tau_1), \text{Im}(\tau_2))\,,
\eeq
where again $Z_0(\tau) = 1/(\sqrt{\text{Im}(\tau)} |\eta(\tau)|^2)$ and we have regularized $\widetilde{F}_{0,0}$ as explained previously. The Fourier coefficients $\widetilde{F}_{s_1,s_2}$ are in general rather complicated. However they simplify enormously in the low temperature limit $\text{Im}(\tau_1)=\beta_1,\text{Im}(\tau_2)=\beta_2\to \infty$ with $\beta_1/\beta_2$ fixed. In that limit one can show from the integral representation~\eqref{E:generalGs} that\footnote{This property is visible from the integrated expression in~\eqref{E:Fs1s2p2}, which holds in the cases when $s_1$ and $s_2$ have opposite sign or one of the spins vanishes.}
\beq
\label{E:tildeFlowT}
	\widetilde{F}_{s_1,s_2}(\beta_1,\beta_2) =e^{-2\pi |s_1|\beta_1-2\pi |s_2|\beta_2}\left(  \frac{\pi \beta_1\beta_2}{\beta_1+\beta_2} \,\delta_{s_1,s_2} +\sum_{n=0}^{\infty}  \sum_{c=1}^{\infty} \frac{\mathcal{I}_{s_1,s_2}^{(n)}\left(c;\frac{\beta_1}{\beta_2}\right)}{\beta_1^n}\right)\,,
\eeq
where $\mathcal{I}_{s_1,s_2}^{(n)}$ are series coefficients which are suppressed by powers of the temperature relative to the indicated, leading order result. For example, the leading low-temperature correction is the $n=0$ term
\beq
	\mathcal{I}_{s_1,s_2}^{(0)}(c) = \frac{S(s_1,s_2,c)}{2c^2} \, J_0\left( \frac{\Theta(s_1s_2)4\pi \sqrt{|s_1s_2|}}{c}\right)\,.
\eeq
Here $\Theta(x)$ is the Heaviside step function.  So there is universal behavior in the low temperature limit, with rather complicated corrections. Taking the low temperature limit zooms in on low-energy physics in the usual way. In this setting the low-energy physics is that of BTZ microstates near threshold.

We would like to interpret the wormhole amplitude as being a good approximation of an ensemble average $\langle Z(\tau_1)Z(\tau_2)\rangle_{\text{ensemble, conn.}}$ in some regime with $Z(\tau)$ a CFT torus partition function. There is a JT limit of AdS$_3$ gravity~\cite{ghosh2019universal} at large spin and low temperature, with a genus expansion parameter which suppresses fluctuation of topology. At least in, and perhaps beyond that regime we expect our result to be a good approximation to the complete one. The torus partition function of a given, fixed CFT is determined by its spectrum on the circle, which is organized into representations of the Virasoro symmetry. The complete spectrum is determined in terms of the spectrum of primaries, which themselves may be labeled by their energy and momentum. The wormhole amplitude $Z_{\mathbb{T}^2\times I}$ may be separated into a contribution from primary states alone, which we call $Z^P$, and contributions from descendants determined by symmetry. From the final form of $Z_{\mathbb{T}^2 \times I}(\tau_1, \tau_2)$ in~\eqref{E:mainResult2}, as well as the unintegrated expression in~\eqref{E:preThere}, we see that the full amplitude comes from non-degenerate representations of Virasoro on the two boundary tori with no vacuum contribution.  So to obtain $Z^P$ we simply strip off the infinite products in the prefactors $Z_0(\tau_1)Z_0(\tau_2)$, giving
\beq
	Z^P(\tau_1,\tau_2) = \frac{1}{2\pi^2}\frac{|q_1 q_2|^{-\frac{1}{12}}}{\sqrt{\text{Im}(\tau_1)\text{Im}(\tau_2)}} \sum_{s_1,s_2=-\infty}^{\infty} e^{ -2\pi i \text{Re}(\tau_1)s_1-2 \pi i \text{Re}(\tau_2)s_2}\widetilde{F}_{s_1,s_2}(\text{Im}(\tau_1),\text{Im}(\tau_2))\,.
\eeq
The prefactor is independent of $\text{Re}(\tau_1)$ and $\text{Re}(\tau_2)$, and so the Fourier coefficients of $Z^P$ are simply proportional to the $\widetilde{F}$'s. We arrive at the leading low-temperature expression
\beq
\label{E:ZP}
	Z^P_{s_1,s_2}(\beta_1,\beta_2) = \frac{1}{2\pi} \frac{\sqrt{\beta_1 \beta_2}}{\beta_1+\beta_2} \, e^{-E_{s_1}\beta_1-E_{s_2}\beta_2} \left( \delta_{s_1,s_2} + O\left(\frac{1}{\beta}\right)\right)\,, \qquad E_s = 2\pi \left( |s|-\frac{1}{12}\right)\,.
\eeq
Here $E_s$ is the threshold energy (with respect to the Hamiltonian $2\pi \left(L_0+\bar{L}_0-\frac{c}{12}\right)$) for a BTZ black hole at spin $s$.

We can extract the spectral form factor for primaries. See Subsection~\ref{S:SFF} for a summary of the spectral form factor in random matrix theory. We simply take our result in~\eqref{E:ZP} and analytically continue $\beta_1 \to \beta + i T$, $\beta_2 \to \beta - i T$, which allows us to study finite temperature correlations of the spectrum at Lorentzian time $T$. Including the first low-temperature correction we have
\begin{align}
\begin{split}
\label{E:ramp0}
	&\hspace{-.5cm}Z_{s_1,s_2}^P(\beta+iT,\beta-iT) =\frac{\sqrt{\beta^2+T^2}}{4\pi \beta}e^{-2\beta E_{s_1}}\delta_{s_1,s_2} 
	\\
	&+ e^{- \beta (E_{s_1} + E_{s_2}) -i T (E_{s_1} - E_{s_2})} \sum_{c=1}^{\infty}\left(   \frac{S(s_1, s_2, c)}{2c^2 \sqrt{\beta^2+T^2}} \,J_0\left(\frac{\Theta(s_1 s_2) 4 \pi \sqrt{|s_1 s_2|}}{c}\right) + O(\beta^{-2})\right)\,,
\end{split}
\end{align}
which for $T\gg \beta$ becomes
\beq
\label{E:ramp1}
	Z_{s_1,s_2}^P(\beta+iT,\beta-iT)  = \frac{T}{4 \pi\beta} \, e^{- 2 \beta E_{s_1}} \delta_{s_1, s_2} + O(T^{-1})\,.
\eeq
The leading correction comes from the expansion of the square root of the first term of~\eqref{E:ramp0}, as well as of from the leading behavior of the second line.

Eq.~\eqref{E:ramp1} has the behavior of a linear ramp, with fluctuations which are power-law suppressed in $1/T$. Guided by the JT limit of AdS$_3$ gravity, we have some reason to believe that the torus times interval amplitude dominates the two-torus partition function at least at low temperature and large spin (and implicitly for times which are not exponentially long compared to $c$). We would then have a spectral form factor containing a ramp with small fluctuations around it, which is a smoking gun of a disordered theory. In the next Subsection we will find a precise connection between the slope of the ramp and random matrix theory.

\subsection{Random matrix statistics and Virasoro symmetry}

In looking for an ensemble dual to pure AdS$_3$ gravity, we notice several features which emulate random matrix theory.  In random matrix theory, nearest-neighbor eigenvalue statistics are controlled by the symmetries of the ensemble.  In this Subsection we are primarily interested in GUE eigenvalue statistics (in the absence of symmetry) and GOE eigenvalue statistics (in the presence of $\mathsf{T}^2 = 1$ time-reversal symmetry).  Operationally, we will consider the late-time behavior of the spectral form factor, whose behavior depends on the symmetry class.  For a recent discussion in the relevant context of double-scaled random matrix theory, see~\cite{stanford2019jt}.

Suppose that we envision AdS$_3$ gravity as being dual to an ensemble of CFTs, inducing an ensemble over CFT Hamiltonians on the circle. CFT Hamiltonians are (infinitely) large matrices, and so perhaps there is a connection between AdS$_3$ gravity and random matrix theory with Virasoro symmetry. We are not aware of a discussion of such matrix models, nor do we know how to construct such an ensemble for AdS$_3$ gravity specifically. However, due to the universality of eigenvalue pair correlations in random matrix theory, we can analyze certain properties of matrix ensembles with Virasoro symmetry on general grounds.

The complete spectrum of such a random matrix ensemble is organized into a sum over representations of the Virasoro symmetry, determined by the spectrum of primaries, which are themselves labeled by an energy and momentum. Let us call the Hamiltonian which labels Virasoro primary states $H$. It commutes with a momentum operator $\mathsf{P}$, which we take to also act only on primary states. Of course the spectrum of primaries completely determines the spectrum of the full Hamiltonian and momentum.  We can block diagonalize each Hamiltonian $H$ in the ensemble into blocks $H_p$ of fixed momentum $p$. Positing that $H_p$ is itself a large random matrix in the absence of additional symmetries, we would expect each $H_p$ to have GUE eigenvalue statistics.  Accordingly, we can define a momentum-block spectral form factor
\beq
	Z^P_{p,q}(\beta + iT, \beta - i T) = \left\langle \text{tr}\big(e^{- (\beta + i T) H_p}\big) \, \text{tr}\big(e^{- (\beta - i T) H_q}\big) \right\rangle_{\text{ensemble, conn.}}
\eeq
We would expect that each $Z^P_{p,p}(\beta + iT, \beta - i T)$ has a GUE ramp due to level repulsion, and that $Z^P_{p,q}(\beta + i T, \beta - i T)$ for $p \not = q$ does not have a ramp since $H_p$ and $H_q$ will have statistically independent eigenvalues.  Our AdS$_3$ computation realizes these expectations, thus supporting the conceptual framework of AdS$_3$ being dual to a ``random'' CFT.  Specifically, we find a ramp exactly matching that of double-scaled GUE random matrix theory.

As an additional piece of evidence for a putative duality with random CFT, suppose we gauge time reversal symmetry in the bulk.  This corresponds to introducing a global time reversal symmetry on the boundary theory, implemented by an anti-unitary operator $\mathsf{T}$ with
\begin{equation}
[\mathsf{P},\mathsf{T}] = 0\,.
\end{equation}
For simplicity we suppose $\mathsf{T}^2 = 1$.  As above, let us block diagonalize an $H$ in the ensemble into blocks $H_p$ of fixed momentum.  Then for a $|\psi\rangle$ in the subspace corresponding to the block $H_p$, we have $\mathsf{P} |\psi\rangle = p |\psi\rangle$.  But we also have $\mathsf{P} (\mathsf{T} |\psi\rangle) = - \mathsf{T} \mathsf{P} |\psi \rangle = - p (\mathsf{T} |\psi\rangle)$, and so $\mathsf{T} |\psi\rangle$ is in the subspace corresponding to the block $H_{-p}$.  More generally, $\mathsf{T} H_p \mathsf{T}^\dagger = H_{-p}$, and so $H_p$ and $H_{-p}$ have identical eigenvalues.  Since $\mathsf{T} H_0 \mathsf{T}^\dagger = H_0$, $H_0$ is a real symmetric matrix in a suitable basis.  Taking all of these considerations into account, we have that $Z^{P}_{p,p}(\beta + iT, \beta - i T)$ and $Z^P_{p,-p}(\beta + iT, \beta - i T)$ for $p \not = 0$ have GUE ramps, $Z^P_{0,0}(\beta + iT, \beta - i T)$ has a GOE ramp, and $Z^P_{p,q}(\beta + iT, \beta - i T)$ for $p \not = q$ does not have any ramp.  These features are mirrored in gravity. After gauging bulk time reversal, the Fourier coefficients become
\begin{equation}
	Z_{s_1, s_2}(\beta_1, \beta_2) = Z_{s_1, s_2}(\beta_1, \beta_2) + Z_{s_2, s_1}(\beta_1, \beta_2)\,.
\end{equation}
Accordingly, we find that for small temperatures and long times,
\begin{align}
\begin{split}
	Z_{0,0}^{P}(\beta + i T, \beta - i T) &=  \frac{T}{2 \pi \beta} \, e^{- 2 \beta E_0} + O(T^{-1})\,, 
\\
 	Z_{s,\pm s}^{P}(\beta + i T, \beta - i T) &=  \frac{T}{4 \pi \beta} \,e^{- 2 \beta E_s}+O(T^{-1})\,,\qquad s \not = 0\,,
	 \\
 	Z_{s_1, s_2}^{P}(\beta + i T, \beta - i T) &=O(T^{-1})\,,\qquad \qquad \qquad\quad \,\,|s_1| \not = |s_2|\,.
\end{split}
\end{align}
 Indeed, this exactly matches expectations from random matrix theory.
 
In the remainder of this Subsection we return our focus to the ungauged model. In fact our gravitational result~\eqref{E:ZP} matches more than the slope of a ramp predicted by random matrix theory. The leading contribution to the connected two-point function of eigenvalues is a universal result in random matrix theory. In the present setting, it implies that in the double-scaling limit
 \beq
 	\left\langle \text{tr}\left( e^{-\beta_1 H_p}\right)\text{tr}\left( e^{-\beta_2H_q}\right)\right\rangle_{\text{ensemble, conn.}} = \frac{1}{2\pi} \frac{\sqrt{\beta_1\beta_2}}{\beta_1+\beta_2} \,e^{-\beta_1 E_p -\beta_2 E_q}\delta_{p,q} + \cdots\,,
 \eeq
 where the dots indicate corrections in the genus expansion.  Here the cut for ensemble averaged density of states of $H_p$ runs from $[E_p,\infty)$. This result precisely matches the leading low-temperature limit of our gravitational result.
 
In summary, we find that the $\mathbb{T}^2 \times I$ contribution to the spectral form factor ramps, and more broadly to the full low-temperature 2-point fluctuation statistics of BTZ microstates, exactly matches the predictions of random matrix theory with Virasoro symmetry.  While this agreement with random matrix theory is totally striking, we note that there may be roadblocks to a straightforward random matrix interpretation.  Specifically, the Maloney-Witten density of states is negative near threshold~\cite{Maloney:2007ud, keller2015poincare, benjamin2019light, alday2019rademacher}. This flatly contradicts the prospect of a random matrix duality, although perhaps the negativity of the Maloney-Witten density of states is cured upon performing a complete non-perturbative path integral analysis (instead of just summing over saddles corresponding to smooth geometries with torus boundary, and geometries continuously connected to these).  An interesting feature of our gravitational result is corrections to the ramps which are suppressed by powers of the temperature.  These are not expected in standard random matrix theory, and so could provide clues as to how AdS$_3$ may go beyond this framework.
 
 In any case, now let us consider the complete form of the wormhole amplitude, restoring the contribution from descendants. The full result is determined by Virasoro symmetry in terms of the contribution from primaries. From the leading low-temperature limit of the Fourier coefficients $\widetilde{F}_{s_1,s_2}$, we find
\beq
	Z_{\mathbb{T}^2\times I}(\tau_1,\tau_2) = \frac{1}{4\pi}\frac{\sqrt{\text{Im}(\tau_1)\text{Im}(\tau_2)}}{\text{Im}(\tau_1)+\text{Im}(\tau_2)} \frac{1}{|\eta(\tau_1)|^2|\eta(\tau_2)|^2}\left( \frac{1+q_1q_2}{1-q_1 q_2} + \frac{1+\bar{q}_1\bar{q}_2}{1-\bar{q}_1\bar{q}_2} + O(\text{Im}(\tau)^{-1})\right)\,.
\eeq
The Fourier coefficients of the full result at fixed spin read
\beq
	Z_{s_1,s_2}(\beta_1,\beta_2) = \frac{1}{2\pi}\frac{\sqrt{\beta_1\beta_2}}{\beta_1+\beta_2} \sum_{n,m=0}^{\infty} e^{-\beta_1 E_{s_1,n}-\beta_2 E_{s_2,m}}\left( \mathcal{C}_{s_1,s_2,n,m} + O\left( \frac{1}{\beta}\right)\right)\,,
\eeq
where $E_{s,n} = 2\pi \left( |s|+2n - \frac{1}{12}\right)$ and $\mathcal{C}_{s_1,s_2,n}$ is an integer determined by the small$-q$ expansion of the Dedekind eta prefactor. In other words, the Fourier coefficients contain an infinite sum of increasingly suppressed exponentials. As an example, when $s_1=s_2=s$ and $n=m=0$ we have
\beq
	\mathcal{C}_{s,s,0,0} = \sum_{k=0}^s p(k)^2\,,
\eeq
with $p(k)$ the partition function of $k$, the number of ways $k$ may be partitioned into integers. As another example, consider $s_1=2$, $s_2=1$, and $n=m=0$. In that case $\mathcal{C}_{2,1,0,0}=3$. Note that there are low temperature correlations for any $s_1$ and $s_2$, not merely when the spins are equal, and that there is an infinite tower of exponentially suppressed corrections. These facts guarantee that the full spectral form factor has the long time form
\beq
	Z_{s_1,s_2}(\beta+i T,\beta-i T) = \frac{T}{4\pi \beta}\sum_{n,m=0}^{\infty}\mathcal{C}_{s_1,s_2,n,m} \, e^{-\beta (E_{s_1,n}+E_{s_2,m}) - i T(E_{s_1,n}-E_{s_2,m})}\left( 1 +O\left( \frac{1}{T}\right)\right)\,.
\eeq
So there is a ramp only when the two energies coincide, with a more complicated slope than before.

These expressions automatically match the prediction from random matrix theory with Virasoro symmetry, which is similarly determined in terms of the primaries by the symmetry. The full Hamiltonian $\mathscr{H}$ block diagonalizes into sectors $\mathscr{H}_p$ of fixed spin $p$, in such a way that the full blocks are determined in terms of the Hamiltonian $H$ and momentum $\mathsf{P}$ on primaries. In particular, the symmetry determines the full correlation function
\beq
\label{E:generalMMtwoPoint}
	\left\langle \text{tr}\left( e^{-\beta_1 \mathscr{H}_p}\right)\text{tr}\left(e^{-\beta_2 \mathscr{H}_q}\right)\right\rangle_{\text{ensemble, conn.}} = \frac{1}{2\pi}\frac{\sqrt{\beta_1\beta_2}}{\beta_1+\beta_2} \sum_{n,m=0}^{\infty} e^{-\beta_1 E_{s_1,n}-\beta_2 E_{s_2,m}} \mathcal{C}_{s_1,s_2,n,m}  + \cdots\,,
\eeq
where the dots indicate genus corrections. 

Let us unpack this with the example of $s_1=2$, $s_1=1$, and $n=m=0$. The full Hamiltonian $\mathscr{H}_2$ has infinitely many blocks, one corresponding to the spin-2 primaries $H_2$, and the others corresponding to descendants. There are three other ``descendant'' blocks whose densities of states starts at the minimum value $E_2$ (the other descendant blocks have densities of states starting at $E_{2,n}$): one comes from acting on the spin-1 primaries $H_1$ with $L_{-1}$, and the other two from acting on the scalar primaries $H_0$ with $L_{-1}^2$ or $L_{-2}$. Similarly at spin-1 the lowest energy contributions come from the primary block $H_1$ and the descendant block $L_{-1}\cdot H_0$. So $\mathscr{H}_2 = H_2 \otimes (L_{-1}\cdot H_1) \otimes( L_{-1}^2 \cdot H_0)\otimes (L_{-2}\cdot H_0) \otimes \hdots$ and $\mathscr{H}_1 = H_1\otimes (L_{-1}\cdot H_0)\otimes \hdots$, where the dots indicate blocks whose cuts begin at larger energies. The leading low-temperature contribution to~\eqref{E:generalMMtwoPoint} simply comes from tracking the effects of eigenvalue repulsion between identical primary blocks. There are three such terms in the two-point function $\langle \text{tr}\left( e^{-\beta_1 \mathscr{H}_2}\right) \text{tr}\left( e^{-\beta_2 \mathscr{H}_1}\right)\rangle_{\text{ensemble, conn.}}$: one where $L_{-1}\cdot H_1$ repels against $H_1$, and two more where $L_{-1}^2\cdot H_0$ and $L_{-2}\cdot H_0$ repel against $L_{-1}\cdot H_0$. So from the matrix ensemble we obtain $\mathcal{C}_{2,1,0,0}=3$, matching what we found from the Fourier analysis of the wormhole amplitude above.

Having gone through this example, we emphasize that more generally the constants $\mathcal{C}_{s_1,s_2,n,m}$ is determined both from the wormhole amplitude and matrix ensemble by the same Virasoro symmetry.  And so given that the statistics of the primaries match, the statistics of the descendants match too.

\subsection{Random CFT}

We have seen that the leading low-temperature limit of the wormhole amplitude $Z_{\mathbb{T}^2\times I}$ is precisely described by double-scaled random matrix theory with Virasoro symmetry. However there are some simple reasons why we expect that the dual to AdS$_3$ gravity is not a random matrix theory. The first is that in Euclidean gravity we may arrange for the boundary to be any genus $g$ surface, and it is not clear how to interpret such an observable in a putative random matrix dual.  Another reason is the power-law suppressed fluctuations of the ramp, which are not expected in standard random matrix theory, although perhaps these can be accommodated for with suitable long-range eigenvalue correlations. Finally, beyond Virasoro symmetry the gravitational amplitude is invariant under independent modular transformations of each boundary torus.  Interpreting the two torus partition function as characterizing the fluctuations of torus partition functions within the dual ensemble, the independent modular invariances tell us that the objects we are averaging over (the partition functions) are themselves modular invariant point by point in the ensemble, and it is not clear how to incorporate this property in random matrix theory.  For these reasons we have made the guess that AdS$_3$ gravity is dual to an ensemble of CFTs, although perhaps when the boundaries are tori there is a non-standard random matrix description. (More precisely, we have in mind an ensemble of CFTs with a large gap to the BTZ threshold.)

But what would this mean? In this Subsection, we sketch a plausible schematic framework, following the usual logic of statistical physics. We stress that this sketch is speculative. We regard it as an organizing principle to keep in mind when trying to make sense of AdS$_3$ gravity beyond the torus times interval. In particular, we are going to ignore the pressing conceptual concern that we do not have any knowledge of specific irrational CFTs at large central charge, much less an ensemble of such theories. Our optimistic hope is that there are many solutions to the modular bootstrap for CFTs that ``look'' gravitational, and that in a sense 3d gravity is an average over this solution space.

With these caveats and declarations out of the way we proceed. Suppose we denote that data of a 2d CFT by $\mathcal{T}$, standing for ``theory.'' For a given theory $\mathcal{T}$ there is a genus $g$ partition function $Z_{\mathcal{T}, \Sigma_g}(\Omega)$, where $\Omega$ is the period matrix of the surface.  Let $d[\mathcal{T}]$ be a measure on 2d CFTs, with which we can consider correlators like
\begin{equation}
\label{E:RandomCFTZs}
	\langle Z_{\mathcal{T}, \Sigma_{g_1}}(\Omega_1)  \cdots Z_{\mathcal{T}, \Sigma_{g_n}}(\Omega_n) \rangle_{\text{ensemble}} \equiv \int d[\mathcal{T}] \, Z_{\mathcal{T}, \Sigma_{g_1}}(\Omega_1) \cdots Z_{\mathcal{T}, \Sigma_{g_n}}(\Omega_n)\,.
\end{equation}
Next we turn to AdS$_3$ quantum gravity.  Let us define
\begin{equation}
\label{E:AdS3bigpartition}
	Z_{\text{AdS}_3}( \Sigma_{g_1}, \Omega_1 ; \dots ; \Sigma_{g_n}, \Omega_n) \equiv  \sum_{\substack{\text{bulk topologies of }\mathcal{M}_3 \\ \partial \mathcal{M}_3 = \Sigma_{g_1} \sqcup \cdots \sqcup \Sigma_{g_n}}}  \int d[g]_{\Omega_1,...,\Omega_n} \, e^{- S_{\text{grav}}[g]}\,,
\end{equation}
where $S_{\text{grav}}$ is the gravitational action.  The path integral in the summand is understood to be the amplitude of Euclidean AdS$_3$ gravity on $\mathcal{M}_3$ with boundary $\partial \mathcal{M}_3 = \Sigma_{g_1} \sqcup \cdots \sqcup \Sigma_{g_n}$ and corresponding period matrices $\Omega_1,...,\Omega_n$.  It is far from clear if the right-hand side of~\eqref{E:AdS3bigpartition} is well-defined.  For example, there is no known analogue of the genus expansion for 3-manifolds, nor do we know the amplitudes of AdS$_3$ gravity on general topologies.  In light of these unknowns, an ambitious conjecture is that there exists an ensemble of 2d CFTs with fixed central charge obeying
\begin{equation}
\label{E:boldconj1}
\langle Z_{\mathcal{T}, \Sigma_{g_1}}(\Omega_1)  \cdots Z_{\mathcal{T}, \Sigma_{g_n}}(\Omega_n) \rangle_{\text{ensemble}} \simeq Z_{\text{AdS}_3}( \Sigma_{g_1}, \Omega_1 ; \dots ; \Sigma_{g_n}, \Omega_n) \,.
\end{equation}
Here ``$\simeq$'' means non-perturbative equivalence to all orders in a putative asymptotic expansion where we might hope that the analogue of the genus expansion parameter of JT gravity is a coupling $\sim e^{-\#/G}$.  This would be an AdS$_3$ generalization of the duality between nearly-AdS$_2$ JT gravity and a double scaled matrix model in~\cite{Saad:2019lba}.  We emphasize that essential refinements of the conjecture in Eqn.~\eqref{E:boldconj1} are required.

A logical possibility is that quantities like those in~\eqref{E:AdS3bigpartition} are somehow pathological in AdS$_3$ gravity.  Even if this is so, it may still be that a correspondence like Eqn.~\eqref{E:boldconj1} holds, with the same pathologies occurring on both sides of the duality. As an example we have in mind the duality~\cite{stanford2019jt} between JT gravity on unorientable manifolds and a double-scaled random matrix theory with time reversal symmetry, both sides of which diverge.

Our approach to~\eqref{E:boldconj1} has been to compute examples of the right-hand side for AdS$_3$ gravity, and suggest the existence of an ensemble average on the left-hand side.  Very recent works~\cite{1800406, 1800422} can be cast in a similar conceptual framework as~\eqref{E:boldconj1}: examples of partition functions in an ensemble average of \textit{free} CFT's are computed, and a bulk dual is suggested (up to the treatment of zero modes it is a Chern-Simons theory with gauge group $\mathbb{R}^{2c}$ -- not Einstein gravity as in our setting).

In this paper, we have investigated a special case of~\eqref{E:AdS3bigpartition}, the $\mathbb{T}^2 \times I$ amplitude. If this amplitude was expressible in terms of an average over an ensemble of CFTs, it would correspond to a contribution to
\begin{equation}
\langle Z_{\mathcal{T}, \mathbb{T}^2}(\tau_1) Z_{\mathcal{T}, \mathbb{T}^2}(\tau_2) \rangle_{\text{ensemble, conn.}}\,.
\end{equation}
We might even hope that the geometry $\mathbb{T}^2 \times I$ provides the leading contribution.  We can suggestively rewrite the above equation as
\begin{equation}
\left\langle \text{tr}\left(e^{- \text{Im}(\tau_1) \mathscr{H}+ i \,\text{Re}(\tau_1) \mathsf{P} } \right) \text{tr}\left(e^{- \text{Im}(\tau_2) \mathscr{H}+ i \,\text{Re}(\tau_2) \mathsf{P} } \right) \right\rangle_{\text{ensemble, conn.}}\,.
\end{equation}
Note that here, the only feature of the CFT ensemble that matters is the induced distributions over Virasoro-invariant Hamiltonians $\mathscr{H}_p$.  Specifically, the distribution $d[\mathcal{T}]$ would induce a measure $d\mathscr{H}$ over infinite-dimensional matrices $\mathscr{H}$ so that
\begin{equation}
\langle Z_{\mathcal{T}, \mathbb{T}^2}(\tau_1) Z_{\mathcal{T}, \mathbb{T}^2}(\tau_2) \rangle_{\text{ensemble}} = \int d\mathscr{H} \, \text{tr}\left(e^{- \text{Im}(\tau_1) \mathscr{H}+ i \,\text{Re}(\tau_1) \mathsf{P} } \right) \text{tr}\left(e^{- \text{Im}(\tau_2) \mathscr{H}+ i \,\text{Re}(\tau_2) \mathsf{P} } \right)\,,
\end{equation}
and more generally
\begin{equation}
\langle Z_{\mathcal{T}, \mathbb{T}^2}(\tau_1) \cdots Z_{\mathcal{T}, \mathbb{T}^2}(\tau_n) \rangle_{\text{ensemble}} = \int d\mathscr{H}  \,\text{tr}\left(e^{- \text{Im}(\tau_1) \mathscr{H}+ i \,\text{Re}(\tau_1) \mathsf{P} } \right) \cdots \text{tr}\left(e^{- \text{Im}(\tau_n) \mathscr{H}+ i \,\text{Re}(\tau_n) \mathsf{P} } \right)\,.
\end{equation}
Then a special case of~\eqref{E:AdS3bigpartition} would be
\begin{equation}
\int d\mathscr{H}  \,\text{tr}\left(e^{- \text{Im}(\tau_1) \mathscr{H}+ i \,\text{Re}(\tau_1) \mathsf{P} } \right) \cdots \text{tr}\left(e^{- \text{Im}(\tau_n) \mathscr{H}+ i \,\text{Re}(\tau_n) \mathsf{P} } \right) \simeq Z_{\text{AdS}_3}(\mathbb{T}^2, \tau_1; ... ; \mathbb{T}^2, \tau_n)\,.
\end{equation}
While we have introduced the above equation as a consequence of~\eqref{E:AdS3bigpartition}, it may be instead viewed as a separate, weaker conjecture about the existence of a matrix model which captures Euclidean AdS$_3$ amplitudes with exclusively torus boundaries.

\section{Discussion}
\label{S:discussion}

In this paper we have computed the path integral of Euclidean AdS$_3$ gravity on the torus times interval. These configurations are Euclidean wormholes, and they represent a non-perturbative effective in three-dimensional quantum gravity.  Our answer encodes the fluctuation statistics of microstates of the BTZ black hole near threshold, and we found that these correlations precisely match those of double-scaled random matrix theory with Virasoro symmetry. The gravitational computation also includes low temperature corrections which differ from what one finds in standard random matrix theory.  For this and other reasons, our computation strongly supports the hypothesis that pure AdS$_3$ gravity is dual to an ensemble of CFTs.  If true this duality would be a higher-dimensional analogue of the duality between Jackiw-Teitelboim gravity and a double-scaled matrix model in \cite{Saad:2019lba}.

Our analysis opens up a path to studying quantum AdS$_3$ gravity on 3-manifolds with the topology $\mathbb{S}^1\times_f \Sigma_{g,n}$ or $\mathbb{R}\times_f \Sigma_{g,n}$. The former would correspond to a class of partition functions in Euclidean AdS$_3$, whereas the latter could be used to study topology-changing amplitudes in Lorentzian AdS$_3$.  On the Euclidean side, it is important to study the finiteness (or lack thereof) of partition functions on these spaces, and more broadly develop an analog of the genus expansion for 3-manifolds. Only then can we know if the $\mathbb{T}^2 \times I$ amplitude is the dominant contribution in some regime.  More ambitious still would be to determine if AdS$_3$ gravity is dual to an ensemble as our results suggest, and to establish an exact duality. 

The torus times interval amplitude suggests how the quantization of AdS$_3$ gravity on these manifolds proceeds. There were ``trumpets'' associated with each asymptotic region, stitched together by an integral over time-dependent moduli. In our example it was simple to integrate out the twist moduli, which enforced that the length moduli $b$ and $\bar{b}$ were constant. For constant $b$ and $\bar{b}$ each trumpet was a Virasoro character stemming from the path integral over two Alekseev-Shatashvili modes on its boundary, which decreases exponentially at large $b$ and $\bar{b}$. As a result the moduli space integral was finite. This finiteness is similar in spirit to how Schwarzian modes in JT gravity regulate the volumes of moduli spaces of Riemann surfaces with asymptotically hyperbolic boundaries (see~\cite{witten2020volumes} for a discussion). In our three-dimensional analysis there were two twist moduli, with a compact moduli space, corresponding to large diffeomorphisms of space and time respectively. The main open question to us is whether the twist moduli space is finite at higher genus.

It is far from clear how to construct an appropriate ensemble of irrational CFTs at large central charge.  Although there are candidate examples of such CFTs, they have not been studied conclusively.  Even if there was much better knowledge of the broader landscape of irrational CFTs at large central charge, one would still be have to construct a measure over them.  Our approach in this paper has been pragmatic, focusing on computations in gravity rather than positing a precise dual framework in which to interpret them.

However, there is a putative example of an ensemble of irrational CFTs which is worth noting, a family of CFT fixed points coming from a two-dimensional version of the supersymmetric SYK model~\cite{murugan2017more}.  Each element of the ensemble is thought to flow to a CFT with large central charge at long distance, albeit with a relatively low twist gap. The ensemble averaged, low-energy spectrum is a well-defined tower of operators. That is, the averaged density of states is a sum of delta functions at $O(1)$ energies.  However, we expect that the high energy part of the averaged spectrum, meaning $h+\bar{h}>c/12$, will exhibit a continuous density of states with some correlations. This feature is reminiscent of the black hole microstates of AdS$_3$ gravity including the fluctuation statistics computed in the present work.

More speculatively, we would like to propose an organizing principle for thinking about dualities between quantum gravity and disordered theories. One perspective in condensed matter physics is that many disordered theories can be viewed as effective theories; for instance, we could imagine augmenting standard effective field theory by an appropriate disorder average over irrelevant operators.  In some circumstances, like the SYK model~\cite{kitaev, Maldacena:2016hyu, Kitaev:2017awl} such a disorder average may simplify computations of long-wavelength physics. 

But in the case of Jackiw-Teitelboim gravity and pure AdS$_3$ gravity, the situation is more peculiar.  These theories are self-consistent in their own right and we do not have to regard them as effective theories.  Yet they are dual to ensembles of matrices and (tentatively) CFTs, respectively.  However, JT gravity in two dimensions and pure gravity in three dimensions only contain boundary gravitons, moduli and topology. From that point of view, we might say that low-dimensional pure quantum gravity is a kind of self-contained, quantum hydrodynamical theory of long-wavelength fluctuations of spacetime.  Then the dual disordered descriptions would provide a way of averaging over different microscopic theories with common long-wavelength gravitational physics to yield a universal \textit{pure} quantum gravity theory.  To borrow terminology from condensed matter physics, it may be appropriate to call such theories ``mesoscopic quantum gravity.''

Beyond AdS$_3$ gravity, our computational techniques generalize to other settings.  In three dimensions one can adapt our methods to study quantum effects with positive and zero cosmological constant \cite{cotler2019low, merbis2020geometric, CJWIP1}. But, perhaps a more promising pathway is to adapt our phase space techniques to study Euclidean wormholes in four and higher dimensions \cite{CJWIP2}.  We anticipate that these methods will open up new horizons in quantum gravity.

\subsection*{Acknowledgements}

We would like to thank Nathan Benjamin, Felipe Hernandez, Nicholas Hunter-Jones, Theodore Jacobson, Alexander Maloney, Semon Rezchikov, Moshe Rozali, Douglas Stanford, Joaquin Turiaci, and Jiaobao Yang for valuable discussions.  JC is supported by the Fannie and John Hertz Foundation and the Stanford Graduate Fellowship program. KJ is supported in part by the Department of Energy under grant number DE-SC 0013682.

\appendix

\section{Pole of Zeta-regularized Poincar\'{e} series}
\label{App:ZetaReg}

A core object in our study of the AdS$_3$ ramp is the Poincar\'{e} series
\begin{equation}
\mathcal{F}(\tau_1, \tau_2) = \sum_{\gamma \in PSL(2;\mathbb{Z})} \frac{\text{Im}(\tau_1) \text{Im}(\gamma \tau_2)}{|\tau_1 + \gamma \tau_2|^2}
\end{equation}
for $\tau_1, \tau_2$ in the fundamental domain of the Poincar\'{e} half-plane $\mathbb{H}$.  Unfortunately, the function $\mathcal{F}(\tau_1, \tau_2)$ is divergent.  In this Appendix, we consider a natural regularization scheme by generalizing $\mathcal{F}(\tau_1, \tau_2)$ to a suitable Zeta function, and isolating the divergent behavior.

We begin by generalizing $\mathcal{F}(\tau_1, \tau_2)$ to a Zeta function
\begin{equation}
\mathcal{F}_s(\tau_1, \tau_2) = \sum_{\gamma \in PSL(2;\mathbb{Z})} \left(\frac{\text{Im}(\tau_1) \text{Im}(\gamma \tau_2)}{|\tau_1 + \gamma \tau_2|^2}\right)^s\,.
\end{equation}
This reduces to $\mathcal{F}(\tau_1, \tau_2)$ for $s = 1$.  We will show that $\mathcal{F}_s(\tau_1, \tau_2)$ has a simple pole at $s = 1$ with a residue which is independent of $\tau_1$ and $\tau_2$.  Then we will regulate $\mathcal{F}_s(\tau_1, \tau_2)$ at $s = 1$ by subtracting out the pole, thus providing a natural regularization of $\mathcal{F}(\tau_1, \tau_2)$.

For our analysis, we require the spectral theory of the non-Euclidean Laplacian $\Delta$ on $SL(2;\mathbb{Z})\setminus \mathbb{H}$.  This Laplacian has an orthonormal basis of eigenfunctions $f_\lambda(z)$ with respect to the inner product
\begin{equation}
\langle f, g \rangle = \int_{SL(2;\mathbb{Z})\setminus \mathbb{H}} f(z) \overline{g(z)} \, \text{Im}(z)^{-2} \, dz
\end{equation}
including a single zero mode eigenfunction $f_0(z) = \sqrt{\frac{3}{\pi}}$.  The Green's function of $\Delta - \lambda$ is $L^2(SL(2;\mathbb{Z})\setminus \mathbb{H})$-integrable with respect to the inner product defined above.

Let us treat $\mathcal{F}_s(\tau_1, \tau_2)$ as a function of $\tau_2$ with $\tau_1$ fixed.  We will consider $s$ in a neighborhood of $s=1$, say $(1-\epsilon, 1+\epsilon)$.  Then as a function of $\tau_2$, it is readily checked that $\mathcal{F}_s(\tau_1, \tau_2)$ is $L^2(SL(2;\mathbb{Z})\setminus \mathbb{H})$ integrable for $s \in (1-\epsilon,1) \cup (1, 1+ \epsilon)$.  Note that $\mathcal{F}_s$ is not $L^2(SL(2;\mathbb{Z})\setminus \mathbb{H})$ integrable at $s = 1$ itself.  On the other hand, $\Delta_{\tau_2} \mathcal{F}_s(\tau_1, \tau_2)$ is $L^2(SL(2;\mathbb{Z})\setminus \mathbb{H})$ integrable for $s \in (1-\epsilon, 1 + \epsilon)$, notably including $s = 1$.  The $L^2(SL(2;\mathbb{Z})\setminus \mathbb{H})$ integrability conditions are checked most easily by noting that $\mathcal{F}_s$ and $\Delta_{\tau_2} \mathcal{F}_s$ are both bounded functions of $\tau_2$, and by using a special case of H\"{o}lder's inequality, namely $\|f\|_2^2 \leq \|f\|_1 \|f\|_\infty$.

Since both $\Delta_{\tau_2} \mathcal{F}_s$ and the Green's function of $\Delta - \lambda$ are $L^2(SL(2;\mathbb{Z})\setminus \mathbb{H})$ integrable, this implies that
\begin{equation}
\mathcal{F}_s(\tau_1, \tau_2) - \langle \mathcal{F}_s\,, f_0 \rangle \,f_0
\end{equation}
has a uniformly and absolutely convergent Roelcke-Selberg expansion (i.e., an expansion in the eigenfunctions of $\Delta$, see \cite{terras2013harmonic}) for $s \in (1-\epsilon, 1 + \epsilon)$.  Furthermore, the fact that $\mathcal{F}_s$ is $L^2(SL(2;\mathbb{Z})\setminus \mathbb{H})$ integrable away from $s = 1$ implies that all of the singular behavior in $\mathcal{F}_s$ is due to $\langle \mathcal{F}_s\,, f_0\rangle \,f_0$ at $s = 1$, i.e.~the singular behavior is contained in the projection onto the zero mode.

It remains to compute $\langle \mathcal{F}_s\,, f_0\rangle \,f_0$ and examine the singular behavior at $s = 1$.  We have
\begin{align}
\begin{split}
	\langle \mathcal{F}_s\,, f_0\rangle \,f_0 &= \left(\int_{SL(2;\mathbb{Z}) \setminus \mathbb{H}}  \mathcal{F}_s(\tau_1, \tau_2) \, \overline{f_0(\tau_2)} \, \text{Im}(\tau_2)^{-2} \,d\tau_2 \right) \, f_0(\tau_2) 
	\\
	&= \frac{3}{\pi} \int_{\mathbb{H}} \left(\frac{\text{Im}(\tau_1) \text{Im}(\tau_2)}{|\tau_1 + \tau_2|^2}\right)^s \, \text{Im}(\tau_2)^{-2} \,d\tau_2 
	\\
	&= \frac{3}{\sqrt{\pi}} \frac{\Gamma(s-1) \Gamma(s-1/2)}{\Gamma(2s-1)}\,.
\end{split}
\end{align}
As claimed, the above has a simple pole at $s = 1$ with residue $3$, clearly independent of $\tau_1$ and $\tau_2$.  Since $\mathcal{F}_s(\tau_1, \tau_2) - \langle \mathcal{F}_s\,, f_0 \rangle \,f_0$ is non-singular at $s = 1$, we see that
\begin{equation}
\text{Res}_{s = 1} \mathcal{F}_s(\tau_1, \tau_2) = 3\,,
\end{equation}
agreeing with the analysis in Subsection~\ref{S:zeroMomentum}.  This motivates the definition of the regularized Zeta function
\begin{equation}
\widetilde{\mathcal{F}}_s(\tau_1, \tau_2) = \mathcal{F}_s(\tau_1, \tau_2) - \frac{3}{s-1}
\end{equation}
which is non-singular at $s = 1$.  Then $\widetilde{\mathcal{F}}_{s=1}(\tau_1, \tau_2)$ provides a natural regularization of $\mathcal{F}(\tau_1, \tau_2)$, as desired.

Note that when performing Zeta regularization, the choice of regularized Zeta function is often ambiguous up to an additive constant.  The present situation is no different.  As an explicit example, suppose instead of the Zeta function $\mathcal{F}_{s}(\tau_1, \tau_2)$ we chose
\begin{equation}
e^{\frac{K}{3}(s-1)} \mathcal{F}_{s}(\tau_1, \tau_2)
\end{equation}
for a constant $K$.  This function still has a simple pole at $s = 1$ with residue $3$, and so we can regularize the behavior at $s = 1$ by the subtraction
\begin{equation}
\widetilde{\widetilde{\mathcal{F}}}_s(\tau_1, \tau_2) = e^{\frac{K}{3}(s-1)} \mathcal{F}_{s}(\tau_1, \tau_2) - \frac{3}{s+1}\,.
\end{equation}
However, the regularizations $\widetilde{\widetilde{\mathcal{F}}}_s$ and $\widetilde{\mathcal{F}}_s$ disagree by an additive constant at $s=1$, in particular
\begin{equation}
\lim_{s \to 1} \left( \widetilde{\widetilde{\mathcal{F}}}_s - \widetilde{\mathcal{F}}_s\right) = K\,.
\end{equation}
As usual, this ambiguity is unphysical.

\section{Double modular sum and $SL(2;\mathbb{Z})$}
\label{App:FourierSum}

Here we derive Eq.~\eqref{E:FourierFsimp2} from Section~\ref{S:ModularSum}.  We begin with
\begin{equation}
\label{Eq:SL2ZsumApp}
\widetilde{F}_{s_1,s_2} = \sum_{\gamma \in PSL(2;\mathbb{Z})} \int_0^1 dz_1 \int_0^1 dw_1\,e^{2\pi i z_1 s_1 +2\pi i w_1 s_2} f(z,\gamma w)
\end{equation}
where
\begin{equation}
f(z,w) = \frac{\text{Im}(z) \text{Im}(w)}{|z + w|^2}\,.
\end{equation}
We have the important property that for any $\gamma \in PSL(2;\mathbb{Z})$, $f(z, w) = f(\gamma z, \gamma^{-1} w)$.  As mentioned in Section~\ref{S:ModularSum}, a general element of $SL(2;\mathbb{Z})$ can be represented by an integer matrix
\begin{equation}
\label{Eq:SL2Zcond1}
\gamma = \begin{pmatrix}
a & & b \\ c & & d
\end{pmatrix} \quad \text{such that } a d - b c = 1\,.
\end{equation}
Then a general element of $PSL(2;\mathbb{Z})$ can be written as the restriction to $c \geq 0$, and when $c = 0$ we have $a = 1$.  The matrix $\gamma$ acts on a $\tau$ in $\mathbb{H}$ by $\gamma \tau = \frac{a \tau + b}{c \tau + d}$.  Notice that this is invariant under $\gamma \to - \gamma$.

Upon inspecting~\eqref{Eq:SL2Zcond1}, we observe that $\text{gcd}(a,b) = \text{gcd}(a,c) = \text{gcd}(b,d) = \text{gcd}(c,d) = 1$.  For example, suppose $\text{gcd}(a,b) = k$.  Then we could write $a = k a'$ and $b = k b'$ for some integers $a',b'$.  But then $ad - bc = k(a' d - b' c)$ has the form $k \cdot (\text{integer})$ which can only equal one if $k = 1$.  There are additional properties of the $a,b,c,d$ that are useful.  For instance, since $ad - bc = 1$, we have $ad \equiv 1\text{ (mod c)}$.  That is, $a$ and $d$ must be multiplicative inverses modulo $c$.  Regarding $a$ as the multiplicative inverse of $d$, we write $a = [d^{-1}]_c + c n$ where $0 \leq [d^{-1}]_c \leq c-1$ and $n$ is an integer.  Then $ad - bc = [d^{-1}]_c \, d + c d n - b c$.  But $[d^{-1}]_c \, d = 1 + c [r]_{c,d}$ for some fixed integer $[r]_{c,d}$\,, and so $a d - b c = 1 + c ([r]_{c,d} + n d - b) = 1$.  Solving for $b$, we find $b = [r]_{c,d} + n d$, and so a general $PSL(2;\mathbb{Z})$ element takes the form
\begin{equation}
\label{E:PSL2Zgeneralform}
\begin{pmatrix}
[d^{-1}]_c + n c & & [r]_{c,d} + n d \\ c & & d
\end{pmatrix}\,.
\end{equation}
That is, we can specify an element of $PSL(2,\mathbb{Z})$ by coprime $c,d$ with $c \geq 0$, and an integer $n$.

It is prudent to further massage~\eqref{E:PSL2Zgeneralform} for our purposes.  Let $(\mathbb{Z}/c\mathbb{Z})^*$ denote the residue classes mod $c$ which are multiplicatively invertible.  Specifically, we take $(\mathbb{Z}/c\mathbb{Z})^*$ to denote the set of integers from $1$ to $c-1$ which are coprime to $c$.  Given coprime $c,d$, we can decompose $d$ as $d = d' + m c$ where $d' \in (\mathbb{Z}/c\mathbb{Z})^*$ and $m$ is an integer.  Noting that $[(d' + m c)^{-1}]_c = [d'^{-1}]_c$ and $[r]_{c,d' + mc} = [r]_{c,d'} + m [d'^{-1}]_c$\,, we can represent a general $PSL(2;\mathbb{Z})$ element by
\begin{equation}
\gamma_{c,d',m,n} = \begin{pmatrix}
[d'^{-1}]_c + n c & & [r]_{c,d'} + m [d'^{-1}]_c + n (d' + m c) \\ c & & d' + m c
\end{pmatrix}
\end{equation}
where $c \geq 0$, $d' \in (\mathbb{Z}/c\mathbb{Z})^*$, and $m,n \in \mathbb{Z}$.

Translations $\tau \to \tau + 1$ are implemented by
\begin{equation}
T = \begin{pmatrix}
1 && 1 \\ 0 && 1
\end{pmatrix}
\end{equation}
and its powers.  In fact, we have
\begin{align}
\gamma_{c,d',m,n} &= T^n \cdot \begin{pmatrix}
[d'^{-1}]_c & & [r]_{c,d'} \\ c & & d' 
\end{pmatrix} \cdot T^m \nonumber \\
&= T^n \cdot \gamma_{c,d',0,0} \cdot T^m\,.
\end{align}
In light of the above decomposition, to ease the notation we denote $\gamma_{c,d',0,0}$ by $\gamma_{c,d'}$.  It is easy to check that the complete set $PSL(2;\mathbb{Z})$ elements (i.e., without duplication) is given by
\begin{equation}
\label{E:PSL2Zset}
PSL(2;\mathbb{Z}) = \{T^n\}_{n=1}^\infty \cup \{T^n \cdot \gamma_{c,d'} \cdot T^m\}_{c \geq 1,\, d' \in (\mathbb{Z}/c\mathbb{Z})^*,\,m,n\in \mathbb{Z}}\,.
\end{equation}
Using~\eqref{E:PSL2Zset}, we have
\begin{align}
\widetilde{F}_{s_1,s_2} &= \sum_{\gamma \in PSL(2;\mathbb{Z})} \int_0^1 dz_1 \int_0^1 dw_1\,e^{2\pi i z_1 s_1 +2\pi i w_1 s_2} f(z,\gamma w) \nonumber \\
&= \sum_{n \in \mathbb{Z}} \int_0^1 dz_1 \int_0^1 dw_1\,e^{2\pi i z_1 s_1 +2\pi i w_1 s_2} f(z,T^n w) \nonumber \\
& \qquad \qquad + \sum_{m,n\in \mathbb{Z}} \sum_{c \geq 1,\, d' \in (\mathbb{Z}/c\mathbb{Z})^*} \int_0^1 dz_1 \int_0^1 dw_1\,e^{2\pi i z_1 s_1 +2\pi i w_1 s_2} f(z,(T^n \cdot \gamma_{c,d} \cdot T^m) w)\,.
\end{align}
Since the first term is equivalent to
\begin{equation}
\sum_{n \in \mathbb{Z}} \int_0^1 dz_1 \int_0^1 dw_1\,e^{2\pi i z_1 s_1 +2\pi i w_1 s_2} f(z,w + n)
\end{equation}
and the second term is equivalent to
\begin{align}
&\sum_{m,n\in \mathbb{Z}} \sum_{c \geq 1,\, d' \in (\mathbb{Z}/c\mathbb{Z})^*} \int_0^1 dz_1 \int_0^1 dw_1\,e^{2\pi i z_1 s_1 +2\pi i w_1 s_2} f(T^{n} z, (\gamma_{c,d} \cdot T^m) w) \nonumber \\
=\,& \sum_{c \geq 1,\, d' \in (\mathbb{Z}/c\mathbb{Z})^*} \int_{-\infty}^\infty dz_1 \int_{-\infty}^\infty dw_1\,e^{2\pi i z_1 s_1 +2\pi i w_1 s_2} f(z, \gamma_{c,d}  w)\,,
\end{align}
we have derived
\begin{align}
\widetilde{F}_{s_1,s_2} &= \sum_{n \in \mathbb{Z}} \int_0^1 dz_1 \int_0^1 dw_1 \, e^{2 \pi i z_1 s_1 + 2\pi i w_1 s_2}f(z, w + n) \nonumber \\
& \qquad + \sum_{c \geq 1,\,\,d \in (\mathbb{Z}/c\mathbb{Z})^*} \int_{-\infty}^\infty dz_1 \int_{-\infty}^{\infty} dw_1\,e^{2\pi i z_1 s_1 +2\pi i w_1 s_2} f(z,\gamma_{c,d} w)
\end{align}
which is Eq.~\eqref{E:FourierFsimp2} from Section~\ref{S:ModularSum}.

\bibliographystyle{JHEP}
\bibliography{refs}

\end{document}